\documentclass[11pt,english,preprint,aps,prd,showpacs,superscriptaddress,nofootinbib,tightenlines]{revtex4}
\usepackage{babel}
\usepackage{array}
\usepackage{multirow}
\usepackage{amsmath}
\usepackage{amssymb}
\usepackage{graphicx}
\usepackage{amsfonts}
\usepackage{mathrsfs}
\usepackage{epstopdf}
\usepackage{bm}
\usepackage{bbm}
\usepackage{color}
\usepackage{diagbox}
\usepackage{float}
\usepackage{braket}
\usepackage{upgreek}
\usepackage{subfigure}
\usepackage{diagbox}
\usepackage{slashed}
\usepackage[figureright]{rotating}

\makeatletter

\providecommand{\tabularnewline}{\\}

\def\OMIT#1{}

\def\hlinew#1{%
  \noalign{\ifnum0=`}\fi\hrule \@height #1 \futurelet
   \reserved@a\@xhline}

\usepackage{tikz}
\usetikzlibrary {shapes.misc}
\usetikzlibrary{arrows.meta}
\usetikzlibrary{calc}
\usetikzlibrary{
  decorations.pathmorphing,
  decorations.pathreplacing,
  decorations.markings
}
\usetikzlibrary{patterns}

\tikzset{
  every picture/.style={semithick, line cap=round},
  scalar/.style={dashed},
  fermion/.default=0.5,
  fermion/.style={postaction={decorate, decoration={
    markings,
    mark=at position #1 with {\arrow{Stealth[angle=30:7pt,inset=1.5pt]}},
    transform={xshift={3.5pt*cos(15)}}
  }}},
  antifermion/.default=0.5,
  antifermion/.style={postaction={decorate, decoration={
    markings,
    mark=at position #1 with {\arrowreversed{Stealth[angle=30:7pt,inset=1.5pt]}},
    transform={xshift={-3.5pt*cos(15)}}
  }}},
  gluon/.default=3pt,
  gluon/.style={decorate, decoration={
    coil,
    amplitude=0.5*#1,
    aspect=1,
    segment length=#1
  }},
  rgluon/.default=3pt,
  rgluon/.style={decorate, decoration={
    coil,
    amplitude=-0.5*#1,
    aspect=-1,
    segment length=#1
  }},
  gluonpre/.default=0pt,
  gluonpre/.style={decorate, decoration={
    coil,
    amplitude=1.5pt,
    aspect=1,
    segment length=3pt,
    pre length=#1
  }},
  rgluonpre/.default=0pt,
  rgluonpre/.style={decorate, decoration={
    coil,
    amplitude=-1.5pt,
    aspect=-1,
    segment length=3pt,
    pre length=#1
  }},
  crossmark/.style={cross out, draw=red, inner sep=2pt},
  counter/.style={path picture={
    \draw (path picture bounding box.south east) --
      (path picture bounding box.north west)
      (path picture bounding box.south west) --
      (path picture bounding box.north east);
  }},
  cnode/.default=8pt,
  cnode/.style={inner sep=0pt, minimum size=#1, circle}
}

\makeatother

\begin{document}

\title{%
Final-state rescattering mechanism of doubly-charmed baryon decays: $\mathcal{B}_{cc}\to\mathcal{B}_{c}V$}

\author{Xiao-Hui Hu}

\thanks{huxiaohui@cumt.edu.cn, corresponding author}

\affiliation{The College of Materials and physics, China University of mining
and technology, Xuzhou 221116, China \vspace{0.2cm}
 }

\affiliation{Lanzhou Center for Theoretical Physics, Key Laboratory of Theoretical Physics of Gansu Province, Key Laboratory of Quantum Theory and Applications of MoE, Gansu Provincial Research Center for Basic Disciplines of Quantum Physics, Lanzhou University, Lanzhou 730000, China
\vspace{0.2cm}
 }

\author{Fu-Sheng Yu}
\thanks{yufsh@lzu.edu.cn, corresponding author}

\affiliation{Frontiers Science Center for Rare Isotopes, and School of Nuclear
Science and Technology, Lanzhou University, Lanzhou 730000, China\vspace{0.2cm}
 }



\author{Ye Xing}

\thanks{xingye\_guang@cumt.edu.cn, corresponding author}

\affiliation{The College of Materials and physics, China University of mining
and technology, Xuzhou 221116, China \vspace{0.2cm}
 }

\date{\today}

\begin{abstract}
We study the non-leptonic weak decays of doubly charmed baryons (${\cal B}_{cc}$) into singly charmed baryons (${\cal B}_c$) and vector mesons ($V$), denoted as ${\cal B}_{cc}\to{\cal B}_{c}V$. The short-distance contributions are calculated within the naive factorization hypothesis, while the long-distance final-state interaction effects are modeled via hadronic triangle diagrams. Unlike previous approaches, which compute only the imaginary part using the Cutkosky cutting rule, we evaluate the complete loop integrals to obtain both the real and imaginary parts of the amplitudes. These provide the nontrivial strong phases essential for CP violation. The model parameters are determined using experimental data. With this improved calculation method, we predict the branching ratios and decay asymmetry parameters for various decay channels, as well as $CP$ violations for short-distance dominated and singly Cabibbo-suppressed channels. This strengthens our theoretical framework for future study of doubly charmed baryons. Certain decays, primarily driven by long-distance effects, have been calculated; their observation in future experiments could help clarify the role of final-state interactions in charm baryon decays.
Therefore, our calculation of ${\cal B}_{cc}\to{\cal B}_{c}V$ provides crucial predictions for branching ratios, decay asymmetry parameters, and $CP$ violation, which are essential for guiding experimental study at LHCb.

\end{abstract}
\maketitle
\section{Introduction}
The doubly heavy baryons with two heavy quarks ($b$ or $c$ quark), predicted by the quark model, have attracted extensive attention in both theoretical and experimental studies. In 2002, the SELEX collaboration initially reported the detection of $\Xi_{cc}^{+}$ with an unexpected short lifetime and a relatively large production rate~\cite{Mattson:2002vu,SELEX:2004lln}. However, subsequent experiments including FOCUS~\cite{Ratti:2003ez}, BaBar~\cite{Aubert:2006qw}, Belle~\cite{Chistov:2006zj}, and LHCb~\cite{Aaij:2013voa} have not confirmed this discovery, which has remained a longstanding puzzle in experiments. In 2017, the LHCb Collaboration announced the discovery of $\Xi_{cc}^{++}(3621)$ in the invariant mass spectrum of $\Lambda_{c}^{+}K^{-}\pi^{+}\pi^{+}$~\cite{Aaij:2017ueg}. The existence of $\Xi_{cc}^{++}$ was further confirmed through the decay channel $\Xi_{cc}^{++}\to\Xi_{c}^{+}\pi^{+}$~\cite{Aaij:2018gfl}. LHCb has performed precise measurements of its mass and lifetime~\cite{LHCb:2018zpl,LHCb:2019qed}, as well as the branching fraction ratio $\mathcal{B}(\Xi_{cc}^{++}\to\Xi_{c}^{\prime+}\pi^{+})/\mathcal{B}(\Xi_{cc}^{++}\to\Xi_{c}^{+}\pi^{+})=(1.41\pm0.17\pm0.10)$~\cite{LHCb:2022rpd}. Through the continuous search by the LHCb Collaboration~\cite{He:2024ehg,LHCb:2021eaf,LHCb:2019gqy}, the first observation of the singly charged doubly charmed baryon $\Xi_{cc}^{+}$ has been reported through its decay $\Xi_{cc}^{+}\to\Lambda_{c}^{+}K^{-}\pi^{+}$ by LHCb~\cite{LHCb:2026pxn} this year, with a statistical significance exceeding seven standard deviations.
The study of $\Omega_{cc}^{+}$ is also ongoing, with LHCb searching for its decay signals~\cite{LHCb:2021rkb}.
Other doubly heavy baryons, such as $\Xi_{bc}^{+}$~\cite{LHCb:2022fbu}, $\Xi_{bc}^{0}$~\cite{LHCb:2020iko,LHCb:2021xba} and $\Omega_{bc}^{0}$~\cite{LHCb:2021xba}, have also been investigated by the LHCb Collaboration, though no signals have been detected to date. The doubly charmed tetraquark $T_{cc}^{+}(3875)$ was identified by the LHCb Collaboration~\cite{LHCb:2021vvq,LHCb:2021auc}, which may provide insights into the nature of doubly heavy systems.

Prior to the experimental discovery of $\Xi_{cc}^{++}$, theoretical studies had already suggested that the most likely discovery channels for these doubly charmed baryons would be $\Xi_{cc}^{++}\to\Lambda_{c}^{+}K^{-}\pi^{+}\pi^{+}$ and $\Xi_{cc}^{++}\to\Xi_{c}^{+}\pi^{+}$~\cite{Yu:2017zst}. Therefore, pre-theoretical studies of doubly charmed baryon decays are crucial for experimental searches.
Theoretical investigations on doubly heavy baryons have been extensive, covering production mechanisms, mass spectra, and both strong and weak decays~\cite{Kiselev:2001fw,Ebert:1996ec,Tong:1999qs,Ebert:2002ig,Gershtein:2000nx,Roberts:2007ni,Valcarce:2008dr,Lu:2017meb,Yu:2022lel,Li:2017ndo,Yu:2017zst,Ebert:2004ck,Roberts:2008wq,Branz:2010pq,Albertus:2010hi,Qin:2021zqx,Bahtiyar:2018vub,Chen:2016spr,Cheng:2021qpd,Chen:2020aos,Shi:2019fph,Hu:2019bqj,Hu:2022xzu,Aliev:2022tvs,Aliev:2022maw}. However, studies on non-leptonic weak decays remain relatively limited compared to semileptonic decays. This is mainly due to the large number of non-leptonic decay channels and the presence of non-factorizable contributions that are difficult to calculate using QCD methods.

In our previous work, we investigate the decays $\mathcal{B}_{cc}\to\mathcal{B}_{c}P$ the final states interaction approach. In order to study the weak decay of doubly charmed baryon systemly and select out the golden mode for the Research of doubly charmed baryon, in this work, we investigate the decay $\mathcal{B}_{cc}\to\mathcal{B}_{c}V$ with vector mesons in the final state. The vector meson final states are especially important for several reasons: (i) Vector mesons can be reconstructed efficiently through their decay products such as $\rho\to\pi^{+}\pi^{-}$, $K^{*}\to K^{+}\pi^{-}$, $\omega\to\pi^{+}\pi^{-}\pi^{0}$, and $\phi\to K^{+}K^{-}$, which have high detection efficiencies in experiments; (ii) The angular distributions of vector mesons provide additional information about the decay dynamics through decay asymmetry parameters; (iii) Understanding the $\mathcal{B}_{cc}\to\mathcal{B}_{c}V$ decays is essential for predicting the discovery channels of yet-unobserved doubly charmed baryons  $\Omega_{cc}^{+}$.

In this paper, we study the non-leptonic weak decays $\mathcal{B}_{cc}\to\mathcal{B}_{c}V$ for doubly charmed baryons $\mathcal{B}_{cc}=(\Xi_{cc}^{++},\Xi_{cc}^{+},\Omega_{cc}^{+})$, where $\mathcal{B}_c$ denotes singly charmed baryons including both anti-triplet ($\Lambda_c,\Xi_c$) and sextet ($\Sigma_c,\Xi_c',\Omega_c$) states, and $V$ represents light vector mesons ($\rho,K^{*},\omega,\phi$). We calculate branching ratios and decay asymmetry parameters for 49 decay channels.
For the specific decay mode $\mathcal{B}_{cc}\to\mathcal{B}_{c}V$, there is currently one dedicated theoretical study in the literature~\cite{Jiang:2018oak}.
Compared with the previous study~\cite{Jiang:2018oak}, we improve the calculation method for long-distance contributions by utilizing the complete analytical expression of loop integrals rather than only the imaginary parts from the Cutkosky cutting rules. This allows us to obtain both the magnitude and strong phase of the triangle diagrams. The strong phases are essential for calculating decay asymmetry parameters. We adopt the Pauli-Villars scheme with a momentum cutoff to regularize the divergence in loop integrals. 

This paper is organized as follows. In Section~\ref{sec:framework}, we present the topological analysis, the short-distance factorization calculation, and the long-distance rescattering mechanism. Section~\ref{sec:results} contains the numerical results and discussions, including input parameters, branching ratios, and decay asymmetry parameters. A summary is given in Section~\ref{sec:summary}. Details of the decay amplitudes and strong couplings are provided in the Appendices.

\section{Theoretical framework}
\label{sec:framework} 
In this section, we will present a topological
analysis of each nonleptonic weak decay channel of doubly charmed
baryons, as detailed in Sec.~\ref{subsec:topo}. Subsequently,
we present the general framework for calculating the short-distance
and long-distance dynamics of all topological diagrams in Sec.~\ref{subsec:short}
and Sec.~\ref{subsec:long}, respectively.

\subsection{Topological analysis}

\label{subsec:topo} 
\begin{figure}[htp]
\includegraphics[width=1\textwidth]{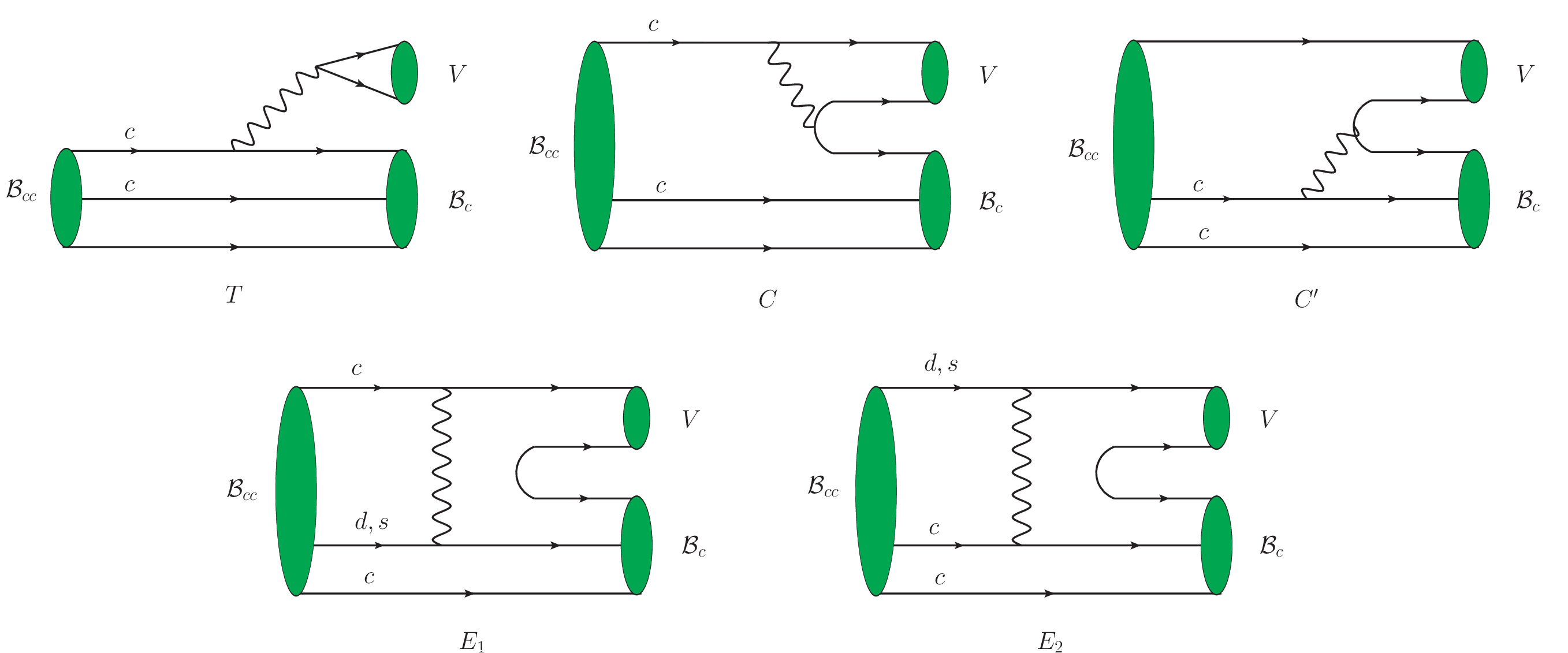} 
\caption{The five tree level topological diagrams for two body non leptonic
decays ${\cal B}_{cc}\to{\cal B}_{c}V$ of the doubly charmed baryons
${\cal B}_{cc}=(\Xi_{cc}^{++},\Xi_{cc}^{+},\Omega_{cc}^{+})$. }
\label{fig:treetopo} 
\end{figure}

In this work, we will investigate the exclusive channels of nonleptonic
weak decay of doubly charmed baryons $\mathcal{B}_{cc}\to\mathcal{B}_{c}V$.
The initial state is characterized by a doubly charmed baryon triplet
$\mathcal{B}_{cc}=(\Xi_{cc}^{++},\Xi_{cc}^{+},\Omega_{cc}^{+})$,
while the final state encompasses the singly charmed baryon $\mathcal{B}_{c}=(\mathcal{B}_{\bar{3}},\mathcal{B}_{6})$,
and the light vector mesons $V=(\rho,K^{*},\omega,\phi)$, as shown
in Appendix~\ref{app:strong}. Based on the symmetry analysis of the
initial and final states of particles, we will study forty-nine decay
processes.

The tree-level contributions to $\mathcal{B}_{cc}\to\mathcal{B}_cV$ decays (Fig.~\ref{fig:treetopo}) consist of both short-distance and long-distance parts. The five diagrams are classified into three types: $T$ (color-favored external $W$-emission), $C$ and $C'$ (color-suppressed internal $W$-emission), and $E_1$ and $E_2$ (two $W$-exchange topologies). The distinction between $C$ and $C'$ lies in whether both quarks of the final light meson originate from the weak vertex (only in $C$). In $E_1$, the light quark from the charmed quark decay goes to the final light meson; in $E_2$, it goes to the final charmed baryon.

The topological amplitudes for each channel are summarized in Tab.~\ref{tab:topologies}. Those involving sextet single-charmed baryons are denoted with a tilde (e.g., $\tilde{T}$). Short-distance dominated modes are listed first; the remaining three parts correspond to long-distance dominated modes, grouped by CKM matrix elements: Cabibbo-favored ($c\to su\bar{d}$), singly Cabibbo-suppressed ($c\to du\bar{d}$ or $c\to su\bar{s}$), and doubly Cabibbo-suppressed ($c\to du\bar{s}$). The $T$ diagram, dominated by factorizable short-distance contributions~\cite{Lu:2009cm}, can be computed using factorization. The $C(C')$ diagram is strongly suppressed at the charm scale due to the color factor $a_2(m_c)=C_1(m_c)+C_2(m_c)/N_c$~\cite{Yu:2017zst}. For $E_1(E_2)$, long-distance contributions exceed short-distance ones, which are suppressed by at least an order of magnitude~\cite{Lu:2009cm}. However, non-factorizable long-distance effects may significantly influence $C(C')$ and thus the branching ratios, warranting further study. Our model incorporates such effects via final-state interactions mediated by loop mechanisms, allowing improved agreement with experiment. The calculation of factorizable short-distance and non-factorizable long-distance contributions is detailed in the following subsections.

\begin{table}
\caption{The topological analysis of each nonleptonic weak decay channel.}
\label{tab:topologies} %
\tiny
\begin{tabular}{|llllll|}
\hline \hline 
Decay  & Topology  & Decay  & Topology  & Decay  & Topology \tabularnewline
\hline 
$\Xi_{cc}^{++}\to\Xi_{c}^{+}\rho^{+}$  & $\lambda_{sd}(T+C^{\prime})$  & $\Xi_{cc}^{+}\to\Xi_{c}^{0}\rho^{+}$  & $\lambda_{sd}(T-E_{2})$  & $\Omega_{cc}^{+}\to\Omega_{c}^{0}\rho^{+}$  & $\lambda_{sd}\tilde{T}$ \tabularnewline
$\Xi_{cc}^{++}\to\Xi_{c}^{\prime+}\rho^{+}$  & $\frac{1}{\sqrt{2}}\lambda_{sd}\big(\tilde{T}+\tilde{C}^{\prime}\big)$  & $\Xi_{cc}^{+}\to\Xi_{c}^{\prime0}\rho^{+}$  & $\frac{1}{\sqrt{2}}\lambda_{sd}\big(\tilde{T}+\tilde{E}_{2}\big)$  & $\Omega_{cc}^{+}\to\Xi_{c}^{0}\rho^{+}$  & $-\lambda_{d}T-\lambda_{s}E_{2}$ \tabularnewline
$\Xi_{cc}^{++}\to\Sigma_{c}^{+}\rho^{+}$  & $\frac{1}{\sqrt{2}}\lambda_{d}\big(\tilde{T}+\tilde{C}^{\prime}\big)$  & $\Xi_{cc}^{+}\to\Sigma_{c}^{0}\rho^{+}$  & $\lambda_{d}(\tilde{T}+\tilde{E}_{2})$  & $\Omega_{cc}^{+}\to\Xi_{c}^{\prime0}\rho^{+}$  & $\frac{1}{\sqrt{2}}\big(\lambda_{d}\tilde{T}+\lambda_{s}\tilde{E}_{2}\big)$ \tabularnewline
$\Xi_{cc}^{++}\to\Lambda_{c}^{+}\rho^{+}$  & $\lambda_{d}(T+C^{\prime})$  & $\Xi_{cc}^{+}\to\Xi_{c}^{\prime0}K^{*+}$  & $\frac{1}{\sqrt{2}}\big(\lambda_{s}\tilde{T}+\lambda_{d}\tilde{E}_{2}\big)$  & $\Omega_{cc}^{+}\to\Omega_{c}^{0}K^{*+}$  & $\lambda_{s}\big(\tilde{T}+\tilde{E}_{2}\big)$ \tabularnewline
$\Xi_{cc}^{++}\to\Xi_{c}^{\prime+}K^{*+}$  & $\frac{1}{\sqrt{2}}\lambda_{s}\big(\tilde{T}+\tilde{C}^{\prime}\big)$  & $\Xi_{cc}^{+}\to\Xi_{c}^{0}K^{*+}$  & $\lambda_{s}T+\lambda_{d}E_{2}$  & $\Omega_{cc}^{+}\to\Xi_{c}^{0}K^{*+}$  & $\lambda_{ds}(-T+E_{2})$ \tabularnewline
$\Xi_{cc}^{++}\to\Xi_{c}^{+}K^{*+}$  & $\lambda_{s}(T+C^{\prime})$  & $\Xi_{cc}^{+}\to\Sigma_{c}^{0}K^{*+}$  & $\lambda_{ds}\tilde{T}$  & $\Omega_{cc}^{+}\to\Xi_{c}^{\prime0}K^{*+}$  & $\frac{1}{\sqrt{2}}\lambda_{ds}\big(\tilde{T}+\tilde{E}_{2}\big)$ \tabularnewline
$\Xi_{cc}^{++}\to\Lambda_{c}^{+}K^{*+}$  & $\lambda_{ds}(T+C^{\prime})$  & $\Xi_{cc}^{++}\to\Sigma_{c}^{+}K^{*+}$  & $\frac{1}{\sqrt{2}}\lambda_{ds}\big(\tilde{T}+\tilde{C}^{\prime}\big)$  &  & \tabularnewline
\hline 
$\Xi_{cc}^{++}\to\Sigma_{c}^{++}\bar{K}^{*0}$  & $\tilde{C}$  & $\Xi_{cc}^{+}\to\Xi_{c}^{\prime+}\rho^{0}$  & $\frac{1}{2}(-\tilde{C}^{\prime}+\tilde{E}_{2})$  & $\Xi_{cc}^{+}\to\Xi_{c}^{+}\omega$  & $\frac{1}{\sqrt{6}}(C^{\prime}-2E_{1}-E_{2})$ \tabularnewline
$\Xi_{cc}^{+}\to\Omega_{c}^{0}K^{*+}$  & $\tilde{E}_{2}$  & $\Xi_{cc}^{+}\to\Xi_{c}^{\prime+}\phi$  & $\frac{1}{\sqrt{6}}(\tilde{C}^{\prime}+\tilde{E}_{1}+\tilde{E}_{2})$  & $\Omega_{cc}^{+}\to\Xi_{c}^{\prime+}\bar{K}^{*0}$  & $\frac{1}{\sqrt{2}}(\tilde{C}+\tilde{C}^{\prime})$ \tabularnewline
$\Xi_{cc}^{+}\to\Sigma_{c}^{+}\bar{K}^{*0}$  & $\frac{1}{\sqrt{2}}\big(\tilde{C}+\tilde{E}_{1}\big)$  & $\Xi_{cc}^{+}\to\Xi_{c}^{\prime+}\omega$  & $\frac{1}{2\sqrt{3}}(\tilde{C}^{\prime}-2\tilde{E}_{1}+\tilde{E}_{2})$  & $\Omega_{cc}^{+}\to\Xi_{c}^{+}\bar{K}^{*0}$  & $-C+C^{\prime}$ \tabularnewline
$\Xi_{cc}^{+}\to\Lambda_{c}^{+}\bar{K}^{*0}$  & $-C+E_{1}$  & $\Xi_{cc}^{+}\to\Xi_{c}^{+}\rho^{0}$  & $-\frac{1}{\sqrt{2}}(C^{\prime}+E_{2})$  &  & \tabularnewline
$\Xi_{cc}^{+}\to\Sigma_{c}^{++}K^{*-}$  & $\tilde{E}_{1}$  & $\Xi_{cc}^{+}\to\Xi_{c}^{+}\phi$  & $\frac{1}{\sqrt{3}}(C^{\prime}+E_{1}-E_{2})$  &  & \tabularnewline
\hline 
{$\Xi_{cc}^{++}\to\Sigma_{c}^{++}\rho^{0}$ }  & {$-\frac{1}{\sqrt{2}}\tilde{C}$}  & {$\Xi_{cc}^{+}\to\Xi_{c}^{\prime+}K^{*0}$ }  & {$\frac{1}{\sqrt{2}}(\lambda_{s}\tilde{C}^{\prime}\!\!+\!\!\lambda_{d}\tilde{E}_{1})$
}  & {$\Omega_{cc}^{+}\to\Sigma_{c}^{++}K^{*-}$ }  & {$\lambda_{s}\tilde{E}_{1}$ }\tabularnewline
{$\Xi_{cc}^{++}\to\Sigma_{c}^{++}\phi$ }  & {$\frac{1}{\sqrt{3}}(\lambda_{d}\!\!+\!\!\lambda_{s})\tilde{C}$}  & {$\Xi_{cc}^{+}\to\Xi_{c}^{+}K^{*0}$ }  & {$\lambda_{s}C^{\prime}\!\!+\!\!\lambda_{d}E_{1}$ }  & {$\Omega_{cc}^{+}\to\Xi_{c}^{\prime+}\rho^{0}$ }  & {$\frac{1}{2}(-\lambda_{d}\tilde{C}\!\!+\!\!\lambda_{s}\tilde{E}_{2})$ }\tabularnewline
{$\Xi_{cc}^{++}\to\Sigma_{c}^{++}\omega$ }  & {$\frac{1}{\sqrt{6}}(\lambda_{d}-2\lambda_{s})\tilde{C}$ }  & {$\Xi_{cc}^{+}\to\Lambda_{c}^{+}\rho^{0}$ }  & {$\frac{\lambda_{d}}{\sqrt{2}}(C\!\!-\!\!C^{\prime}\!\!-\!\!E_{1}\!\!-\!\!E_{2})$
}  & {$\Omega_{cc}^{+}\to\Xi_{c}^{+}\rho^{0}$ }  & {$\frac{1}{\sqrt{2}}(\lambda_{d}C-\lambda_{s}E_{2})$ }\tabularnewline
\multirow{2}{*}{{$\Xi_{cc}^{+}\to\Sigma_{c}^{+}\rho^{0}$ } } & \multirow{2}{*}{$\frac{\lambda_{d}}{2}(\!\!-\!\!\tilde{C}\!\!-\!\!\tilde{C}^{\prime}\!\!+\!\!\tilde{E}_{1}\!\!+\!\!\tilde{E}_{2})$}  & \multirow{2}{*}{{$\Xi_{cc}^{+}\to\Lambda_{c}^{+}\phi$ } } & {$\frac{1}{\sqrt{3}}[\lambda_{d}(C^{\prime}\!\!-\!\!C\!\!+\!\!E_{1}\!\!-\!\!E_{2})$}  & \multirow{2}{*}{{$\Omega_{cc}^{+}\to\Xi_{c}^{\prime+}\phi$ } } & {$\frac{1}{\sqrt{6}}[\lambda_{s}(\tilde{C}\!\!+\!\!\tilde{C}^{\prime}\!\!+\!\!\tilde{E}_{1}\!\!+\!\!\tilde{E}_{2})$ }\tabularnewline
 &  &  & {$\qquad-{\lambda_{s}}C${]}}  &  & {$\qquad+{\lambda_{d}}\tilde{C}${]}} \tabularnewline
\multirow{2}{*}{{$\Xi_{cc}^{+}\to\Sigma_{c}^{+}\phi$ } } & {$\frac{1}{\sqrt{6}}[\lambda_{d}(\tilde{C}\!\!+\!\!\tilde{C}^{\prime}\!\!+\!\!\tilde{E}_{1}\!\!+\!\!\tilde{E}_{2})$}  & \multirow{2}{*}{{$\Xi_{cc}^{+}\to\Lambda_{c}^{+}\omega$ } } & {$\frac{1}{\sqrt{6}}[\lambda_{d}(C^{\prime}\!\!-\!\!C\!\!+\!\!E_{1}\!\!-\!\!E_{2})$
}  & \multirow{2}{*}{{$\Omega_{cc}^{+}\to\Xi_{c}^{\prime+}\omega$ } } & {$\frac{1}{2\sqrt{3}}[\lambda_{s}({\tilde{E}_{2}}\!\!-\!\!2\tilde{C}\!\!-\!\!2\tilde{C}^{\prime}\!\!-\!\!2\tilde{E}_{1})$ }\tabularnewline
 & {$\qquad+{\lambda_{s}}\tilde{C}${]}}  &  & {$\qquad+{2\lambda_{s}}C${]}}  &  & {$\qquad+\lambda_{d}\tilde{C}${]}}\tabularnewline
\multirow{2}{*}{{$\Xi_{cc}^{+}\to\Sigma_{c}^{+}\omega$ } } & {$\frac{1}{2\sqrt{3}}[\lambda_{d}(\tilde{C}\!\!+\!\!\tilde{C}^{\prime}\!\!+\!\!\tilde{E}_{1}\!\!+\!\!\tilde{E}_{2})$
}  & \multirow{2}{*}{{$\Omega_{cc}^{+}\to\Xi_{c}^{+}\omega$ } } & {$\frac{1}{\sqrt{6}}[\lambda_{s}(2C\!\!-\!\!2C^{\prime}\!\!+\!\!2E_{1}\!\!-\!\!E_{2})$ } & \multirow{2}{*}{{$\Omega_{cc}^{+}\to\Xi_{c}^{+}\phi$ } } & {$\frac{1}{\sqrt{3}}[\lambda_{s}(C^{\prime}\!\!-\!\!C\!\!+\!\!E_{1}\!\!-\!\!E_{2})$}\tabularnewline
 & {$\qquad-2{\lambda_{s}}\tilde{C}${]}}  &  &{$ \qquad-{\lambda_{d}}C${]}} &  & {$\qquad-{\lambda_{d}}C${]}}\tabularnewline
{$\Xi_{cc}^{+}\to\Sigma_{c}^{++}\rho^{-}$ }  & {$\lambda_{d}\tilde{E}_{1}$ }  & {$\Omega_{cc}^{+}\to\Sigma_{c}^{+}\bar{K}^{*0}$ }  & {$\frac{1}{\sqrt{2}}(\lambda_{d}\tilde{C}^{\prime}\!\!+\!\!\lambda_{s}\tilde{E}_{1})$
}  & {$\Omega_{cc}^{+}\to\Lambda_{c}^{+}\bar{K}^{*0}$ }  & {$\lambda_{d}C^{\prime}\!\!+\!\!\lambda_{s}E_{1}$ } \tabularnewline\hline
{$\Xi_{cc}^{++}\to\Sigma_{c}^{++}K^{*0}$ }  & {$\tilde{C}$ }  & {$\Omega_{cc}^{+}\to\Sigma_{c}^{+}\omega$ }  & {$\frac{1}{2\sqrt{3}}\big(-2\tilde{C}^{\prime}+\tilde{E}_{1}+\tilde{E}_{2}\big)$
}  & {$\Omega_{cc}^{+}\to\Sigma_{c}^{++}\rho^{-}$ }  & {$\tilde{E}_{1}$ }\tabularnewline
{$\Xi_{cc}^{+}\to\Sigma_{c}^{+}K^{*0}$ }  & {$\frac{1}{\sqrt{2}}\big(\tilde{C}+\tilde{C}^{\prime}\big)$
}  & {$\Omega_{cc}^{+}\to\Lambda_{c}^{+}\rho^{0}$ }  & {$-\frac{1}{\sqrt{2}}(E_{1}+E_{2})$ }  & {$\Omega_{cc}^{+}\to\Xi_{c}^{\prime+}K^{*0}$ }  & {$\frac{1}{\sqrt{2}}\big(\tilde{C}+\tilde{E}_{1}\big)$ }\tabularnewline
{$\Xi_{cc}^{+}\to\Lambda_{c}^{+}K^{*0}$ }  & {$-C+C^{\prime}$ }  & {$\Omega_{cc}^{+}\to\Lambda_{c}^{+}\phi$ }  & {$\frac{1}{\sqrt{3}}(C^{\prime}+E_{1}-E_{2})$ }  & {$\Omega_{cc}^{+}\to\Xi_{c}^{+}K^{*0}$ }  & {$-C+E_{1}$ }\tabularnewline
{$\Omega_{cc}^{+}\to\Sigma_{c}^{+}\rho^{0}$ }  & {$\frac{1}{2}(-\tilde{E}_{1}+\tilde{E}_{2})$ }  & {$\Omega_{cc}^{+}\to\Lambda_{c}^{+}\omega$ }  & {$-\frac{1}{\sqrt{6}}(2C^{\prime}-E_{1}+E_{2})$ }  &  & \tabularnewline
{$\Omega_{cc}^{+}\to\Sigma_{c}^{+}\phi$ }  & {$\frac{1}{\sqrt{6}}\big(\tilde{C}^{\prime}+\tilde{E}_{1}+\tilde{E}_{2}\big)$
}  & {$\Omega_{cc}^{+}\to\Sigma_{c}^{0}\rho^{+}$ }  & {$\tilde{E}_{2}$ }  &  & \tabularnewline
\hline 
\hline
\end{tabular}
\end{table}

\subsection{Calculation of short-distance amplitudes under the factorization
hypothesis}
\label{subsec:short}

At the tree level, the non-leptonic weak decay of doubly charmed baryons
are induced by the decay of the charm quark. The effective Hamiltonian
can be represented as follows, 
\begin{align}
\mathcal{H}_{eff}=\frac{G_{F}}{\sqrt{2}}\sum_{q^{\prime}=d,s}^ {}V_{cq^{\prime}}^{\ast}V_{uq}[C_{1}(\mu)O_{1}(\mu)+C_{2}(\mu)O_{2}(\mu)]+h.c..\label{eq:hamilton}
\end{align}
Here $V_{cq^{\prime}}$ and $V_{uq}$ represent the Cabibbo-Kobayashi-Maskawa
(CKM) matrix elements. And the four-fermion operators $O_{1}$ and
$O_{2}$ can be expressed as, 
\begin{align}
O_{1}=(\bar{u}_{\alpha}q_{\beta})_{V-A}(\bar{q}^{\prime}_{\beta}c_{\alpha})_{V-A},\quad O_{2}=(\bar{u}_{\alpha}q_{\alpha})_{V-A}(\bar{q}^{\prime}_{\beta}c_{\beta})_{V-A},
\end{align}
with the color indices $\alpha$ and $\beta$. And $C_{1,2}(\mu)$
are the relevant Wilson coefficients. After inserting the effective
Hamiltonian as Eq.~(\ref{eq:hamilton}), the amplitude of the non-leptonic
weak decay of doubly charmed baryons $\mathcal{B}_{cc}\to\mathcal{B}_{c}V$
can be evaluated with the hadronic matrix element, 
\begin{align}
\langle\mathcal{B}_{c}V|\mathcal{H}_{eff}|\mathcal{B}_{cc}\rangle=\frac{G_{F}}{\sqrt{2}}V_{cq^{\prime}}^{\ast}V_{uq}\sum_{i=1,2}C_{i}\langle\mathcal{B}_{c}V|O_{i}|\mathcal{B}_{cc}\rangle.\label{eq:hme}
\end{align}
In this work, we neglect the penguin operators for the strong suppression
of CKM matrix elements in charmed hadron decays.
According to the factorization hypothesis, the matrix elements $\langle\mathcal{B}_{c}M|O_{i}|\mathcal{B}_{cc}\rangle$
in Eq.~(\ref{eq:hme}) can be factorized into two parts. The first
part is parameterized with the decay constant of the emitted meson,
while the second part can be evaluated using the heavy-light transition
form factors. The short-distance contribution of the $T$ diagram
can be expressed as: 
\begin{align}
\langle\mathcal{B}_{c}M|\mathcal{H}_{eff}|\mathcal{B}_{cc}\rangle_{SD}^{T}=\frac{G_{F}}{\sqrt{2}}V_{cq^{\prime}}^{\ast}V_{uq}a_{1}(\mu)\langle M|\bar{u}\gamma^{\mu}(1-\gamma_{5})q|0\rangle\langle\mathcal{B}_{c}|\bar{q}^{\prime}\gamma_{\mu}(1-\gamma_{5})c|\mathcal{B}_{cc}\rangle,\label{eq:hadronic1}
\end{align}
here $a_{1}(\mu)=C_{1}(\mu)+C_{2}(\mu)/3$ is the effective Wilson
coefficients. For color suppressed $C$ diagram, its short-distance
contribution can be given via its relation to the $T$ diagram after
Fierz transformation. And under the charm scale $\mu=m_{c}$, the
Wilson coefficients are taken as $C_{1}(\mu)=1.21$ and $C_{2}(\mu)=-0.42$~\cite{Li:2012cfa}.
$M$ represents both pseudoscalar meson($P$) and vector meson($V$).
The vector meson also contributes to long-distance dynamics as an
intermediate state in the next subsections.

By utilizing the heavy-light transition form factors $f_{1,2,3}$
and $g_{1,2,3}$, the transition matrix elements of ${\cal B}_{cc}\to{\cal B}_{c}$
can be effectively parameterized as, 
\begin{eqnarray}
 &  & \langle{\cal B}_{c}(p^{\prime},s_{z}^{\prime})|\bar{q}^{\prime}\gamma_{\mu}(1-\gamma_{5})c|{\cal B}_{cc}(p,s_{z})\rangle\nonumber \\
 &  & =\bar{u}(p^{\prime},s_{z}^{\prime})\Big[\gamma_{\mu}f_{1}(q^{2})+i\sigma_{\mu\nu}\frac{q^{\nu}}{M_{\mathcal{B}_{cc}}}f_{2}(q^{2})+\frac{q^{\mu}}{M_{\mathcal{B}_{cc}}}f_{3}(q^{2})\Big]u(p,s_{z})\nonumber \\
 &  & -\bar{u}(p^{\prime},s_{z}^{\prime})\Big[\gamma_{\mu}g_{1}(q^{2})+i\sigma_{\mu\nu}\frac{q^{\nu}}{M_{\mathcal{B}_{cc}}}g_{2}(q^{2})+\frac{q^{\mu}}{M_{\mathcal{B}_{cc}}}g_{3}(q^{2})\Big]\gamma_{5}u(p,s_{z}).\label{eq:ff}
\end{eqnarray}
Here $q=p-p^{\prime}$, $M_{\mathcal{B}_{cc}}$ is the mass of doubly
charmed baryons.

The first matrix element in Eq.~(\ref{eq:hadronic1}) are defined
using the decay constants of the emitted mesons $P$ and $V$, denoted
by $f_{P}$ and $f_{V}$, respectively. 
\begin{align}
\langle P(p)|\bar{u}\gamma^{\mu}(1-\gamma_{5})q|0\rangle & =-if_{P}p^{\mu},\label{eq:Pdecay}\\
\langle V(p)|\bar{u}\gamma^{\mu}(1-\gamma_{5})q|0\rangle & =m_{V}f_{V}\epsilon^{\ast\mu},\label{eq:Vdecay}
\end{align}
where $\epsilon^{\mu}$ represents the polarization vector of the
vector meson.

Conventionally, the amplitude of two-body non-leptonic
charmed baryon decays is given in terms of partial wave scheme
\begin{equation} \label{eq:NF_BcBP}
\mathcal{A}(\mathcal{B}_{cc}\to\mathcal{B}_{c}P)=i\bar{u}_{f}(A+B\gamma_{5})u_{i} \,,
\end{equation}
specifically, the invariant amplitudes $A$ and $B$ are expressed as the following factorization form by using the form factor $f_{i}$ and $g_{i}$ 
\begin{equation}
A=\lambda f_{P}(m_{i}-m_{f})f_{1}(m^{2}) \,, \quad B=\lambda f_{P}(m_{i}+m_{f})g_{1}(m^{2}) \,.
\end{equation} 
For vector meson, the amplitudes are
\begin{align} \label{eq:NF_BcBV}
\mathcal{A}(\mathcal{B}_{cc}\to\mathcal{B}_{c}V)=\bar{u}_{f}\Big(A_{1}\gamma_{\mu}\gamma_{5}+A_{2}\frac{p_{f\mu}}{m_{i}}\gamma_{5}+B_{1}\gamma_{\mu}+B_{2}\frac{p_{f\mu}}{m_{i}}\Big)\varepsilon^{\ast \mu}u_{i} \,,
\end{align}
where
\begin{align}
 & A_{1}=-\lambda f_{V}m\Big(g_{1}(m^{2})+g_{2}(m^{2})\frac{m_{i}-m_{f}}{m_{i}}\Big),\quad A_{2}=-2\lambda f_{V}mg_{2}(m^{2})\,,\\
 & B_{1}=\lambda f_{V}m\Big(f_{1}(m^{2})-f_{2}(m^{2})\frac{m_{i}+m_{f}}{m_{i}}\Big),\quad B_{2}=2\lambda f_{V}mf_{2}(m^{2})\,,
\end{align}
with $\lambda=\frac{G_{F}}{\sqrt{2}}V^{*}_{cq^{\prime}}V_{uq}a_{1,2}(\mu)$ for simplicity, $m$ is the mass of final meson. 
As we mentioned before, the factorization formula in Eq.\eqref{eq:NF_BcBP} and \eqref{eq:NF_BcBV} are also essential ingredients for calculating the long-distance final-state interaction effects.

\subsection{Calculation of long-distance contributions using the final states
rescattering mechanism}

\label{subsec:long}


\begin{figure}[htbp]
  \centering
  \subfigure[]{\includegraphics[width=0.45\textwidth]{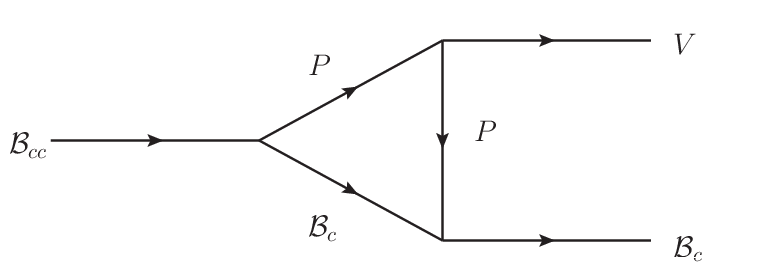}}
  \hfill
  \subfigure[]{\includegraphics[width=0.45\textwidth]{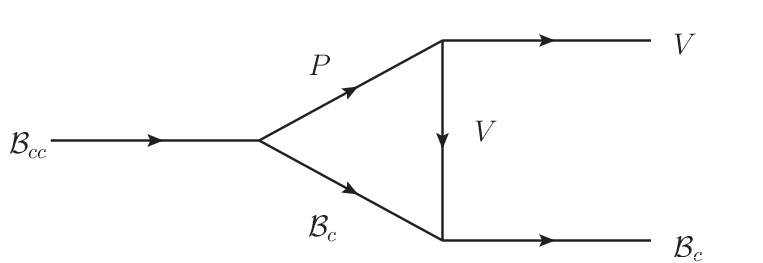}}
      \subfigure[]{\includegraphics[width=0.45\textwidth]{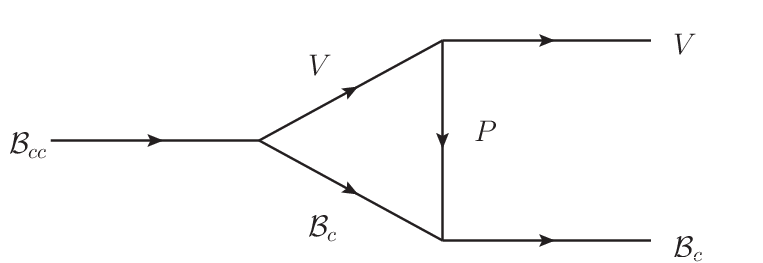}}
  \hfill
  \subfigure[]{\includegraphics[width=0.45\textwidth]{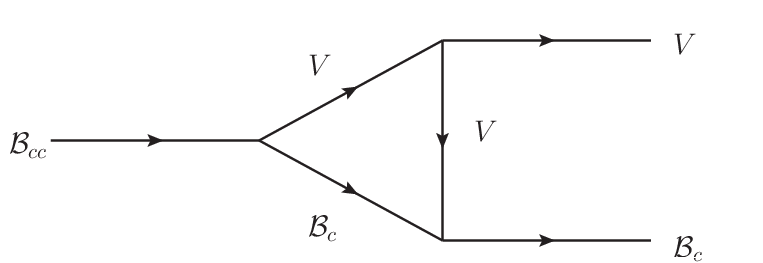}}
    \subfigure[]{\includegraphics[width=0.45\textwidth]{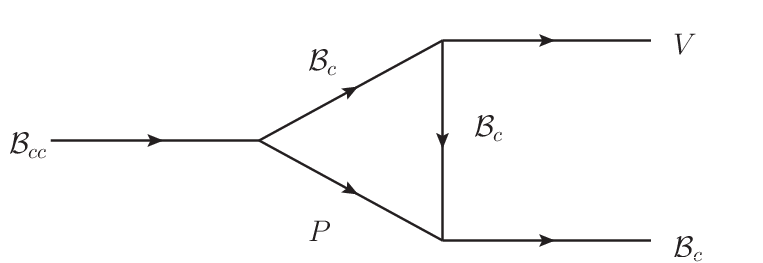}}
  \hfill
  \subfigure[]{\includegraphics[width=0.45\textwidth]{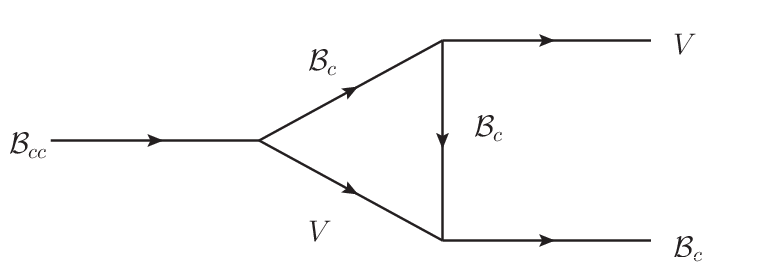}}
      \subfigure[]{\includegraphics[width=0.45\textwidth]{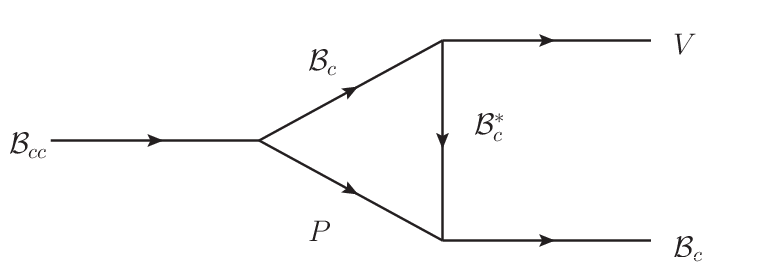}}
  \hfill
  \subfigure[]{\includegraphics[width=0.45\textwidth]{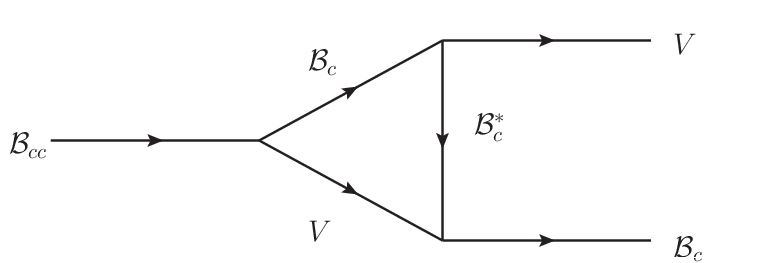}}
\caption{The long-distance rescattering contributions to ${\cal {B}}_{cc}\to{\cal {B}}_{c}P$ manifested at hadron level via singly particle exchange, generating distinct triangle diagrams with intermediate states: (a) $\{P,{\cal {B}}_{c};P\}$,(b) $\{P,{\cal {B}}_{c};V\}$, (c) $\{V,{\cal {B}}_{c};P\}$, (d) $\{V,{\cal {B}}_{c};V\}$, (e) $\{{\cal {B}}_{c},P;{\cal {B}}_{c}\}$, (f) $\{{\cal {B}}_{c},V;{\cal {B}}_{c}\}$, (g) $\{{\cal {B}}_{c},P;{\cal {B}}_{c}^{*}\}$, (h) $\{{\cal {B}}_{c},V;{\cal {B}}_{c}^{*}\}$. And the corresponding amplitudes of these triangle diagrams have been given by Eqs.~(\ref{eq:PBP})-(\ref{eq:VBD}).}
 \label{fig:trianglesev}
\end{figure}

The long-distance contributions are large and difficult to evaluate. 
In this work, we employ the final-state rescattering mechanism, as in Ref.~\cite{Yu:2017zst}, to compute these contributions. 
This mechanism proceeds via the rescattering of two intermediate particles, illustrated in Fig.~\ref{fig:trianglesev}. 
We detail our calculation of the nonleptonic decay amplitudes $\mathcal{B}_{cc}\to\mathcal{B}_{c}V$.
At the quark level, the first weak vertex in the triangle diagram is induced by the external $W$-emission diagram $T$, dominated by the factorizable contribution. 
To avoid double counting, the two-body weak transition amplitudes of doubly charmed baryons include only these short-distance factorizable parts (introduced in the previous section).
The rescattering can occur via $s$-channel or $t/u$-channel exchange. 
The $s$-channel would be significant only if the exchanged particle mass were close to $m_{{\cal B}_{cc}}$. 
However, the heaviest known singly charmed baryon is $\sim 500$ MeV lighter than $m_{\Xi_{cc}^{+}}$; thus the $s$-channel is strongly suppressed by off-shell effects and neglected. 
We therefore consider only the $t/u$-channel triangle diagram. 
Several methods exist for evaluating triangle amplitudes~\cite{Li:2002pj,Ablikim:2002ep,Li:1996cj,Dai:1999cs,Locher:1993cc,Cheng:2004ru,Lu:2005mx}, differing mainly in the treatment of loop integrations. For instance, Refs.~\cite{Han:2021azw,Cheng:2004ru} apply the optical theorem and Cutkosky cutting rules to compute only the absorptive (imaginary) part; the real part, obtainable via dispersion relations, suffers from large ambiguities and is often omitted. To obtain the complete amplitude reliably, we directly evaluate the loop integrals. We use the Passarino-Veltman method to reduce tensor integrals and integration by parts to express results in terms of master integrals.

To compute the triangle diagrams, we combine the weak transition vertex (weak effective Hamiltonian with naive factorization) with the hadronic scattering amplitudes. The two strong-interaction vertices in that diagram are evaluated using the hadronic effective Lagrangians~\cite{Aliev:2010yx,Yan:1992gz,Casalbuoni:1996pg,Meissner:1987ge,Li:2012bt,Aliev:2010nh} given in App.~\ref{app:la}. 
Specifically, we determine the amplitude for ${\cal {B}}_{cc}(p_{i})\to{\cal {B}}_{c}(p_{4})V(p_{3})$. 
Fig.~\ref{fig:trianglesev}(a) shows a triangle diagram with intermediate states $\{P,{\cal {B}}_{c};P\}$: a weak vertex ${\cal {B}}_{cc}(p_{i})\to{\cal {B}}_{c}(p_{2})P(p_{1})$ and a rescattering amplitude ${\cal {B}}_{c}(p_{2})P(p_{1})\to{\cal {B}}_{c}(p_{4})V(p_{3})$. 
We denote a triangle amplitude by $\mathcal{M}[P_{1},P_{2};P_{k}]$. 
We unify momentum and spin-polarization notation as follows:
${\cal B}_{cc}(p_{i},s_{i})\to{\cal B}_{c}(p_{3},s_{3})V(p_{4},\varepsilon_{4})$
with rescattering particles ${\cal B}_{c}(p_{1},s_{1})P(p_{2})/V(p_{2},\varepsilon_{2})$ and exchanged particle $P(k), V(k,\varepsilon_{k}), {\cal B}_{c}/{\cal B}_{c}^{*}(k,s_{k})$.

Under this convention and applying the factorization formula and the hadronic Feynman rules derived accordingly (App.~\ref{app:la}), we obtain analytical expressions for the eight hadronic loop diagrams shown in Fig.~\ref{fig:trianglesev}, with the loop integral variable $p_{1}$, 
\begin{align}
\mathcal{\mathcal{M}}[P,\mathcal{B}_c;P] 
		 &= -i\int\frac{d^{4}p_{1}}{(2\pi)^{4}}g_{P{\mathcal{B}}_{c}{\mathcal{B}}_{c}}\cdot g_{VPP}\bar{u}(p_{4},s_{4})\gamma_{5}(\slashed{p}_2+m_{2})(A_{LD}+B_{LD}\gamma_{5})u(p_{i},s_{i}) \nonumber \\
 &\times\varepsilon^{*}_{\alpha}(p_{3},\lambda_{3})(p_{1}+k)^{\alpha}
 {\cal {PF}}\,,\label{eq:PBP}\\
\mathcal{\mathcal{M}}[P,\mathcal{B}_c;V] 
		&= i\int\frac{d^{4}p_{1}}{(2\pi)^{4}}\frac{g_{VVP}}{\sqrt{m_{V_{1}}m_{V_{2}}}}\bar{u}(p_{4},s_{4})\left[(f_{1V{\cal {B}}_{c}{\cal {B}}_{c}}\gamma_{\nu}-\frac{if_{2V{\cal {B}}_{c}{\cal {B}}_{c}}}{m_2+m_4}\sigma_{\mu\nu}k^{\mu})\varepsilon^{\nu}(k,\lambda_{k})\right] (\slashed{p}_2+m_{2})\nonumber \\
&\times(A_{LD}\gamma_{5}+B_{LD})u(p_{i},s_{i})\varepsilon^{\alpha\beta\rho\tau}\varepsilon_{\beta}^{*}(p_{3},\lambda_{3})\epsilon^{*}_{\tau}(k,\lambda_{k})p_{3\alpha}p_{1\rho}{\cal {PF}}\,, \\
	\mathcal{\mathcal{M}}[V,\mathcal{B}_c;P]
&= \int\frac{d^{4}p_{1}}{(2\pi)^{4}}\frac{g_{VVP}}{\sqrt{m_{V_{1}}m_{V_{2}}}}g_{P{\cal {B}}_{c}{\cal {B}}_{c}}\bar{u}(p_{4},s_{4})\gamma_{5}(\slashed{p}_2+m_{2}) \nonumber \\
&\times\left[A_{LD}^{1}\gamma_{\mu}\gamma_{5}+A_{LD}^{2}\frac{p_{2\mu}}{m_{i}}\gamma_{5}+B_{LD}^{1}\gamma_{\mu}+B_{LD}^{2}\frac{p_{2\mu}}{m_{i}}\right] u(p_{i},s_{i}) \nonumber \\
&\times\epsilon^{\alpha\beta\rho\tau}\epsilon_{\beta}^{*}(p_{3})\epsilon_{\tau}(p_{1})p_{3\alpha}k_{\rho}\epsilon^{*\mu}(p_{1}){\cal {PF}}\,,\\
\mathcal{\mathcal{M}}[V,\mathcal{B}_c;V]
		&= \!\! \int \frac{d^4p_{\text{\tiny $1$}}}{(2\pi)^4}		\frac{g_{VVV}}{\sqrt{2}}[\epsilon^{\alpha}(p_{1})\epsilon^{*\beta}(p_{3})\epsilon^{*}_{\beta}(k)(p_{3\alpha}-k_{\alpha})\nonumber\\
		&\quad -\epsilon^{*\alpha}(k)\epsilon^{\beta}(p_{1})\epsilon^{*}_{\beta}(p_{3})(p_{1\alpha}+p_{3\alpha})
		+\epsilon^{*\alpha}(p_{3})\epsilon^{\beta}(p_{1})\epsilon^{*}_{\beta}(k)(k_{\alpha}+p_{1\alpha})]\nonumber\\
		&\bar{u}(p_4,s_4)(f_{1V{\cal {B}}_{c}{\cal {B}}_{c}}\gamma_{\nu}-\frac{if_{2V{\cal {B}}_{c}{\cal {B}}_{c}}}{m_2+m_4}\sigma_{\mu\nu}k^{\mu})\varepsilon^{\nu}(k,\lambda_{k})(\slashed p_{2}+m_{2})\nonumber\\
		&\quad \times \epsilon^{\rho}(p_{1})\left[A_{LD}^{1}\gamma_{\rho}\gamma_{5}+A_{LD}^{2}\frac{p_{2\rho}}{m_{i}}\gamma_{5}+B_{LD}^{1}\gamma_{\rho}+B_{LD}^{2}\frac{p_{2\rho}}{m_{i}}\right]u(p_{i},s_{i})\times{\cal {PF}},
		\end{align}

\begin{align}
\mathcal{\mathcal{M}}[\mathcal{B}_c,P;\mathcal{B}_c]
& =-\int\frac{d^{4}p_{1}}{(2\pi)^{4}}g_{P{\cal B}_{c}{\cal B}_{c}}\bar{u}(p_{4},s_{4})\gamma_{5}(\slashed{k}+m_{k})(f_{1V{\cal {B}}_{c}{\cal {B}}_{c}}\gamma_{\nu}+\frac{if_{2V{\cal {B}}_{c}{\cal {B}}_{c}}}{m_1+m_k}\sigma_{\mu\nu}p_{3}^{\mu}) \nonumber \\
&\times\varepsilon^{\nu*}(p_{3},\lambda_3)(\slashed{p}_1+m_{1})(A+B\gamma_{5})u(p_{i},s_{i}) \times{\cal {PF}}\,,
\\
\mathcal{\mathcal{M}}[\mathcal{B}_c,V;\mathcal{B}_c]
& =-i\int\frac{d^{4}p_{1}}{(2\pi)^{4}}\bar{u}(p_{4},s_{4})(f_{1V{\cal {B}}_{c}{\cal {B}}_{c}}\gamma_{\nu}-\frac{if_{2V{\cal {B}}_{c}{\cal {B}}_{c}}}{m_4+m_k}\sigma_{\mu\nu}p_{2}^{\mu})\varepsilon^{*\nu}(p_{2},\lambda_{2})(\slashed{k}+m_{k}) \nonumber\\
&\times(f_{1V{\cal {B}}_{c}{\cal {B}}_{c}}\gamma_{\beta}+\frac{if_{2V{\cal {B}}_{c}{\cal {B}}_{c}}}{m_1+m_k}\sigma_{\alpha\beta}p_{3}^{\alpha})\varepsilon^{*\nu}(p_{3},\lambda_3)(\slashed{p}_1+m_{1})\nonumber\\
&\times\epsilon^{\rho}(p_{2},\lambda_{2})
(A_{1}\gamma_{\rho}\gamma_{5}+A_{2}\frac{p_{1\rho}}{m_{i}}\gamma_{5}+B_{1}\gamma_{\rho}+B_{2}\frac{p_{1\rho}}{m_{i}})u(p_{i},s_{i}) \times{\cal {PF}} \,,\\
 \mathcal{M}[\mathcal{B}_{c},P;\mathcal{B}_{c}^{*}]
&=  ig_{P{\cal {B}}_{c}{\cal {B}}_{c}^{*}}g_{P{\cal {B}}_{c}{\cal {B}}_{c}^{*}}\int\frac{d^{4}p_{1}}{(2\pi)^{4}}\bar{u}(p_{4},\lambda_{4})p_{2\mu}u_{\nu}(k,\lambda_{k})\bar{u}^{\rho}(k,\lambda_{k})\times{\cal {PF}}\,,\nonumber\\
&\quad\times\gamma_{5}\gamma^{\sigma}[p_{3\rho}\varepsilon_{\sigma}^{*}(p_{3},\lambda_{3})-p_{3\sigma}\varepsilon_{\rho}^{*}(p_{3},\lambda_{3})](\slashed p_{1}+m_{1})(A_{LD}+B_{LD}\gamma_{5})u(p_{i},s_{i})\\
 \mathcal{M}[\mathcal{B}_{c},V;\mathcal{B}_{c}^{*}]
 = &-i g_{V{\cal {B}}_{c}{\cal {B}}_{c}^{*}}g_{V{\cal {B}}_{c}{\cal {B}}_{c}^{*}}\int\frac{d^{4}p_{1}}{(2\pi)^{4}}\bar{u}(p_{4},\lambda_{4})\gamma_{5}\gamma^{\beta}[p_{2\alpha}\varepsilon_{\beta}^{*}(p_{2},\lambda_{2})-p_{2\beta}\varepsilon_{\alpha}^{*}(p_{2},\lambda_{2})]\nonumber\\
		&\quad \times u^{\alpha}(k,\lambda_{k})\bar{u}^{\rho}(k,\lambda_{k})\gamma_{5}\gamma^{\sigma}[p_{3\rho}\varepsilon_{\sigma}^{*}(p_{3},\lambda_{3})-p_{3\sigma}\varepsilon_{\rho}^{*}(p_{3},\lambda_{3})](\slashed p_{1}+m_{1})\nonumber\\
		&\quad \times \epsilon^{\mu}(p_{2},\lambda_{2})\left[A_{LD}^{1}\gamma_{\mu}\gamma_{5}+A_{LD}^{2}\frac{p_{1\mu}}{m_{i}}\gamma_{5}+B_{LD}^{1}\gamma_{\mu}+B_{LD}^{2}\frac{p_{1\mu}}{m_{i}}\right]u(p_{i},s_{i})\times{\cal {PF}}\,,\label{eq:VBD}
\end{align}
where the note ${\cal {PF}}$ represents the multiply of the propagators
and form factor, 
\begin{eqnarray}
{\cal {PF}} & = & \frac{1}{(p_{1}^{2}-m_{1}^{2}+i\epsilon)(p_{2}^{2}-m_{2}^{2}+i\epsilon)(p_{k}^{2}-m_{k}^{2}+i\epsilon)}\left(\frac{\Lambda_{k}^{2}-m_{k}^{2}}{\Lambda_{k}^{2}-p_{k}^{2}}\right)^{n}.\label{eq:TF}
\end{eqnarray}

The strong coupling constants (e.g., $g_{VPP}$, $f_{1V{\cal B}_c{\cal B}_c}$, $f_{2V{\cal B}_c{\cal B}_c}$) in the equation are calculated on-shell, but their reliability is compromised because the exchanged particle is generally off-shell. To account for off-shell effects and ensure theoretical self-consistency, a form factor $F(p_j, m_j)$ is introduced~\cite{Cheng:2004ru}. Moreover, a proper regularization scheme is required to handle the divergences in the triangle diagram master integrals. The form factor in Eq.\eqref{eq:TF} is consistent with the Pauli-Villars regularization scheme.

\begin{align}
F(p_{k},m_{k})=\left(\frac{\Lambda_{k}^{2}-m_{k}^{2}}{\Lambda_{k}^{2}-p_{k}^{2}}\right)^{n}.\label{eq:Ffactor}
\end{align}
The cutoff $\Lambda_{k}$ can be given as 
\begin{align}
\Lambda_{k}=m_{k}+\eta\Lambda_{{\rm QCD}}.
\end{align}
With $\Lambda_{\rm QCD}=330~\text{MeV}$ for charm quark decays, the phenomenological parameter $\eta$ is determined by experimental data, not from first principles. Multiple strong vertices would require extensive data to determine $\eta$ individually, but in our previous work on $\mathcal{B}_{cc}\to\mathcal{B}_{c}V$~\cite{Hu:2024uia} we fixed $\eta$ using the experimental ratio $\mathcal{B}(\Xi_{cc}^{++}\to\Xi_{c}^{\prime+}\pi^{+})/\mathcal{B}(\Xi_{cc}^{++}\to\Xi_{c}^{+}\pi^{+})=(1.41\pm0.17\pm0.10)$~\cite{LHCb:2022rpd}. Since the same charm quark decay is involved, we adopt that value. The form factor in Eq.~(\ref{eq:Ffactor}) typically exhibits monopole or dipole behavior with exponent $n=1$ or $2$. As the branching ratios of $B$ meson decays~\cite{Cheng:2004ru} are similar for both choices, we adopt $n=2$.

After gathering all the fragments, the amplitude of decay ${\cal {B}}_{cc}\to{\cal {B}}_{c}V$
can be expressed as: 
\begin{align}
\mathcal{A}({\cal {B}}_{cc}\to{\cal {B}}_{c}V) & =\mathcal{M}_{SD}({\cal {B}}_{cc}\to{\cal {B}}_{c}V)\nonumber \\
&+\mathcal{M}[P,{\cal {B}}_{c};P]+\mathcal{M}[P,{\cal {B}}_{c};V]+\mathcal{M}[V,{\cal {B}}_{c};P]+\mathcal{M}[V,{\cal {B}}_{c};V]\nonumber \\
 & +\mathcal{M}[{\cal {B}}_{c},P;{\cal {B}}_{c}]+\mathcal{M}[{\cal {B}}_{c},V;{\cal {B}}_{c}]+\mathcal{M}[{\cal {B}}_{c},P;{\cal {B}}_{c}^{*}]+\mathcal{M}[{\cal {B}}_{c},V;{\cal {B}}_{c}^{*}],
\end{align}
where $\mathcal{M}_{SD}$ labels its short-distance contributions {including the factorization contributions of $T$ and $C$ diagrams}. The amplitudes
for all channels can be found in Appendix~\ref{app:amp}.

\subsection{Decay asymmetry parameters and CP violation}
As the amplitude for the two body nonleptonic weak decay of doubly-charmed baryon ${\cal {B}}_{cc}\to{\cal {B}}_{c}V$ can be deduced as 
\begin{align}
  \mathcal{M}({\cal {B}}_{cc}\to{\cal {B}}_{c}V) & =\epsilon^{\ast\mu}\bar{u}_{{{\cal {B}}_{c}}}\left[A_{1}\gamma_{\mu}\gamma_{5}+A_{2}\frac{p_{f\mu}}{m_{i}}\gamma_{5}+B_{1}\gamma_{\mu}+B_{2}\frac{p_{f\mu}}{m_{i}}\right]u_{{{\cal {B}}_{cc}}},\label{eq:B2BPc}
\end{align} 
here $p_{f\mu}$ is the momentum of singly charmed baryon ${\cal {B}}_{c}$ in the rest frame of the mother particle ${{\cal {B}}_{cc}}$. The ${{\cal {B}}_{cc}}$ spin has not yet been measured but the quark model
 prediction is spin $1/2$. The ${{\cal {B}}_{cc}}\to{{\cal {B}}_{c}}V$ mode is therefore the decay of a spin $1/2$ particle into a spin $1$ and a spin $1/2$ particle. In the
 helicity formalism, the decay can be described by four $H_{\lambda_{f},\lambda_{V}}^{\lambda_{i}}$ 
helicity amplitudes ($H_{-{1\over 2},0}^{{1\over 2}}$, $H_{{1\over 2},0}^{-{1\over 2}}$, $H_{-{1\over 2},-1}^{-{1\over 2}}$ and $H_{{1\over 2},1}^{{1\over 2}}$), where
 $\lambda_{i},\lambda_{f},\lambda_{V}$ are the helicity of ${{\cal {B}}_{cc}}$, ${{\cal {B}}_{c}}$ and $V$, respectively. In this decay process, the helicity amplitudes can be obtained in the Pauli-Dirac representation, 
\begin{eqnarray}
 & HV_{\lambda_{f},\lambda_{V}}^{\lambda_{i}} & =\langle{\cal B}_{f}(\lambda_{f})|\bar{c}\gamma^{\mu}b|{\cal B}_{i}(\lambda_{i})\rangle\epsilon_{\mu}^{*}(\lambda_{V}),\nonumber \\
 & HA_{\lambda_{f},\lambda_{V}}^{\lambda_{i}} & =\langle{\cal B}_{f}(\lambda_{f})|\bar{c}\gamma^{\mu}\gamma_{5}b|{\cal B}_{i}(\lambda_{i})\rangle\epsilon_{\mu}^{*}(\lambda_{V}),\nonumber \\
 & H_{\lambda_{f},\lambda_{V}}^{\lambda_{i}}= & HV_{\lambda_{f},\lambda_{V}}^{\lambda_{i}}-HA_{\lambda_{f},\lambda_{V}}^{\lambda_{i}}.\label{eq:helicity}
\end{eqnarray}
The helicity amplitudes are related to the transition form factors as follows:
\begin{eqnarray}
HV_{-\frac{1}{2},0}^{\frac{1}{2}} & = & HV_{\frac{1}{2},0}^{-\frac{1}{2}}=-i\sqrt{\frac{Q_{-}}{q^{2}}}\big((m_{i}+m_{f})f_{1}-\frac{q^2}{m_{i}}f_{2}\big),\label{eq:helicityV12012}\\
HV_{\frac{1}{2},1}^{\frac{1}{2}} & = & HV_{-\frac{1}{2},-1}^{-\frac{1}{2}}=i\sqrt{2Q_{-}}(-f_{1}+\frac{m_{i}+m_{f}}{m_{i}}f_{2}),\label{eq:helicityV12112}\\
HA_{-\frac{1}{2},0}^{\frac{1}{2}} & = & -HA_{\frac{1}{2},0}^{-\frac{1}{2}}=-i\sqrt{\frac{Q_{+}}{q^{2}}}((m_{i}-m_{f})g_{1}+\frac{q^2}{m_{i}}g_{2}),\label{eq:helicityA12012}\\
HA_{\frac{1}{2},1}^{\frac{1}{2}} & = & -HA_{-\frac{1}{2},-1}^{-\frac{1}{2}}=i\sqrt{2Q_{+}}(-g_{1}-\frac{m_{i}-m_{f}}{m_{i}}g_{2}).\label{eq:helicityA12112}
\end{eqnarray}
Then the decay width can be given as following,
\begin{align}
  \Gamma({\cal {B}}_{cc}\to{\cal {B}}_{c}V)
  &=\frac{|\vec{p}_{c}|(E_{f}+m_{f})}{4\pi m_{i}} \Big[ 2 \big( |\mathcal{S}|^{2}+|\mathcal{P}_{2}|^2 \big)+\frac{E_{V}^{2}}{m_{V}^{2}} \big( |\mathcal{S}+\mathcal{D}|^{2}+|\mathcal{P}_{1}|^{2} \big) \Big]\nonumber \\
  &=\frac{|\vec{p}_{c}|}{8\pi m_i^2}\frac{1}{2}\sum_{\lambda_{i},\lambda_{f},\lambda_{V}}|H_{\lambda_{f},\lambda_{V}}^{\lambda_{i}}({\cal {B}}_{cc}\to{\cal {B}}_{c}V)|^2,\label{eq:widthc}
  \end{align}
here $|\vec{p}_{c}|$ is the magnitude of three-momentum of ${{\cal {B}}_{c}}$
in the rest frame of ${{\cal {B}}_{cc}}$. 
The relations between $A_1,~A_2,~B_1,~B_2$ and $\mathcal{S},~\mathcal{P}_{1},~\mathcal{P}_{2},~\mathcal{D}$ can be given as
  \begin{align}
	\mathcal{S} & =-A_{1} \,, \nonumber\\
	\mathcal{P}_{1} & =-\frac{|\vec{p}_{c}|}{E_{V}}\left(\frac{m_{i}+m_{f}}{E_{f}+m_{f}}B_{1}+B_{2}\right) \,,\nonumber\\
	\mathcal{P}_{2} & =\frac{|\vec{p}_{c}|}{E_{f}+m_{f}}B_{1} \,,\nonumber\\
	\mathcal{D} & =-\frac{|\vec{p}_{c}|^{2}}{E_{V}(E_{f}+m_{f})}(A_{1}-A_{2}) \,.
	\end{align}

Specifically, the decay process ${\cal {B}}_{cc}\to{\cal {B}}_{c}V$ involves four distinct non-zero helicity amplitudes ($H_{-{1\over 2},0}^{{1\over 2}}$, $H_{{1\over 2},0}^{-{1\over 2}}$, $H_{-{1\over 2},-1}^{-{1\over 2}}$ and $H_{{1\over 2},1}^{{1\over 2}}$). Armed with these helicity amplitudes given in Eqs.~(\ref{eq:helicity}), we can also proceed to perform the calculations of the decay branching ratios and various asymmetry parameters, which are defined as~\cite{LHCb:2013hzx,Gutsche:2018utw},
\begin{eqnarray}\label{eq:alp}
\mathcal{BR}&=&\Gamma({\cal {B}}_{cc}\to{\cal {B}}_{c}V)\cdot\tau_{\Xi_{bc}},\nonumber\\
\alpha_b &=&-|\hat{H}_{\frac{1}{2},1}^{{1\over 2}}|^2+|\hat{H}_{-\frac{1}{2},-1}^{-{1\over 2}}|^2+|\hat{H}_{\frac{1}{2},0}^{-{1\over 2}}|^2-|\hat{H}_{-\frac{1}{2},0}^{{1\over 2}}|^2, \nonumber\\
\alpha_2&=&|\hat{H}_{\frac{1}{2},1}^{{1\over 2}}|^2-|\hat{H}_{-\frac{1}{2},-1}^{-{1\over 2}}|^2+|\hat{H}_{\frac{1}{2},0}^{-{1\over 2}}|^2-|\hat{H}_{-\frac{1}{2},0}^{{1\over 2}}|^2, \nonumber\\
r_0 &=&|\hat{H}_{\frac{1}{2},1}^{{1\over 2}}|^2+|\hat{H}_{-\frac{1}{2},0}^{{1\over 2}}|^2, \nonumber\\
r_1 &=&|\hat{H}_{\frac{1}{2},1}^{{1\over 2}}|^2-|\hat{H}_{-\frac{1}{2},0}^{{1\over 2}}|^2,
\end{eqnarray}
here we introduce the moduli squared of normalized helicity amplitudes
$|\hat{H}_{\lambda_f,\lambda_V}^{\lambda_{i}}|^2=|H_{\lambda_f,\lambda_V}^{\lambda_{i}}|^2/H_N$ as doing in Ref.~\cite{Gutsche:2018utw},  with
\begin{eqnarray}\label{eq:hn}
H_N&=&\sum_{\lambda_{i},\lambda_{f},\lambda_{V}}|H_{\lambda_{f},\lambda_{V}}^{\lambda_{i}}({\Xi_{bc}}\to{\Xi_{c}}J/\Psi)|^2, \nonumber\\
&=&|H_{\frac{1}{2},1}^{{1\over 2}}|^2+|H_{-\frac{1}{2},-1}^{-{1\over 2}}|^2+|H_{\frac{1}{2},0}^{-{1\over 2}}|^2+|H_{-\frac{1}{2},0}^{{1\over 2}}|^2.
\end{eqnarray}
And $\alpha_b$ is the parity-violating asymmetry parameter, 
$\alpha_2$ represents the longitudinal polarization of the daughter ${\Xi_{c}}$ baryon,
and $r_0 (r_1)$ corresponds to the longitudinal unpolarized (polarized) parameter.

As our calculation can give the strong phase of the decay amplitudes, the information of $CP$ violation can be derived.
The direct $CP$ asymmetry is defined as:
\begin{align}\label{eq:direct_CPV}
    A^\text{dir}_{CP}(\mathcal{B}_{cc}\to\mathcal{B}_{c}V) 
    & = \frac{\Gamma(\mathcal{B}_{cc}\to\mathcal{B}_{c}V)-\overline{\Gamma}(\mathcal{B}_{cc}\to\mathcal{B}_{c}V)}{\Gamma(\mathcal{B}_{cc}\to\mathcal{B}_{c}V) + \overline{\Gamma}(\mathcal{B}_{cc}\to\mathcal{B}_{c}V)} \nonumber\\
    & = \frac{2r\sin\Delta\delta\sin\Delta\phi}{1+r^2+2r\cos\Delta\delta\cos\Delta\phi} \,,   
\end{align}
where $r$ denotes the amplitude ratio. Furthermore, $\Delta\delta$ and $\Delta\phi$ symbolize the strong and weak phase differences respectively.

The violation of $CP$ arises from the parameters associated with decay asymmetry.
\begin{align} 
\alpha_{CP}  =\frac{\alpha-\bar{\alpha}}{\alpha+\bar{\alpha}},\qquad
r_{CP} & =\frac{r-\bar{r}}{r+\bar{r}}.
\end{align}

\section{Numerical results and discussions}
\label{sec:results}
In this section, we present the numerical results for the branching ratios and decay asymmetry parameters of the $\mathcal{B}_{cc}\to\mathcal{B}_{c}V$ decays.
\subsection{Input parameters}
The calculation requires inputs including initial and final state masses, decay constants of pseudoscalar and vector mesons, strong couplings, transition form factors, and the lifetimes of the doubly charmed baryons. In the following, we will classify and discuss the input parameter values used in this work.
\begin{itemize}
  \item[(i)] 
 The LHCb collaboration has measured the mass and lifetime of the $\Xi_{cc}^{++}$ baryon as $m_{\Xi_{cc}^{++}}=3.621\ \rm{GeV}$ and $\tau_{\Xi_{cc}^{++}}=256\ \rm{fs}$~\cite{LHCb:2018zpl,LHCb:2019qed}, along with the mass of the $\Xi_{cc}^{+}$ baryon: $m_{\Xi_{cc}^{+}}=3.619\ \rm{GeV}$~\cite{LHCb:2026pxn}. 
From Refs.~\cite{Yu:2018com,Berezhnoy:2018bde,Cheng:2026mlv}, we adopt $\tau_{\Xi_{cc}^{+}}=47\ \rm{fs}$, as well as $m_{\Omega_{cc}^{+}}=3.738\ \rm{GeV}$ and $\tau_{\Omega_{cc}^{+}}=179\ \rm{fs}$. The masses of the final states including singly heavy baryons and light mesons are taken from the Particle Data Group~\cite{ParticleDataGroup:2024cfk}.
  \item[(ii)] Decay constants of pseudoscalar and vector mesons are obtained from Refs.~\cite{ParticleDataGroup:2024cfk,Choi:2015ywa,Feldmann:1998vh} and are summarized in Table~\ref{table:decayconstants}.
  The form factors for the transition $\mathcal{B}_{cc}\to\mathcal{B}_{c}$ are taken from the light-front quark model calculations~\cite{Wang:2017mqp}, which have been successfully used to predict the discovery channel of $\Xi_{cc}^{++}$ in Ref.~\cite{Yu:2017zst}.
  \item[(iv)] The non-perturbative strong coupling constants are crucial for our calculation. Most of them are taken from theoretical works~\cite{Cheng:2004ru,Nakayama:2006ps,Aliev:2010yx,Aliev:2010nh,Aliev:2010ev,Lin:2017mtz,Yan:1992gz,Casalbuoni:1996pg,Meissner:1987ge,Li:2012bt}, which have been listed in Tab.~\ref{table:decayconstants}. The remaining couplings are inferred from $SU(3)$ flavor symmetry.
\end{itemize}

		\begin{table}[b]
		\centering \caption{{Decay constants (in units of MeV) of the light mesons~\cite{ParticleDataGroup:2024cfk,Choi:2015ywa,Feldmann:1998vh} and strong couplings~\cite{Cheng:2004ru,Nakayama:2006ps,Aliev:2010yx,Aliev:2010nh,Aliev:2010ev,Yan:1992gz,Casalbuoni:1996pg,Meissner:1987ge,Li:2012bt,Oset:2010tof} used in this work.}}
		\label{table:decayconstants} %
		\begin{tabular}{ccccccccccc}
		\hline 
		$f_{\pi}$~  & ~$f_{\eta_{1}}$~  & ~$f_{\rho}$~  & ~$f_{\omega}$~  & ~$f_{\eta_{8}}$~  & ~$f_{\phi}$~  & ~$f_{K}$~  
		& ~$f_{K^{*}}$\tabularnewline
		\hline 
		$130.2\pm1.2$  & $151\pm2.6$  & $216\pm5$  & $195\pm3$  & $169\pm2.6$  & $233\pm4.6$  & $155.7\pm3$  & $217\pm7$ \tabularnewline
		\hline 
		\end{tabular}\\
		\begin{tabular}{ccccccccccccccc}
		  \hline
		  $g_{\rho \rho \rho}$& $g_{\rho\pi\pi}$~  &~$g_{\Xi_{c}^{+}\Xi_{c}^{+}\pi^{0}}$~  & ~$g_{\Xi_{c}^{\prime+}\Xi_{c}^{+}\pi^{0}}$~  & ~$g_{\Sigma_{c}^{+}\Sigma_{c}^{0}\pi^{+}}$~  & ~$g_{\Sigma_{c}^{*0}\Lambda_{c}^{+}\pi^{-}}$~&$g_{\Sigma_{c}^{*+}\Sigma_{c}^{0}\pi^{+}}$ \tabularnewline
		\hline 
		$4.17\pm1$&$6.05\pm0.02$& $0.70\pm0.22$  & $3.1\pm1.1$  & $8.0\pm2.8$  & $3.9\pm0.6$ & $4.3\pm0.4$ \tabularnewline
		\hline 
		 $g_{\omega \rho \pi}$&$g_{\Sigma_{c}^{*+}\Lambda_{c}^{+}\rho^{0}}$~  &~$g_{\Sigma_{c}^{*0}\Sigma_{c}^{+}\rho^{-}}$~  & ~$g_{\Xi_{c}^{0}\Lambda_{c}^{+}K^{*-}}$~  & $g_{\Sigma_{c}^{0}\Lambda_{c}^{+}\rho^{-}}$& $g_{\Sigma_{c}^{+}\Sigma_{c}^{0}\rho^{+}}$\tabularnewline\hline
		  $-10\pm1$& $10\pm1.8$  &$5.77\pm0.5$  & $\{4.6\pm1.5,6\pm2\}$  & $\{2.6\pm0.9,16\pm5.3\}$&$\{4\pm1.3,27\pm9\}$ \tabularnewline\hline 
		\end{tabular}
		\end{table}

\subsection{Branching ratios and decay asymmetry parameters}

\begin{sidewaystable}
\caption{{Branching ratios and decay asymmetry parameters for the short-distance dynamics dominated modes. 
The ``CF\char`\"{}, ``SCS\char`\"{} and ``DCS\char`\"{} represent
CKM favored, singly CKM suppressed and doubly CKM suppressed processes,
respectively. And ${\cal B}_{SD}$, ${\cal B}_{LD}$ and ${\cal B}_{tot}$ denote the branching ratio of short-distance contribution, long-distance contribution and total one respectively. The uncertainty is due to the variation of the phenomenological parameter $\eta=0.3\pm0.05$.}}
\label{tab:result1} %
\footnotesize
\begin{tabular}{l|c|c|c|c|c|c|c|c|c|c|c}
\hline\hline
channels &${\cal B}_{SD}[10^{-2}]$ &${\cal B}_{LD}[10^{-2}]$ &${\cal B}_{Tot}[10^{-2}]$ &${\alpha_{b}}$&${\alpha_{2}}$&${r_{0}}$&${r_{1}}$ & $H^{1/2}_{1/2,1}$& $H^{-1/2}_{-1/2,-1}$& $H^{-1/2}_{1/2,0}$ & $H^{1/2}_{-1/2,0}$\\ \hline
$\Xi_{cc}^{++}\to\Xi_{c}^{+}\rho^{+} $ &$11.8$ 
& $1.08_{-0.63}^{+1.06}$ 
& $12.4_{-0.4}^{+0.8}$ 
& $-0.58_{-0.02}^{+0.02}$ 
& $-0.36_{-0.04}^{+0.05}$ 
& $0.89_{-0.01}^{+0.01}$ 
& $-0.47_{-0.03}^{+0.04}$ 
& $0.11_{-0.01}^{+0.01}$ 
& $0.0021_{-0.0008}^{+0.0013}$ 
& $0.21_{-0.01}^{+0.01}$ 
& $0.68_{-0.02}^{+0.02}$\\
$\Xi_{cc}^{++}\to\Xi_{c}^{\prime+}\rho^{+} $ 
&$15.9$ & $0.201_{-0.109}^{+0.169}$ 
& $18.9_{-1.0}^{+1.1}$ 
& $-0.75_{-0.01}^{+0.01}$ 
& $-0.100_{-0.010}^{+0.009}$ 
& $0.51_{-0.00}^{+0.00}$ 
& $-0.43_{-0.00}^{+0.00}$ 
& $0.41_{-0.01}^{+0.01}$ 
& $0.084_{-0.006}^{+0.007}$ 
& $0.040_{-0.000}^{+0.001}$ 
& $0.47_{-0.00}^{+0.00}$\\
$\Xi_{cc}^{++}\to\Sigma_{c}^{+}\rho^{+} $ &$0.962$ &
$0.0303_{-0.0173}^{+0.0280}$ &
$1.18_{-0.07}^{+0.09}$ &
$-0.82_{-0.03}^{+0.04}$ &
$-0.14_{-0.01}^{+0.01}$ &
$0.55_{-0.00}^{+0.00}$ &
$-0.48_{-0.01}^{+0.01}$ &
$0.39_{-0.01}^{+0.01}$ &
$0.052_{-0.009}^{+0.011}$ &
$0.036_{-0.004}^{+0.006}$ &
$0.52_{-0.00}^{+0.00}$\\
$\Xi_{cc}^{++}\to\Lambda_{c}^{+}\rho^{+} $ &$0.901$ &
$0.0931_{-0.0550}^{+0.0938}$ &
$0.850_{-0.001}^{+0.025}$ &
$-0.61_{-0.00}^{+0.02}$ &
$-0.42_{-0.01}^{+0.02}$ &
$0.88_{-0.03}^{+0.02}$ &
$-0.52_{-0.01}^{+0.02}$ &
$0.11_{-0.01}^{+0.01}$ &
$0.015_{-0.008}^{+0.013}$ &
$0.18_{-0.00}^{+0.01}$ &
$0.70_{-0.03}^{+0.01}$\\
$\Xi_{cc}^{++}\to\Xi_{c}^{\prime+}K^{*+} $ &$0.770$ & $0.00552_{-0.00306}^{+0.00476}$ & $0.793_{-0.006}^{+0.012}$ & $-0.69_{-0.01}^{+0.01}$ & $-0.0073_{-0.0172}^{+0.0153}$ & $0.43_{-0.00}^{+0.00}$ & $-0.35_{-0.01}^{+0.01}$ & $0.45_{-0.00}^{+0.00}$ & $0.11_{-0.00}^{+0.00}$ & $0.042_{-0.005}^{+0.005}$ & $0.39_{-0.01}^{+0.01}$\\
$\Xi_{cc}^{++}\to\Xi_{c}^{+}K^{*+} $ &$0.496$ & $0.0301_{-0.0175}^{+0.0289}$ & $0.717_{-0.082}^{+0.099}$ & $-0.71_{-0.01}^{+0.01}$ & $-0.40_{-0.01}^{+0.02}$ & $0.82_{-0.01}^{+0.01}$ & $-0.56_{-0.00}^{+0.00}$ & $0.17_{-0.01}^{+0.01}$ & $0.011_{-0.000}^{+0.000}$ & $0.13_{-0.01}^{+0.01}$ & $0.69_{-0.01}^{+0.01}$\\
$\Xi_{cc}^{++}\to\Lambda_{c}^{+}K^{*+} $ &$0.0454$ & $0.00156_{-0.00093}^{+0.00163}$ & $0.0592_{-0.0052}^{+0.0064}$ & $-0.69_{-0.02}^{+0.02}$ & $-0.42_{-0.00}^{+0.01}$ & $0.86_{-0.01}^{+0.01}$ & $-0.55_{-0.01}^{+0.01}$ & $0.14_{-0.01}^{+0.01}$ & $0.0019_{0.0001}^{+0.0008}$ & $0.15_{-0.01}^{+0.01}$ & $0.71_{-0.00}^{+0.00}$\\
$\Xi_{cc}^{++}\to\Sigma_{c}^{+}K^{*+} $ &$0.0532$ & $0.000831_{-0.000484}^{+0.000720}$ & $0.0515_{-0.0006}^{+0.0008}$ & $-0.85_{-0.00}^{+0.00}$ & $-0.042_{-0.031}^{+0.025}$ & $0.49_{-0.00}^{+0.01}$ & $-0.44_{-0.01}^{+0.01}$ & $0.46_{-0.01}^{+0.01}$ & $0.055_{-0.003}^{+0.005}$ & $0.022_{-0.004}^{+0.004}$ & $0.47_{-0.01}^{+0.01}$\\
$\Xi_{cc}^{+}\to\Xi_{c}^{0}\rho^{+} $ &$2.13$ & $0.0822_{-0.0471}^{+0.0768}$ & $2.78_{-0.25}^{+0.31}$ & $-0.70_{-0.03}^{+0.02}$ & $-0.48_{-0.01}^{+0.01}$ & $0.88_{-0.01}^{+0.01}$ & $-0.59_{-0.02}^{+0.02}$ & $0.11_{-0.01}^{+0.01}$ & $0.0088_{-0.0008}^{+0.0009}$ & $0.14_{-0.01}^{+0.01}$ & $0.73_{-0.00}^{+0.00}$\\
$\Xi_{cc}^{+}\to\Xi_{c}^{\prime0}\rho^{+} $ &$2.91$ & $0.0253_{-0.0148}^{+0.0251}$ & $2.54_{-0.14}^{+0.13}$ & $-0.81_{-0.01}^{+0.01}$ & $-0.10_{-0.01}^{+0.01}$ & $0.52_{-0.00}^{+0.00}$ & $-0.46_{-0.01}^{+0.01}$ & $0.42_{-0.00}^{+0.00}$ & $0.066_{-0.002}^{+0.001}$ & $0.030_{-0.004}^{+0.004}$ & $0.49_{-0.01}^{+0.01}$\\
$\Xi_{cc}^{+}\to\Sigma_{c}^{0}\rho^{+} $ &$0.352$ & $0.00745_{-0.00429}^{+0.00705}$ & $0.343_{-0.006}^{+0.005}$ & $-0.89_{-0.00}^{+0.00}$ & $-0.12_{-0.01}^{+0.01}$ & $0.55_{-0.00}^{+0.00}$ & $-0.50_{-0.01}^{+0.00}$ & $0.42_{-0.00}^{+0.00}$ & $0.034_{-0.001}^{+0.001}$ & $0.022_{-0.003}^{+0.002}$ & $0.53_{-0.00}^{+0.00}$\\
$\Xi_{cc}^{+}\to\Xi_{c}^{\prime0}K^{*+} $ &$0.140$ & $0.00686_{-0.00404}^{+0.00694}$ & $0.111_{-0.011}^{+0.010}$ & $-0.70_{-0.01}^{+0.01}$ & $0.0022_{-0.0168}^{+0.0145}$ & $0.43_{-0.00}^{+0.00}$ & $-0.35_{-0.01}^{+0.01}$ & $0.46_{-0.00}^{+0.00}$ & $0.11_{-0.00}^{+0.00}$ & $0.042_{-0.005}^{+0.005}$ & $0.39_{-0.01}^{+0.01}$\\
$\Xi_{cc}^{+}\to\Xi_{c}^{0}K^{*+} $ &$0.0893$ & $0.000921_{-0.000546}^{+0.000953}$ & $0.0802_{-0.0032}^{+0.0031}$ & $-0.61_{-0.03}^{+0.04}$ & $-0.41_{-0.01}^{+0.02}$ & $0.88_{-0.01}^{+0.01}$ & $-0.51_{-0.02}^{+0.03}$ & $0.11_{-0.01}^{+0.01}$ & $0.012_{-0.001}^{+0.001}$ & $0.18_{-0.02}^{+0.02}$ & $0.69_{-0.01}^{+0.01}$\\
$\Xi_{cc}^{+}\to\Sigma_{c}^{0}K^{*+} $ &$0.364$ & $-$ & $0.364$ & $-0.84$ & $0.020$ & $0.48$ & $-0.41$ & $0.48$ & $0.048$ & $0.034$ & $0.44$\\
$\Omega_{cc}^{+}\to\Omega_{c}^{0}\rho^{+} $ &$22.5$ & $-$ & $22.5$ & $-0.79$ & $-0.073$ & $0.51$ & $-0.43$ & $0.42$ & $0.066$ & $0.039$ & $0.47$\\
$\Omega_{cc}^{+}\to\Xi_{c}^{0}\rho^{+} $ &$0.661$ & $0.0186_{-0.0108}^{+0.0178}$ & $0.786_{-0.045}^{+0.053}$ & $-0.76_{-0.05}^{+0.05}$ & $-0.30_{-0.02}^{+0.02}$ & $0.75_{-0.01}^{+0.01}$ & $-0.53_{-0.04}^{+0.03}$ & $0.24_{-0.01}^{+0.01}$ & $0.010_{-0.003}^{+0.004}$ & $0.11_{-0.02}^{+0.02}$ & $0.64_{-0.01}^{+0.01}$\\
$\Omega_{cc}^{+}\to\Xi_{c}^{\prime0}\rho^{+} $ &$0.610$ & $0.0271_{-0.0158}^{+0.0265}$ & $0.804_{-0.071}^{+0.086}$ & $-0.66_{-0.05}^{+0.06}$ & $-0.19_{-0.02}^{+0.03}$ & $0.56_{-0.01}^{+0.01}$ & $-0.43_{-0.04}^{+0.05}$ & $0.34_{-0.01}^{+0.00}$ & $0.10_{-0.01}^{+0.01}$ & $0.066_{-0.014}^{+0.018}$ & $0.49_{-0.03}^{+0.02}$\\
$\Omega_{cc}^{+}\to\Omega_{c}^{0}K^{*+} $ &$1.09$ & $0.00509_{-0.00294}^{+0.00484}$ & $1.10_{-0.01}^{+0.02}$ & $-0.71_{-0.02}^{+0.01}$ & $0.031_{-0.001}^{+0.000}$ & $0.43_{-0.00}^{+0.00}$ & $-0.34_{-0.01}^{+0.01}$ & $0.47_{-0.00}^{+0.00}$ & $0.100_{-0.004}^{+0.004}$ & $0.047_{-0.003}^{+0.003}$ & $0.38_{-0.00}^{+0.00}$\\
$\Omega_{cc}^{+}\to\Xi_{c}^{0}K^{*+} $ &$0.0338$ & $0.0000687_{-0.0000389}^{+0.0000647}$ & $0.0359_{-0.0007}^{+0.0008}$ & $-0.68_{-0.02}^{+0.01}$ & $-0.20_{-0.01}^{+0.01}$ & $0.71_{-0.00}^{+0.00}$ & $-0.44_{-0.01}^{+0.01}$ & $0.26_{-0.00}^{+0.00}$ & $0.024_{-0.001}^{+0.000}$ & $0.13_{-0.01}^{+0.01}$ & $0.58_{-0.00}^{+0.01}$\\
$\Omega_{cc}^{+}\to\Xi_{c}^{\prime0}K^{*+} $ &$0.0337$ & $0.000118_{-0.000073}^{+0.000112}$ & $0.0327_{-0.0004}^{+0.0005}$ & $-0.76_{-0.02}^{+0.02}$ & $-0.13_{-0.01}^{+0.01}$ & $0.51_{-0.00}^{+0.00}$ & $-0.44_{-0.01}^{+0.01}$ & $0.40_{-0.01}^{+0.01}$ & $0.089_{-0.008}^{+0.008}$ & $0.033_{-0.003}^{+0.003}$ & $0.48_{-0.01}^{+0.01}$\\
\hline\hline
\end{tabular}
\end{sidewaystable}

\begin{sidewaystable}
  \caption{Branching ratios and decay asymmetry parameters for the long-distance dominated Cabibbo-favored ($\lambda_{sd}$)
  modes.}\label{tab:result2}
  \footnotesize
\begin{tabular}{l|c|c|c|c|c|c|c|c|c|c|c}
\hline\hline
channels &${\cal B}_{SD}[10^{-5}]$ &${\cal B}_{LD}[10^{-2}]$ &${\cal B}_{Tot}[10^{-2}]$ &${\alpha_{b}}$&${\alpha_{2}}$&${r_{0}}$&${r_{1}}$ & $H^{1/2}_{1/2,1}$& $H^{-1/2}_{-1/2,-1}$& $H^{-1/2}_{1/2,0}$ & $H^{1/2}_{-1/2,0}$\\ \hline
$\Xi_{cc}^{++}\to\Sigma_{c}^{++}\bar{K}^{*0} $ &$10.9$ & $3.27_{-1.91}^{+3.31}$ & $3.54_{-2.01}^{+3.44}$ & $-0.23_{-0.04}^{+0.04}$ & $-0.023_{-0.005}^{+0.003}$ & $0.35_{-0.00}^{+0.00}$ & $-0.13_{-0.02}^{+0.02}$ & $0.38_{-0.01}^{+0.01}$ & $0.27_{-0.01}^{+0.01}$ & $0.11_{-0.01}^{+0.01}$ & $0.24_{-0.01}^{+0.01}$\\
$\Xi_{cc}^{+}\to\Omega_{c}^{0}K^{*+} $ &$-$ & $0.0364_{-0.0213}^{+0.0355}$ & $0.0364_{-0.0213}^{+0.0355}$ & $0.028_{-0.015}^{+0.018}$ & $0.100_{-0.003}^{+0.002}$ & $0.35_{-0.00}^{+0.00}$ & $0.064_{-0.006}^{+0.008}$ & $0.34_{-0.01}^{+0.00}$ & $0.31_{-0.00}^{+0.00}$ & $0.21_{-0.00}^{+0.00}$ & $0.15_{-0.00}^{+0.00}$\\
$\Xi_{cc}^{+}\to\Sigma_{c}^{+}\bar{K}^{*0} $ &$0.999$ & $0.919_{-0.539}^{+0.923}$ & $0.951_{-0.551}^{+0.938}$ & $-0.26_{-0.03}^{+0.04}$ & $-0.036_{-0.010}^{+0.009}$ & $0.32_{-0.00}^{+0.00}$ & $-0.15_{-0.01}^{+0.01}$ & $0.40_{-0.01}^{+0.01}$ & $0.29_{-0.01}^{+0.01}$ & $0.085_{-0.006}^{+0.007}$ & $0.23_{-0.01}^{+0.01}$\\
$\Xi_{cc}^{+}\to\Lambda_{c}^{+}\bar{K}^{*0} $ &$0.493$ & $1.22_{-0.73}^{+1.27}$ & $1.23_{-0.73}^{+1.28}$ & $0.28_{-0.04}^{+0.03}$ & $-0.14_{-0.02}^{+0.02}$ & $0.21_{-0.00}^{+0.00}$ & $0.070_{-0.009}^{+0.007}$ & $0.29_{-0.01}^{+0.02}$ & $0.50_{-0.01}^{+0.01}$ & $0.14_{-0.01}^{+0.00}$ & $0.069_{-0.003}^{+0.003}$\\
$\Xi_{cc}^{+}\to\Sigma_{c}^{++}K^{*-} $ &$-$ & $0.433_{-0.251}^{+0.417}$ & $0.433_{-0.251}^{+0.417}$ & $-0.45_{-0.04}^{+0.04}$ & $-0.050_{-0.010}^{+0.010}$ & $0.34_{-0.00}^{+0.00}$ & $-0.25_{-0.01}^{+0.01}$ & $0.43_{-0.01}^{+0.01}$ & $0.23_{-0.01}^{+0.01}$ & $0.044_{-0.007}^{+0.006}$ & $0.29_{-0.01}^{+0.01}$\\
$\Xi_{cc}^{+}\to\Xi_{c}^{\prime+}\rho^{0} $ &$-$ & $0.0475_{-0.0272}^{+0.0448}$ & $0.0475_{-0.0272}^{+0.0448}$ & $0.19_{-0.02}^{+0.02}$ & $-0.21_{-0.02}^{+0.02}$ & $0.45_{-0.01}^{+0.01}$ & $-0.0070_{-0.0197}^{+0.0213}$ & $0.17_{-0.00}^{+0.00}$ & $0.38_{-0.00}^{+0.00}$ & $0.22_{-0.01}^{+0.01}$ & $0.23_{-0.01}^{+0.01}$\\
$\Xi_{cc}^{+}\to\Xi_{c}^{\prime+}\phi $ &$-$ & $0.0422_{-0.0245}^{+0.0410}$ & $0.0422_{-0.0245}^{+0.0410}$ & $-0.16_{-0.01}^{+0.01}$ & $-0.017_{-0.004}^{+0.004}$ & $0.33_{-0.00}^{+0.00}$ & $-0.090_{-0.004}^{+0.004}$ & $0.37_{-0.00}^{+0.00}$ & $0.30_{-0.00}^{+0.00}$ & $0.12_{-0.00}^{+0.00}$ & $0.21_{-0.00}^{+0.00}$\\
$\Xi_{cc}^{+}\to\Xi_{c}^{\prime+}\omega $ &$-$ & $0.0468_{-0.0268}^{+0.0440}$ & $0.0468_{-0.0268}^{+0.0440}$ & $0.19_{-0.02}^{+0.02}$ & $-0.21_{-0.02}^{+0.02}$ & $0.45_{-0.01}^{+0.01}$ & $-0.0061_{-0.0191}^{+0.0208}$ & $0.18_{-0.00}^{+0.00}$ & $0.38_{-0.00}^{+0.00}$ & $0.22_{-0.01}^{+0.01}$ & $0.23_{-0.01}^{+0.01}$\\
$\Xi_{cc}^{+}\to\Xi_{c}^{+}\rho^{0} $ &$-$ & $0.0801_{-0.0470}^{+0.0797}$ & $0.0801_{-0.0470}^{+0.0797}$ & $-0.55_{-0.00}^{+0.00}$ & $-0.24_{-0.01}^{+0.01}$ & $0.69_{-0.00}^{+0.00}$ & $-0.39_{-0.01}^{+0.01}$ & $0.23_{-0.00}^{+0.00}$ & $0.077_{-0.000}^{+0.001}$ & $0.15_{-0.00}^{+0.00}$ & $0.54_{-0.01}^{+0.01}$\\
$\Xi_{cc}^{+}\to\Xi_{c}^{+}\phi $ &$-$ & $0.108_{-0.064}^{+0.111}$ & $0.108_{-0.064}^{+0.111}$ & $0.35_{-0.01}^{+0.01}$ & $-0.19_{-0.01}^{+0.01}$ & $0.29_{-0.00}^{+0.00}$ & $0.077_{-0.002}^{+0.002}$ & $0.22_{-0.01}^{+0.01}$ & $0.49_{-0.01}^{+0.01}$ & $0.18_{-0.00}^{+0.00}$ & $0.11_{-0.00}^{+0.00}$\\
$\Xi_{cc}^{+}\to\Xi_{c}^{+}\omega $ &$-$ & $0.0779_{-0.0457}^{+0.0775}$ & $0.0779_{-0.0457}^{+0.0775}$ & $-0.55_{-0.00}^{+0.00}$ & $-0.24_{-0.01}^{+0.01}$ & $0.69_{-0.00}^{+0.00}$ & $-0.39_{-0.01}^{+0.01}$ & $0.23_{-0.00}^{+0.00}$ & $0.078_{-0.001}^{+0.001}$ & $0.15_{-0.00}^{+0.00}$ & $0.54_{-0.01}^{+0.01}$\\
$\Omega_{cc}^{+}\to\Xi_{c}^{\prime+}\bar{K}^{*0} $ &$3.68$ & $2.29_{-1.34}^{+2.30}$ & $2.40_{-1.39}^{+2.36}$ & $-0.044_{-0.036}^{+0.038}$ & $-0.042_{-0.003}^{+0.003}$ & $0.37_{-0.00}^{+0.00}$ & $-0.043_{-0.019}^{+0.021}$ & $0.31_{-0.01}^{+0.01}$ & $0.31_{-0.01}^{+0.01}$ & $0.16_{-0.01}^{+0.01}$ & $0.21_{-0.01}^{+0.01}$\\
$\Omega_{cc}^{+}\to\Xi_{c}^{+}\bar{K}^{*0} $ &$2.24$ & $2.04_{-1.20}^{+2.10}$ & $2.14_{-1.24}^{+2.14}$ & $0.066_{-0.029}^{+0.034}$ & $-0.073_{-0.006}^{+0.006}$ & $0.34_{-0.00}^{+0.00}$ & $-0.0034_{-0.0119}^{+0.0140}$ & $0.29_{-0.01}^{+0.01}$ & $0.36_{-0.01}^{+0.01}$ & $0.17_{-0.01}^{+0.01}$ & $0.17_{-0.01}^{+0.01}$\\
\hline\hline
\end{tabular}
  \end{sidewaystable}

\begin{sidewaystable}
\caption{{Same as Table.\ref{tab:result2} but for the long-distance dominated singly Cabibbo-suppressed modes.}}
\label{tab:result3} 
\footnotesize
\begin{tabular}{l|c|c|c|c|c|c|c|c|c|c|c}
\hline\hline
channels &${\cal B}_{SD}[10^{-7}]$ &${\cal B}_{LD}[10^{-3}]$ &${\cal B}_{Tot}[10^{-3}]$ &${\alpha_{b}}$&${\alpha_{2}}$&${r_{0}}$&${r_{1}}$ & $H^{1/2}_{1/2,1}$& $H^{-1/2}_{-1/2,-1}$& $H^{-1/2}_{1/2,0}$ & $H^{1/2}_{-1/2,0}$\\ \hline
$\Xi_{cc}^{++}\to\Sigma_{c}^{++}\rho^{0} $ &$50.0$ & $1.24_{-0.74}^{+1.30}$ & $1.14_{-0.70}^{+1.25}$ & $-0.25_{-0.05}^{+0.05}$ & $-0.097_{-0.011}^{+0.013}$ & $0.49_{-0.00}^{+0.00}$ & $-0.17_{-0.03}^{+0.03}$ & $0.29_{-0.01}^{+0.01}$ & $0.21_{-0.01}^{+0.01}$ & $0.16_{-0.02}^{+0.02}$ & $0.33_{-0.02}^{+0.01}$\\
$\Xi_{cc}^{++}\to\Sigma_{c}^{++}\phi $ &$69.9$ & $1.64_{-0.97}^{+1.64}$ & $1.81_{-1.03}^{+1.71}$ & $-0.14_{-0.04}^{+0.03}$ & $-0.025_{-0.007}^{+0.011}$ & $0.33_{-0.00}^{+0.00}$ & $-0.084_{-0.014}^{+0.013}$ & $0.36_{-0.01}^{+0.01}$ & $0.30_{-0.01}^{+0.01}$ & $0.12_{-0.01}^{+0.01}$ & $0.21_{-0.01}^{+0.01}$\\
$\Xi_{cc}^{++}\to\Sigma_{c}^{++}\omega $ &$50.5$ & $0.422_{-0.238}^{+0.386}$ & $0.381_{-0.222}^{+0.366}$ & $-0.64_{-0.02}^{+0.03}$ & $0.059_{-0.014}^{+0.014}$ & $0.32_{-0.01}^{+0.00}$ & $-0.29_{-0.01}^{+0.01}$ & $0.52_{-0.01}^{+0.01}$ & $0.17_{-0.01}^{+0.01}$ & $0.013_{-0.00}^{+0.00}$ & $0.30_{-0.01}^{+0.01}$\\
$\Xi_{cc}^{+}\to\Sigma_{c}^{+}\rho^{0} $ &$4.59$ &
$0.538_{-0.315}^{+0.533}$ &
$0.522_{-0.308}^{+0.525}$ &
$-0.27_{-0.03}^{+0.04}$ &
$-0.11_{-0.01}^{+0.01}$ &
$0.44_{-0.00}^{+0.00}$ &
$-0.19_{-0.02}^{+0.02}$ &
$0.32_{-0.01}^{+0.01}$ &
$0.24_{-0.00}^{+0.01}$ &
$0.13_{-0.01}^{+0.01}$ &
$0.32_{-0.01}^{+0.01}$\\
$\Xi_{cc}^{+}\to\Sigma_{c}^{+}\phi $ &$6.42$ & $0.151_{-0.089}^{+0.150}$ & $0.167_{-0.095}^{+0.157}$ & $-0.14_{-0.04}^{+0.03}$ & $-0.026_{-0.007}^{+0.011}$ & $0.33_{-0.00}^{+0.00}$ & $-0.083_{-0.014}^{+0.013}$ & $0.36_{-0.01}^{+0.01}$ & $0.30_{-0.01}^{+0.01}$ & $0.13_{-0.01}^{+0.01}$ & $0.21_{-0.01}^{+0.01}$\\
$\Xi_{cc}^{+}\to\Sigma_{c}^{+}\omega $ &$4.63$ & $0.294_{-0.169}^{+0.279}$ & $0.289_{-0.167}^{+0.276}$ & $-0.41_{-0.03}^{+0.03}$ & $-0.11_{-0.00}^{+0.00}$ & $0.39_{-0.00}^{+0.00}$ & $-0.26_{-0.02}^{+0.01}$ & $0.38_{-0.01}^{+0.01}$ & $0.23_{-0.01}^{+0.01}$ & $0.063_{-0.008}^{+0.007}$ & $0.32_{-0.01}^{+0.01}$\\
$\Xi_{cc}^{+}\to\Lambda_{c}^{+}\rho^{0} $ &$2.59$ &
$0.468_{-0.279}^{+0.488}$ &
$0.464_{-0.278}^{+0.486}$ &
$0.50_{-0.02}^{+0.02}$ &
$-0.045_{-0.011}^{+0.015}$ &
$0.31_{-0.00}^{+0.00}$ &
$0.23_{-0.00}^{+0.00}$ &
$0.21_{-0.01}^{+0.01}$ &
$0.48_{-0.01}^{+0.01}$ &
$0.27_{-0.00}^{+0.00}$ &
$0.044_{-0.001}^{+0.001}$\\
$\Xi_{cc}^{+}\to\Lambda_{c}^{+}\phi $ &$2.98$ & $0.239_{-0.144}^{+0.252}$ & $0.244_{-0.145}^{+0.254}$ & $0.23_{-0.04}^{+0.03}$ & $-0.033_{-0.015}^{+0.016}$ & $0.28_{-0.00}^{+0.00}$ & $0.096_{-0.010}^{+0.006}$ & $0.30_{-0.01}^{+0.01}$ & $0.42_{-0.01}^{+0.01}$ & $0.19_{-0.01}^{+0.00}$ & $0.092_{-0.001}^{+0.003}$\\
$\Xi_{cc}^{+}\to\Lambda_{c}^{+}\omega $ &$2.59$ & $0.211_{-0.125}^{+0.216}$ & $0.204_{-0.122}^{+0.212}$ & $0.14_{-0.02}^{+0.02}$ & $-0.56_{-0.01}^{+0.01}$ & $0.28_{-0.00}^{+0.00}$ & $-0.21_{-0.00}^{+0.01}$ & $0.18_{-0.01}^{+0.01}$ & $0.54_{-0.01}^{+0.01}$ & $0.034_{-0.001}^{+0.002}$ & $0.24_{-0.00}^{+0.00}$\\
$\Xi_{cc}^{+}\to\Sigma_{c}^{++}\rho^{-} $ &$-$ & $0.251_{-0.146}^{+0.244}$ & $0.251_{-0.146}^{+0.244}$ & $-0.40_{-0.02}^{+0.02}$ & $-0.13_{-0.01}^{+0.01}$ & $0.35_{-0.00}^{+0.00}$ & $-0.26_{-0.00}^{+0.01}$ & $0.39_{-0.01}^{+0.01}$ & $0.26_{-0.01}^{+0.01}$ & $0.045_{-0.003}^{+0.004}$ & $0.31_{-0.00}^{+0.00}$\\
$\Xi_{cc}^{+}\to\Xi_{c}^{\prime+}K^{*0} $ &$-$ & $0.0117_{-0.0070}^{+0.0122}$ & $0.0117_{-0.0070}^{+0.0122}$ & $-0.14_{-0.04}^{+0.02}$ & $0.26_{-0.02}^{+0.03}$ & $0.37_{-0.03}^{+0.03}$ & $0.058_{-0.029}^{+0.025}$ & $0.41_{-0.01}^{+0.02}$ & $0.22_{-0.02}^{+0.01}$ & $0.21_{-0.03}^{+0.03}$ & $0.16_{-0.00}^{+0.00}$\\
$\Xi_{cc}^{+}\to\Xi_{c}^{+}K^{*0} $ &$-$ & $0.0515_{-0.0307}^{+0.0511}$ & $0.0515_{-0.0307}^{+0.0511}$ & $-0.14_{-0.08}^{+0.05}$ & $-0.20_{-0.01}^{+0.02}$ & $0.33_{-0.02}^{+0.03}$ & $-0.17_{-0.03}^{+0.02}$ & $0.32_{-0.01}^{+0.01}$ & $0.35_{-0.04}^{+0.02}$ & $0.083_{-0.002}^{+0.002}$ & $0.25_{-0.02}^{+0.03}$\\
$\Omega_{cc}^{+}\to\Sigma_{c}^{+}\bar{K}^{*0} $ &$-$ & $0.967_{-0.578}^{+1.010}$ & $0.967_{-0.578}^{+1.010}$ & $0.40_{-0.02}^{+0.02}$ & $-0.39_{-0.01}^{+0.01}$ & $0.28_{-0.00}^{+0.00}$ & $0.0065_{-0.0047}^{+0.0047}$ & $0.16_{-0.01}^{+0.00}$ & $0.56_{-0.01}^{+0.01}$ & $0.14_{-0.00}^{+0.00}$ & $0.14_{-0.00}^{+0.00}$\\
$\Omega_{cc}^{+}\to\Lambda_{c}^{+}\bar{K}^{*0} $ &$-$ & $1.49_{-0.88}^{+1.54}$ & $1.49_{-0.88}^{+1.54}$ & $0.23_{-0.05}^{+0.05}$ & $-0.72_{-0.01}^{+0.01}$ & $0.31_{-0.01}^{+0.01}$ & $-0.25_{-0.02}^{+0.02}$ & $0.11_{-0.01}^{+0.01}$ & $0.58_{-0.02}^{+0.02}$ & $0.030_{-0.004}^{+0.004}$ & $0.28_{-0.01}^{+0.01}$\\
$\Omega_{cc}^{+}\to\Sigma_{c}^{++}K^{*-} $ &$-$ & $1.04_{-0.62}^{+1.06}$ & $1.04_{-0.62}^{+1.06}$ & $0.067_{-0.043}^{+0.036}$ & $-0.44_{-0.00}^{+0.00}$ & $0.31_{-0.00}^{+0.00}$ & $-0.19_{-0.02}^{+0.02}$ & $0.22_{-0.01}^{+0.01}$ & $0.47_{-0.01}^{+0.01}$ & $0.059_{-0.008}^{+0.006}$ & $0.25_{-0.01}^{+0.01}$\\
$\Omega_{cc}^{+}\to\Xi_{c}^{\prime+}\rho^{0} $ &$-$ & $0.981_{-0.578}^{+0.990}$ & $0.981_{-0.578}^{+0.990}$ & $-0.15_{-0.05}^{+0.05}$ & $-0.037_{-0.006}^{+0.010}$ & $0.47_{-0.00}^{+0.00}$ & $-0.093_{-0.027}^{+0.030}$ & $0.29_{-0.01}^{+0.01}$ & $0.24_{-0.01}^{+0.01}$ & $0.19_{-0.02}^{+0.02}$ & $0.28_{-0.01}^{+0.01}$\\
$\Omega_{cc}^{+}\to\Xi_{c}^{\prime+}\phi $ &$-$ & $2.47_{-1.45}^{+2.44}$ & $2.47_{-1.45}^{+2.44}$ & $0.048_{-0.021}^{+0.020}$ & $-0.12_{-0.00}^{+0.00}$ & $0.33_{-0.00}^{+0.00}$ & $-0.037_{-0.009}^{+0.010}$ & $0.29_{-0.01}^{+0.01}$ & $0.38_{-0.01}^{+0.01}$ & $0.15_{-0.00}^{+0.00}$ & $0.18_{-0.00}^{+0.00}$\\
$\Omega_{cc}^{+}\to\Xi_{c}^{\prime+}\omega $ &$-$ & $0.0880_{-0.0488}^{+0.0766}$ & $0.0880_{-0.0488}^{+0.0766}$ & $-0.53_{-0.04}^{+0.05}$ & $0.45_{-0.02}^{+0.02}$ & $0.50_{-0.01}^{+0.01}$ & $-0.040_{-0.031}^{+0.035}$ & $0.50_{-0.01}^{+0.01}$ & $0.0055_{-0.0017}^{+0.0022}$ & $0.23_{-0.02}^{+0.02}$ & $0.27_{-0.01}^{+0.01}$\\
$\Omega_{cc}^{+}\to\Xi_{c}^{+}\rho^{0} $ &$-$ & $0.785_{-0.472}^{+0.831}$ & $0.785_{-0.472}^{+0.831}$ & $0.61_{-0.01}^{+0.02}$ & $-0.22_{-0.00}^{+0.00}$ & $0.35_{-0.01}^{+0.01}$ & $0.20_{-0.01}^{+0.01}$ & $0.12_{-0.01}^{+0.01}$ & $0.53_{-0.00}^{+0.00}$ & $0.27_{-0.01}^{+0.01}$ & $0.078_{-0.001}^{+0.001}$\\
$\Omega_{cc}^{+}\to\Xi_{c}^{+}\phi $ &$-$ & $3.09_{-1.84}^{+3.16}$ & $3.09_{-1.84}^{+3.16}$ & $0.26_{-0.02}^{+0.02}$ & $-0.26_{-0.01}^{+0.01}$ & $0.28_{-0.00}^{+0.00}$ & $-0.013_{-0.0064}^{+0.0066}$ & $0.23_{-0.01}^{+0.01}$ & $0.49_{-0.01}^{+0.01}$ & $0.14_{-0.00}^{+0.00}$ & $0.14_{-0.00}^{+0.00}$\\
$\Omega_{cc}^{+}\to\Xi_{c}^{+}\omega $ &$-$ & $0.165_{-0.097}^{+0.162}$ & $0.165_{-0.097}^{+0.162}$ & $0.030_{-0.043}^{+0.042}$ & $-0.47_{-0.03}^{+0.03}$ & $0.25_{-0.01}^{+0.01}$ & $-0.22_{-0.01}^{+0.01}$ & $0.25_{-0.01}^{+0.01}$ & $0.50_{-0.02}^{+0.02}$ & $0.016_{-0.002}^{+0.002}$ & $0.24_{-0.01}^{+0.01}$\\
\hline\hline
\end{tabular}
\end{sidewaystable}

\begin{sidewaystable}
\caption{{Same as Tab.~\ref{tab:result2} but for the long-distance dominated doubly Cabibbo-suppressed modes.}}
\label{tab:result4} 
\footnotesize
\begin{tabular}{l|c|c|c|c|c|c|c|c|c|c|c}
\hline\hline
channels &${\cal B}_{SD}[10^{-8}]$ &${\cal B}_{LD}[10^{-5}]$ &${\cal B}_{Tot}[10^{-5}]$ &${\alpha_{b}}$&${\alpha_{2}}$&${r_{0}}$&${r_{1}}$ & $H^{1/2}_{1/2,1}$& $H^{-1/2}_{-1/2,-1}$& $H^{-1/2}_{1/2,0}$ & $H^{1/2}_{-1/2,0}$\\ \hline
$\Xi_{cc}^{++}\to\Sigma_{c}^{++}K^{*0} $ &$31.0$ & $7.14_{-4.06}^{+7.37}$ & $7.85_{-4.30}^{+7.67}$ & $-0.43_{-0.03}^{+0.06}$ & $0.011_{-0.023}^{+0.005}$ & $0.39_{-0.00}^{+0.00}$ & $-0.21_{-0.01}^{+0.02}$ & $0.42_{-0.02}^{+0.01}$ & $0.20_{-0.01}^{+0.02}$ & $0.089_{-0.007}^{+0.007}$ & $0.30_{-0.01}^{+0.01}$\\
$\Xi_{cc}^{+}\to\Sigma_{c}^{+}K^{*0} $ &$2.84$ & $1.68_{-0.96}^{+1.68}$ & $1.76_{-0.99}^{+1.71}$ & $-0.23_{-0.03}^{+0.05}$ & $0.017_{-0.018}^{+0.000}$ & $0.41_{-0.00}^{+0.00}$ & $-0.11_{-0.01}^{+0.01}$ & $0.36_{-0.02}^{+0.01}$ & $0.23_{-0.01}^{+0.02}$ & $0.15_{-0.01}^{+0.01}$ & $0.26_{-0.01}^{+0.01}$\\
$\Xi_{cc}^{+}\to\Lambda_{c}^{+}K^{*0} $ &$1.40$ & $1.56_{-0.91}^{+1.61}$ & $1.60_{-0.92}^{+1.63}$ & $-0.060_{-0.025}^{+0.052}$ & $-0.036_{-0.016}^{+0.012}$ & $0.35_{-0.00}^{+0.00}$ & $-0.048_{-0.019}^{+0.018}$ & $0.33_{-0.02}^{+0.00}$ & $0.32_{-0.00}^{+0.02}$ & $0.15_{-0.01}^{+0.01}$ & $0.20_{-0.01}^{+0.01}$\\
$\Omega_{cc}^{+}\to\Sigma_{c}^{+}\rho^{0} $ &$-$ & $0.939_{-0.549}^{+0.921}$ & $0.939_{-0.549}^{+0.921}$ & $-0.37_{-0.02}^{+0.02}$ & $-0.30_{-0.01}^{+0.01}$ & $0.50_{-0.00}^{+0.00}$ & $-0.33_{-0.01}^{+0.01}$ & $0.27_{-0.01}^{+0.01}$ & $0.23_{-0.01}^{+0.01}$ & $0.086_{-0.003}^{+0.002}$ & $0.42_{-0.00}^{+0.00}$\\
$\Omega_{cc}^{+}\to\Sigma_{c}^{+}\phi $ &$-$ & $0.786_{-0.450}^{+0.731}$ & $0.786_{-0.450}^{+0.731}$ & $0.038_{-0.001}^{+0.001}$ & $0.081_{-0.012}^{+0.013}$ & $0.42_{-0.00}^{+0.00}$ & $0.059_{-0.006}^{+0.006}$ & $0.30_{-0.00}^{+0.00}$ & $0.28_{-0.01}^{+0.01}$ & $0.24_{-0.01}^{+0.01}$ & $0.18_{-0.00}^{+0.00}$\\
$\Omega_{cc}^{+}\to\Sigma_{c}^{+}\omega $ &$-$ & $2.92_{-1.73}^{+2.97}$ & $2.92_{-1.73}^{+2.97}$ & $0.12_{-0.02}^{+0.02}$ & $-0.27_{-0.01}^{+0.01}$ & $0.34_{-0.00}^{+0.00}$ & $-0.078_{-0.008}^{+0.007}$ & $0.23_{-0.01}^{+0.01}$ & $0.43_{-0.01}^{+0.01}$ & $0.13_{-0.00}^{+0.00}$ & $0.21_{-0.00}^{+0.00}$\\
$\Omega_{cc}^{+}\to\Lambda_{c}^{+}\rho^{0} $ &$-$ & $1.78_{-1.06}^{+1.85}$ & $1.78_{-1.06}^{+1.85}$ & $0.68_{-0.02}^{+0.02}$ & $-0.58_{-0.02}^{+0.02}$ & $0.14_{-0.01}^{+0.01}$ & $0.050_{-0.001}^{+0.000}$ & $0.11_{-0.00}^{+0.00}$ & $0.74_{-0.01}^{+0.01}$ & $0.096_{-0.005}^{+0.005}$ & $0.046_{-0.004}^{+0.004}$\\
$\Omega_{cc}^{+}\to\Lambda_{c}^{+}\phi $ &$-$ & $1.37_{-0.80}^{+1.35}$ & $1.37_{-0.80}^{+1.35}$ & $-0.77_{-0.01}^{+0.01}$ & $0.16_{-0.01}^{+0.01}$ & $0.39_{-0.01}^{+0.01}$ & $-0.30_{-0.01}^{+0.01}$ & $0.53_{-0.00}^{+0.00}$ & $0.071_{-0.002}^{+0.003}$ & $0.045_{-0.002}^{+0.002}$ & $0.35_{-0.01}^{+0.01}$\\
$\Omega_{cc}^{+}\to\Lambda_{c}^{+}\omega $ &$-$ & $2.94_{-1.74}^{+3.00}$ & $2.94_{-1.74}^{+3.00}$ & $-0.020_{-0.041}^{+0.041}$ & $-0.43_{-0.02}^{+0.02}$ & $0.32_{-0.00}^{+0.00}$ & $-0.23_{-0.01}^{+0.01}$ & $0.24_{-0.01}^{+0.01}$ & $0.44_{-0.02}^{+0.02}$ & $0.046_{-0.004}^{+0.004}$ & $0.27_{-0.01}^{+0.01}$\\
$\Omega_{cc}^{+}\to\Sigma_{c}^{0}\rho^{+} $ &$-$ & $1.32_{-0.78}^{+1.36}$ & $1.32_{-0.78}^{+1.36}$ & $0.50_{-0.00}^{+0.00}$ & $0.062_{-0.006}^{+0.007}$ & $0.55_{-0.00}^{+0.00}$ & $0.28_{-0.00}^{+0.00}$ & $0.11_{-0.00}^{+0.00}$ & $0.33_{-0.00}^{+0.00}$ & $0.42_{-0.00}^{+0.00}$ & $0.14_{-0.00}^{+0.00}$\\
$\Omega_{cc}^{+}\to\Sigma_{c}^{++}\rho^{-} $ &$-$ & $6.44_{-3.80}^{+6.46}$ & $6.44_{-3.80}^{+6.46}$ & $-0.10_{-0.03}^{+0.03}$ & $-0.35_{-0.01}^{+0.01}$ & $0.35_{-0.00}^{+0.00}$ & $-0.23_{-0.01}^{+0.01}$ & $0.26_{-0.01}^{+0.01}$ & $0.39_{-0.01}^{+0.01}$ & $0.059_{-0.004}^{+0.004}$ & $0.29_{-0.00}^{+0.01}$\\
$\Omega_{cc}^{+}\to\Xi_{c}^{\prime+}K^{*0} $ &$10.5$ & $8.17_{-4.72}^{+8.33}$ & $8.53_{-4.85}^{+8.50}$ & $-0.24_{-0.03}^{+0.04}$ & $-0.14_{-0.02}^{+0.01}$ & $0.35_{-0.00}^{+0.00}$ & $-0.19_{-0.01}^{+0.01}$ & $0.35_{-0.01}^{+0.01}$ & $0.30_{-0.01}^{+0.02}$ & $0.080_{-0.004}^{+0.005}$ & $0.27_{-0.01}^{+0.00}$\\
$\Omega_{cc}^{+}\to\Xi_{c}^{+}K^{*0} $ &$6.37$ & $11.9_{-7.1}^{+12.6}$ & $12.2_{-7.2}^{+12.7}$ & $0.59_{-0.01}^{+0.01}$ & $-0.41_{-0.02}^{+0.01}$ & $0.21_{-0.00}^{+0.00}$ & $0.093_{-0.002}^{+0.002}$ & $0.14_{-0.01}^{+0.01}$ & $0.65_{-0.01}^{+0.01}$ & $0.15_{-0.00}^{+0.00}$ & $0.059_{-0.000}^{+0.000}$\\
\hline\hline
\end{tabular}
\end{sidewaystable}

\begin{figure}
  \centering
  \subfigure[]{\includegraphics[width=0.5\textwidth]{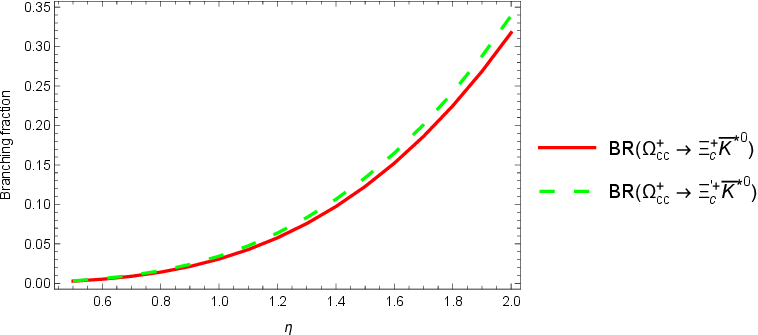}}
  \hfill
  \subfigure[]{\includegraphics[width=0.4\textwidth]{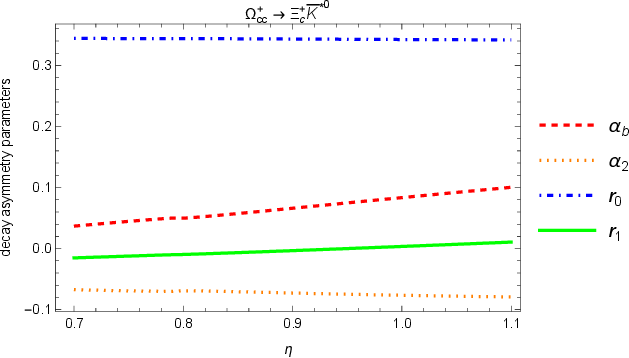}}
  \caption{The dependence of the branching ratios and decay parameters on the model parameter $\eta$.}
  \label{fig:dependence}
\end{figure}

We present our numerical results for the branching ratios, helicity amplitudes and four decay asymmetry parameters for all 49 decay channels. The results are summarized in four tables: Tab.~\ref{tab:result1} for short-distance dominated modes, Tab.~\ref{tab:result2} for long-distance dominated Cabibbo-favored modes, Tab.~\ref{tab:result3} for singly Cabibbo-suppressed modes, and Tab.~\ref{tab:result4} for doubly Cabibbo-suppressed modes.
In our previous study of the $\mathcal{B}_{cc}\to\mathcal{B}_{c}P$ process~\cite{Hu:2024uia}, the model parameter $\eta$ in the form factor Eq.~(\ref{eq:Ffactor}) was determined to be $\eta=0.9\pm0.2$ by fitting to the experimental data $\mathcal{B}(\Xi_{cc}^{++}\to\Xi_{c}^{\prime+}\pi^{+})/\mathcal{B}(\Xi_{cc}^{++}\to\Xi_{c}^{+}\pi^{+})=(1.41\pm0.17\pm0.10)$~\cite{LHCb:2022rpd}. Since the process $\mathcal{B}_{cc}\to\mathcal{B}_{c}V$ also involves a charm quark decay, we adopt the same value as in our previous work~\cite{Hu:2024uia}. We also analyze the dependence of the branching ratios and decay parameters on the model parameter $\eta$, as shown in Fig.~\ref{fig:dependence}.

The main features of our results are:
\begin{itemize}
\item The branching ratios for the short-distance dominated modes are typically at the level of $10^{-4}$ to $10^{-2}$. 
\item The long-distance contributions are significant for the non-factorizable channels, often comparable to or even larger than the short-distance contributions.
\item The decay asymmetry parameters $\alpha_b$, $\alpha_2$, $r_0$, and $r_1$ show rich patterns across different channels, reflecting the underlying dynamics of the decay processes.
\item One can see that the branching ratios are moderately sensitive to this parameter, while the decay asymmetry parameters show relatively smaller sensitivity. This feature is advantageous because the asymmetry parameters can be measured more precisely in experiments, providing stronger constraints on the theoretical framework.
\end{itemize}

\subsection{CP violations}
\begin{figure}
  \includegraphics[width=0.5\textwidth]{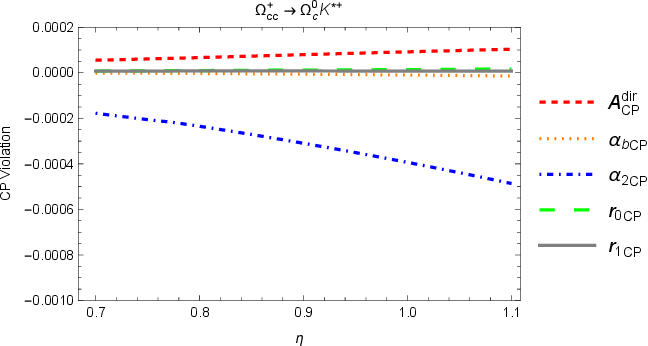}
  \caption{The dependence of the CP violations on the model parameter $\eta$.}
  \label{fig:CPVde}
\end{figure}

We present our numerical results for CP violations. The results are summarized in two tables: Tab.~\ref{tab:hecp1} for short-distance dominated modes, Tab.~\ref{tab:hecp3} for singly Cabibbo-suppressed modes. We also analyze the dependence of CP violations on the model parameter $\eta$ as shown in Fig.~\ref{fig:CPVde}.

The main features of our results are:
\begin{itemize}
\item The CP violations for the short-distance dominated modes are typically at the level of $10^{-5}$ to $10^{-4}$, and the ones of singly Cabibbo-suppressed modes are at the level of $10^{-7}$ to $10^{-3}$. 
\item One can see that the CP violations are relatively smaller sensitive to this parameter $\eta$. 
\end{itemize}

\begin{table}
  \caption{CP violations for short-distance dominated modes.}
\label{tab:hecp1} 
\begin{tabular}{l|c|c|c|c|c}
\hline\hline
channels &${A_{CP}^{\rm dir}}[10^{-5}]$&${\alpha_{b}}_{CP}[10^{-5}]$ &${\alpha_{2}}_{CP}[10^{-5}]$&${r_{0}}_{CP}[10^{-5}]$&${r_{1}}_{CP}[10^{-5}]$\\ \hline
$\Xi_{cc}^{++}\to\Sigma_{c}^{+}\rho^{+} $ &
$0.91_{-0.27}^{+0.23}$ &
$0.86_{-0.42}^{+0.69}$ &
$-2.1_{-0.2}^{+0.6}$ &
$-0.31_{0.10}^{+0.10}$ &
$0.29_{-0.15}^{+0.26}$\\
$\Xi_{cc}^{++}\to\Lambda_{c}^{+}\rho^{+} $ &
$2.1_{-0.7}^{+0.8}$ &
$-2.2_{-1.2}^{+0.9}$ &
$-1.1_{-0.3}^{+0.3}$ &
$0.35_{-0.16}^{+0.25}$ &
$-1.7_{-0.8}^{+0.7}$\\
$\Xi_{cc}^{++}\to\Xi_{c}^{\prime+}K^{*+} $ &
$-0.14_{-0.09}^{+0.07}$ &
$-0.39_{-0.22}^{+0.17}$ &
$26_{-40}^{+14}$ &
$0.18_{-0.09}^{+0.06}$ &
$-0.11_{-0.07}^{+0.06}$\\
$\Xi_{cc}^{++}\to\Xi_{c}^{+}K^{*+} $ &
$-1.1_{-0.3}^{+0.3}$ &
$1.00_{-0.31}^{+0.32}$ &
$0.60_{-0.32}^{+0.07}$ &
$-0.44_{-0.32}^{+0.15}$ &
$0.85_{-0.25}^{+0.24}$\\
$\Xi_{cc}^{+}\to\Sigma_{c}^{0}\rho^{+} $ &
$-0.68_{-0.34}^{+0.26}$ &
$0.24_{-0.08}^{+0.08}$ &
$-1.4_{-0.3}^{+0.4}$ &
$-0.16_{-0.34}^{+0.06}$ &
$0.043_{-0.015}^{+0.022}$\\
$\Xi_{cc}^{+}\to\Xi_{c}^{\prime0}K^{*+} $ &
$0.38_{-0.14}^{+0.13}$ &
$-0.017_{-0.007}^{+0.027}$ &
$64_{-59}^{+59}$ &
$-0.17_{-0.07}^{+0.07}$ &
$-0.22_{-0.04}^{+0.07}$\\
$\Xi_{cc}^{+}\to\Xi_{c}^{0}K^{*+} $ &
$-0.30_{-0.11}^{+0.10}$ &
$-0.57_{-0.31}^{+0.22}$ &
$-0.42_{-0.11}^{+0.17}$ &
$0.23_{-0.11}^{+0.09}$ &
$-0.51_{-0.29}^{+0.20}$\\
$\Omega_{cc}^{+}\to\Xi_{c}^{0}\rho^{+} $ &
$0.0094_{-0.0115}^{+0.0153}$ &
$-0.80_{-0.13}^{+0.18}$ &
$-0.82_{-0.02}^{+0.21}$ &
$-0.0077_{0.0100}^{+0.0100}$ &
$-0.81_{-0.13}^{+0.19}$\\
$\Omega_{cc}^{+}\to\Xi_{c}^{\prime0}\rho^{+} $ &
$-1.1_{-0.3}^{+0.3}$ &
$0.51_{-0.09}^{+0.03}$ &
$-0.16_{-0.29}^{+0.13}$ &
$0.071_{-0.291}^{+0.001}$ &
$0.36_{-0.07}^{+0.04}$\\
$\Omega_{cc}^{+}\to\Omega_{c}^{0}K^{*+} $ &
$0.80_{-0.25}^{+0.24}$ &
$-0.060_{-0.079}^{+0.048}$ &
$-3.1_{-0.2}^{+1.3}$ &
$0.12_{-0.04}^{+0.05}$ &
$0.080_{-0.010}^{+0.008}$\\
\hline\hline
\end{tabular}
\end{table}

\begin{table}
  \caption{CP violations for singly Cabibbo-suppressed modes.}
\label{tab:hecp3} 
\begin{tabular}{l|c|c|c|c|c}
\hline\hline
channels &${A_{CP}^{\rm dir}}[10^{-4}]$&${\alpha_{b}}_{CP}[10^{-4}]$ &${\alpha_{2}}_{CP}[10^{-4}]$&${r_{0}}_{CP}[10^{-4}]$&${r_{1}}_{CP}[10^{-4}]$\\ \hline
$\Xi_{cc}^{++}\to\Sigma_{c}^{++}\rho^{0} $ &
$0.46_{-0.04}^{+0.05}$ &
$1.6_{-0.3}^{+0.5}$ &
$-0.26_{-0.04}^{+0.01}$ &
$-0.038_{-0.039}^{+0.009}$ &
$1.1_{-0.2}^{+0.3}$\\
$\Xi_{cc}^{++}\to\Sigma_{c}^{++}\omega $ &
$-1.1_{-0.1}^{+0.0}$ &
$-0.41_{-0.07}^{+0.07}$ &
$-1.2_{-0.0}^{+0.7}$ &
$-0.36_{-0.01}^{+0.00}$ &
$-0.33_{-0.00}^{+0.00}$\\
$\Xi_{cc}^{+}\to\Sigma_{c}^{+}\rho^{0} $ &
$0.040_{-0.004}^{+0.006}$ &
$-0.072_{-0.015}^{+0.013}$ &
$0.24_{-0.04}^{+0.04}$ &
$-0.0080_{-0.0041}^{+0.0005}$ &
$0.0098_{-0.0025}^{+0.0020}$\\
$\Xi_{cc}^{+}\to\Sigma_{c}^{+}\omega $ &
$-0.30_{-0.03}^{+0.02}$ &
$-0.40_{-0.07}^{+0.06}$ &
$0.25_{-0.02}^{+0.02}$ &
$-0.017_{-0.009}^{+0.009}$ &
$-0.26_{-0.04}^{+0.04}$\\
$\Xi_{cc}^{+}\to\Lambda_{c}^{+}\rho^{0} $ &
$-0.0032_{-0.0015}^{+0.0002}$ &
$-0.29_{-0.04}^{+0.04}$ &
$1.2_{-0.0}^{+0.1}$ &
$-0.037_{-0.002}^{+0.001}$ &
$-0.89_{-0.09}^{+0.12}$\\
$\Xi_{cc}^{+}\to\Lambda_{c}^{+}\omega $ &
$0.068_{-0.015}^{+0.007}$ &
$-1.3_{-0.1}^{+0.1}$ &
$0.25_{0.01}^{+0.01}$ &
$0.19_{-0.00}^{+0.00}$ &
$0.78_{-0.04}^{+0.05}$\\
$\Xi_{cc}^{+}\to\Sigma_{c}^{++}\rho^{-} $ &
$-0.066_{-0.006}^{+0.005}$ &
$-0.37_{-0.05}^{+0.05}$ &
$0.61_{-0.00}^{+0.00}$ &
$0.012_{-0.001}^{+0.001}$ &
$-0.13_{-0.02}^{+0.02}$\\
$\Xi_{cc}^{+}\to\Xi_{c}^{\prime+}K^{*0} $ &
$-0.13_{-0.02}^{+0.02}$ &
$-8.4_{-1.4}^{+1.7}$ &
$-2.7_{-0.0}^{+0.4}$ &
$-0.26_{0.02}^{+0.03}$ &
$4.3_{-1.1}^{+3.7}$\\
$\Xi_{cc}^{+}\to\Xi_{c}^{+}K^{*0} $ &
$-0.41_{-0.06}^{+0.04}$ &
$-4.1_{-2.4}^{+1.8}$ &
$1.2_{-0.0}^{+0.1}$ &
$-0.29_{0.04}^{+0.06}$ &
$-0.93_{-0.16}^{+0.19}$\\
$\Omega_{cc}^{+}\to\Sigma_{c}^{+}\bar{K}^{*0} $ &
$-0.068_{-0.001}^{+0.001}$ &
$-0.31_{-0.02}^{+0.02}$ &
$-0.12_{-0.00}^{+0.00}$ &
$0.047_{-0.001}^{+0.001}$ &
$-6.0_{-16.1}^{+2.6}$\\
$\Omega_{cc}^{+}\to\Lambda_{c}^{+}\bar{K}^{*0} $ &
$-0.00085_{-0.00412}^{+0.00422}$ &
$-0.54_{-0.14}^{+0.10}$ &
$-0.11_{-0.01}^{+0.01}$ &
$0.038_{-0.004}^{+0.004}$ &
$0.089_{-0.011}^{+0.012}$\\
$\Omega_{cc}^{+}\to\Sigma_{c}^{++}K^{*-} $ &
$-0.28_{-0.04}^{+0.03}$ &
$2.7_{-0.9}^{+4.7}$ &
$0.17_{-0.02}^{+0.02}$ &
$0.0040_{-0.0070}^{+0.0070}$ &
$-0.29_{-0.06}^{+0.07}$\\
$\Omega_{cc}^{+}\to\Xi_{c}^{\prime+}\rho^{0} $ &
$0.13_{-0.01}^{+0.01}$ &
$0.57_{-0.14}^{+0.28}$ &
$-0.81_{-0.01}^{+0.08}$ &
$-0.0084_{-0.0102}^{+0.0003}$ &
$0.30_{-0.08}^{+0.15}$\\
$\Omega_{cc}^{+}\to\Xi_{c}^{\prime+}\phi $ &
$0.10_{-0.01}^{+0.01}$ &
$-1.3_{-1.0}^{+0.4}$ &
$-0.037_{-0.008}^{+0.007}$ &
$0.0087_{-0.0075}^{+0.0009}$ &
$0.78_{-0.17}^{+0.29}$\\
$\Omega_{cc}^{+}\to\Xi_{c}^{\prime+}\omega $ &
$-1.5_{-0.0}^{+0.0}$ &
$-0.73_{-0.20}^{+0.17}$ &
$0.83_{-0.01}^{+0.01}$ &
$0.17_{-0.04}^{+0.04}$ &
$-9.6_{-67.7}^{+4.8}$\\
$\Omega_{cc}^{+}\to\Xi_{c}^{+}\rho^{0} $ &
$0.096_{-0.006}^{+0.007}$ &
$-0.23_{-0.02}^{+0.02}$ &
$-0.043_{-0.020}^{+0.016}$ &
$-0.0015_{-0.0064}^{+0.0023}$ &
$-0.33_{-0.02}^{+0.02}$\\
$\Omega_{cc}^{+}\to\Xi_{c}^{+}\phi $ &
$0.0028_{-0.0003}^{+0.0005}$ &
$0.24_{-0.02}^{+0.03}$ &
$-0.17_{-0.01}^{+0.00}$ &
$0.0037_{-0.0007}^{+0.0008}$ &
$-40_{-33}^{+50}$\\
$\Omega_{cc}^{+}\to\Xi_{c}^{+}\omega $ &
$-0.45_{-0.01}^{+0.01}$ &
$13_{-41}^{+8}$ &
$0.36_{-0.01}^{+0.04}$ &
$-0.18_{-0.01}^{+0.02}$ &
$-0.50_{-0.06}^{+0.07}$\\
\hline\hline
\end{tabular}
\end{table}

\subsection{Analysis of $\Omega_{cc}$}
As shown in Tabs.~\ref{tab:result1} and ~\ref{tab:result2}, there are three decays,
all of which have large decay branching ratio. 
\begin{eqnarray}
  &&{\cal B}(\Omega_{cc}^{+}\to\Omega_{c}^{0}\rho^{+}) =22.52\%, \\
 && {\cal B}(\Omega_{cc}^{+}\to\Xi_{c}^{\prime+}\bar{K}^{*0})= (2.40_{-1.39}^{+2.36})\%, \\
  &&{\cal B}(\Omega_{cc}^{+}\to\Xi_{c}^{+}\bar{K}^{*0})=(2.14_{-1.24}^{+2.14})\%
\end{eqnarray}  
In this work, the value of lifetime of $\Omega_{cc}$ is taken as $176 {\rm fs}$~\cite{Cheng:2026mlv}.
In Ref~\cite{Jiang:2018oak}, the decay widths of these decays of $\Omega_{cc}$ have been calculated, in order to compared with them, we calculate the decay width of them, and list them in Tab.~\ref{tab:decaywidth}. 

For the first process $\Omega_{cc}^{+}\to\Omega_{c}^{0}\rho^{+} $, the two results are similar because this process only involves tree-level contributions (ignoring the s-channel contribution). Therefore, the slight difference between the two studies arises from the input parameters, such as decay constants or form factors.

For the other two processes $\Omega_{cc}^{+}\to\Xi_{c}^{\prime+}\bar{K}^{*0} $ and $\Omega_{cc}^{+}\to\Xi_{c}^{+}\bar{K}^{*0} $, we can see our results smaller than the ones of Ref~\cite{Jiang:2018oak}. This is because, different from Ref.~\cite{Jiang:2018oak}, the parameter in Eq.~\ref{eq:Ffactor}, we take the value $n=2$. This choice is more effective at suppressing the contribution of loop diagrams, leading to more reasonable results. Different from Ref.~\cite{Jiang:2018oak} computes loop diagrams using the cutting rule and only evaluates the imaginary part, we have calculated the complete loop diagrams. Furthermore, we have additionally taken into account the contributions from two other kinds of loop diagrams $\{{\cal {B}}_{c},P;{\cal {B}}_{c}^{*}\}$ and $\{{\cal {B}}_{c},V;{\cal {B}}_{c}^{*}\}$, as shown by Fig.~\ref{fig:trianglesev} (g) and (h).

At LHCb Run 3, an integrated luminosity of $23 \rm{fb}^{-1}$
is anticipated~\cite{Qin:2021zqx}, with a projected $\Omega_{cc}$
production cross-section of $60 \sim 1800 \rm{nb}$ in the LHCb detector~\cite{Berezhnoy:1998aa,Kiselev:2001fw,Ma:2003zk,Chang:2006xp,Chang:2006eu,Zhang:2011hi,Chang:2005bf,LHCb:2021rkb}.
 The corresponding production events of $\Omega_{cc}$, $N_{p}(\Omega_{cc})$ is approximately $10^9\sim10^{10}$. Considering the subsequent decay chain involving $Br(\Xi_{c}^{\prime+}\to \Xi_{c}^{+}\gamma)=100\%$, $Br(\Xi_{c}^+\to pK^{-}\pi^{+})=6.2\times10^{-3}$\cite{Belle:2019bgi,LHCb:2020gge}, $Br(\bar{K}^{*0}\to K^{+}\pi^{-})=100\%$~\cite{ParticleDataGroup:2024cfk}, $Br(\Omega_{c}^{0}\to \Omega^{-}\pi^{+})=5.1\times10^{-3}\sim3.43\times10^{-2}$~\cite{Zeng:2024yiv,Hsiao:2020gtc}, $Br(\Omega^{-}\to \Lambda K^{-})=67.8\%$~\cite{ParticleDataGroup:2024cfk}, $Br(\Lambda\to p \pi^{-})=63.9\%$~\cite{ParticleDataGroup:2024cfk}, $Br(\rho^{+}\to \pi^{+}\pi^{0})=100\%$~~\cite{ParticleDataGroup:2024cfk}, combined with detection efficiencies for final states $\epsilon_{p, K^{\pm},\pi^{\pm}} = 90 \sim 95\%$~\cite{LHCbRICHGroup:2012mgd,LHCb:2014nio}, $\epsilon_{\pi^{0}} = 20\sim 25\%$~\cite{Belyaev:2015nlt} and $\epsilon_{\gamma} \sim 1\%$~\cite{Govorkova:2015vqa}
the expected signal events $Ns$ for three decay channel can be estimated as follows,
\begin{eqnarray}
  Ns&=&Np(\Omega_{cc})Br(\Omega_{cc}^{+}\to\Omega_{c}^{0}\rho^{+})Br(\Omega_{c}^{0}\to \Omega^{-}\pi^{+})Br(\rho^{+}\to \pi^{+}\pi^{0})\nonumber\\
  &&\quad\times Br(\Omega^{-}\to \Lambda K^{-})Br(\Lambda\to p \pi^{-})(90\%)^5(25\%)\sim(7.2\times10^{4}-4.7\times10^6),\\
  Ns&=&Np(\Omega_{cc})Br(\Omega_{cc}^{+}\to\Xi_{c}^{\prime+}\bar{K}^{*0})Br(\Xi_{c}^{\prime+}\to \Xi_{c}^{+}\gamma)Br(\Xi_{c}^+\to pK^{-}\pi^{+})\nonumber\\
  &&\quad Br(\bar{K}^{*0}\to K^{+}\pi^{-})(90\%)^5(1\%)=(8.8\times10^{2}\sim8.8\times10^3),\\
  Ns&=&Np(\Omega_{cc})Br(\Omega_{cc}^{+}\to\Xi_{c}^{+}\bar{K}^{*0})Br(\Xi_{c}^+\to pK^{-}\pi^{+})Br(\bar{K}^{*0}\to K^{+}\pi^{-})(90\%)^5\nonumber\\
  &=&(7.8\times10^{4}-7.8\times10^5).
\end{eqnarray}


\begin{table}
  \caption{Comparison of decay widths for three decays $\mathcal{B}_{cc}\to\mathcal{B}_{c}V$ (${\Gamma}_{Tot}[10^{-14}]$)}
  \label{tab:decaywidth}
\begin{tabular}{l|c|c}
\hline\hline
channels  &This work &Ref.~\cite{Jiang:2018oak} \\ \hline
$\Omega_{cc}^{+}\to\Omega_{c}^{0}\rho^{+} $ &
$82.8$& $87.5$\\
\hline
$\Omega_{cc}^{+}\to\Xi_{c}^{\prime+}\bar{K}^{*0} $ &
$8.82_{-5.09}^{+8.66}$& $26.4_{-17.9}^{+27.2}$\\
$\Omega_{cc}^{+}\to\Xi_{c}^{+}\bar{K}^{*0} $ &
$7.86_{-4.57}^{+7.88}$&$13.8_{-9.5}^{+14.9}$\\
\hline\hline
\end{tabular} 
\end{table}

\section{Summary}
\label{sec:summary}
In this work, we study the two-body nonleptonic weak decays $\mathcal{B}_{cc} \to \mathcal{B}_c V$ (with $\mathcal{B}_{cc}=(\Xi_{cc}^{++},\Xi_{cc}^{+},\Omega_{cc}^{+})$, $\mathcal{B}_c=(\mathcal{B}_{\bar{3}},\mathcal{B}_6)$, $V=(\rho,K^*,\omega,\phi)$) under the final-state interaction mechanism. As shown in Tab.~\ref{tab:result1}, the factorizable diagram $T$ dominates over the long-distance contributions from $C'$ and $E_2$. Conversely, from Tabs.~\ref{tab:result2}-\ref{tab:result4}, long-distance effects dominate because the short-distance amplitude $C_{SD}$ is suppressed by the effective Wilson coefficients at the charm scale, where $a_2(m_c) = -0.017 < a_1(m_c) = 1.07$; the latter is used for the weak decay vertex. No absolute branching fraction for $\mathcal{B}_{cc} \to \mathcal{B}_c V$ has yet been measured directly. Since this process involves a charm quark decay, we adopt $\eta = 0.9 \pm 0.2$, as determined in our previous work~\cite{Hu:2024uia} by fitting the ratio $\mathcal{B}(\Xi_{cc}^{++}\to\Xi_{c}^{\prime+}\pi^{+})/\mathcal{B}(\Xi_{cc}^{++}\to\Xi_{c}^{+}\pi^{+}) = (1.41\pm0.17\pm0.10)$ from LHCb~\cite{LHCb:2022rpd}.

\begin{enumerate}
  \item[(1)] Using $\eta = 0.9 \pm 0.2$, we predict the branching ratios, decay asymmetry parameters, and CP violation for all $\mathcal{B}_{cc}\to\mathcal{B}_c V$ decays.
  \item[(2)] The decay asymmetry parameters are less sensitive to the model parameter $\eta$ than the branching ratios, as shown in Fig.~\ref{fig:dependence}, regardless of whether short-distance or long-distance contributions dominated.    
  \item[(3)] After the calculation of the decay widths of each decay mode, we conduct a comparative analysis with other theoretical works and analysis the production enevts in the LHCb experiment. Different from other study, our work include both real and imaginary contributions of the amplitudes. 
\end{enumerate}

\begin{acknowledgments}
This work is supported in part by the National Natural Science Foundation of China under Grant No. 12505114, 12005294 and No. 12335003, 
 the National Natural Science Foundation of China (Grant No. 12247101), the Fundamental Research Funds for the Central Universities (Grant No. lzujbky-2025-jdzx07), the Natural Science Foundation of Gansu Province (No.25JRRA799), and the '111 Center' under Grant No. B20063.
\end{acknowledgments}

\textbf{Data Availability Statement:}
No Data associated in the manuscript.

\appendix

\section{Amplitudes of each mode}
\label{app:amp}

The expressions of amplitudes for all the forty-seven $\mathcal{B}_{cc}\to\mathcal{B}_{c}V$
decays considered in this paper are collected in this Appendix. Twenty
amplitudes for the short-distance dominated modes are given as follows.
{\scriptsize{
\begin{equation}
\begin{aligned}
\mathcal A(\Xi_{cc}^{++} \to \Xi_c^{+} \rho^{+})
&= T_{SD}(\Xi_{cc}^{++} \to \Xi_c^{+} \rho^{+})+C_{SD}(\Xi_{cc}^{++} \to \Xi_c^{+} \rho^{+}) \\
&\quad + \mathcal M(\pi^{+}, \Xi_c^{(\prime)+}; \pi^{0}) + \mathcal M(\pi^{+}, \Xi_c^{(\prime)+}; \rho^{0}) + \mathcal M(\rho^{+}, \Xi_c^{(\prime)+}; \pi^{0}) + \mathcal M(\rho^{+}, \Xi_c^{(\prime)+}; \rho^{0}) \\
&\quad + \mathcal M(\Xi_c^{(\prime)+}, \pi^{+}; \Xi_c^{(\prime)0}) + \mathcal M(\Xi_c^{(\prime)+}, \rho^{+}; \Xi_c^{(\prime)0}) + \mathcal M(\Xi_c^{(\prime)+}, \pi^{+}; \Xi_c^{*\prime 0}) + \mathcal M(\Xi_c^{(\prime)+}, \rho^{+}; \Xi_c^{*\prime 0}) .
\end{aligned}
\end{equation}
\begin{equation}
\begin{aligned}
\mathcal A(\Xi_{cc}^{++} \to \Xi_c^{\prime +} \rho^{+})
&= T_{SD}(\Xi_{cc}^{++} \to \Xi_c^{\prime +} \rho^{+}) + C_{SD}(\Xi_{cc}^{++} \to \Xi_c^{\prime +} \rho^{+})\\
&\quad + \mathcal M(\pi^{+}, \Xi_c^{(\prime)+}; \pi^{0}) + \mathcal M(\pi^{+}, \Xi_c^{(\prime)+}; \rho^{0}) + \mathcal M(\rho^{+}, \Xi_c^{(\prime)+}; \pi^{0}) + \mathcal M(\rho^{+}, \Xi_c^{(\prime)+}; \rho^{0}) \\
&\quad + \mathcal M(\Xi_c^{(\prime)+}, \pi^{+}; \Xi_c^{(\prime)0}) + \mathcal M(\Xi_c^{(\prime)+}, \rho^{+}; \Xi_c^{(\prime)0}) + \mathcal M(\Xi_c^{(\prime)+}, \pi^{+}; \Xi_c^{*\prime 0}) + \mathcal M(\Xi_c^{(\prime)+}, \rho^{+}; \Xi_c^{*\prime 0}) .
\end{aligned}
\end{equation}
\begin{equation}
\begin{aligned}
\mathcal A(\Xi_{cc}^{++} \to \Sigma_c^{+} \rho^{+})
&= T_{SD}(\Xi_{cc}^{++} \to \Sigma_c^{+} \rho^{+})+C_{SD}(\Xi_{cc}^{++} \to \Sigma_c^{+} \rho^{+}) \\
&\quad + \mathcal M(\pi^{+}, \Lambda_c^{+}; \pi^{0}) + \mathcal M(K^{+}, \Xi_c^{(\prime)+}; K^{0}) + \mathcal M(\pi^{+}, \Lambda_c^{+}; \rho^{0}) + \mathcal M(K^{+}, \Xi_c^{(\prime)+}; K^{*0}) \\
&\quad + \mathcal M(\rho^{+}, \Lambda_c^{+}; \pi^{0}) + \mathcal M(K^{*+}, \Xi_c^{(\prime)+}; K^{0}) + \mathcal M(\rho^{+}, \Lambda_c^{+}; \rho^{0}) + \mathcal M(K^{*+}, \Xi_c^{(\prime)+}; K^{*0}) \\
&\quad + \mathcal M(\Lambda_c^{+}/\Sigma_c^{+}, \pi^{+}; \Sigma_c^{0}) + \mathcal M(\Xi_c^{(\prime)+}, K^{+}; \Xi_c^{(\prime)0}) + \mathcal M(\Lambda_c^{+}/\Sigma_c^{+}, \rho^{+}; \Sigma_c^{0}) + \mathcal M(\Xi_c^{(\prime)+}, K^{*+}; \Xi_c^{(\prime)0}) \\
&\quad  + \mathcal M(\Lambda_c^{+}/\Sigma_c^{+}, \pi^{+}; \Sigma_c^{*0}) + \mathcal M(\Xi_c^{(\prime)+}, K^{+}; \Xi_c^{*\prime 0}) + \mathcal M(\Lambda_c^{+}/\Sigma_c^{+}, \rho^{+}; \Sigma_c^{*0}) + \mathcal M(\Xi_c^{(\prime)+}, K^{*+}; \Xi_c^{*\prime 0}) .
\end{aligned}
\end{equation}
\begin{equation}
\begin{aligned}
\mathcal A(\Xi_{cc}^{++} \to \Lambda_c^{+} \rho^{+})
&= T_{SD}(\Xi_{cc}^{++} \to \Lambda_c^{+} \rho^{+}) +C_{SD}(\Xi_{cc}^{++} \to \Lambda_c^{+} \rho^{+})\\
&\quad + \mathcal M(\pi^{+}, \Sigma_c^{+}; \pi^{0}) + \mathcal M(K^{+}, \Xi_c^{(\prime)+}; K^{0}) + \mathcal M(\pi^{+}, \Sigma_c^{+}; \rho^{0}) + \mathcal M(K^{+}, \Xi_c^{(\prime)+}; K^{*0}) \\
&\quad + \mathcal M(\rho^{+}, \Sigma_c^{+}; \pi^{0}) + \mathcal M(K^{*+}, \Xi_c^{(\prime)+}; K^{0}) + \mathcal M(\rho^{+}, \Sigma_c^{+}; \rho^{0}) + \mathcal M(K^{*+}, \Xi_c^{(\prime)+}; K^{*0}) \\
&\quad + \mathcal M(\Lambda_c^{+}/\Sigma_c^{+}, \pi^{+}; \Sigma_c^{0}) + \mathcal M(\Xi_c^{(\prime)+}, K^{+}; \Xi_c^{(\prime)0}) + \mathcal M(\Lambda_c^{+}/\Sigma_c^{+}, \rho^{+}; \Sigma_c^{0}) + \mathcal M(\Xi_c^{(\prime)+}, K^{*+}; \Xi_c^{(\prime)0}) \\
&\quad  + \mathcal M(\Lambda_c^{+}/\Sigma_c^{+}, \pi^{+}; \Sigma_c^{*0}) + \mathcal M(\Xi_c^{(\prime)+}, K^{+}; \Xi_c^{*\prime 0}) + \mathcal M(\Lambda_c^{+}/\Sigma_c^{+}, \rho^{+}; \Sigma_c^{*0}) + \mathcal M(\Xi_c^{(\prime)+}, K^{*+}; \Xi_c^{*\prime 0}) .
\end{aligned}
\end{equation}
\begin{equation}
\begin{aligned}
\mathcal A(\Xi_{cc}^{++} \to \Xi_c^{\prime +} K^{*+})
&= T_{SD}(\Xi_{cc}^{++} \to \Xi_c^{\prime +} K^{*+}) +C_{SD}(\Xi_{cc}^{++} \to \Xi_c^{\prime +} K^{*+})\\
&\quad + \mathcal M(K^{+}, \Xi_c^{(\prime)+}; \pi^{0}) + \mathcal M(K^{+}, \Xi_c^{(\prime)+}; \eta_8) + \mathcal M(K^{+}, \Xi_c^{\prime+}; \eta_1) + \mathcal M(\pi^{+}, \Lambda_c^{+}/\Sigma_c^{+}; \bar K^{0}) \\
&\quad + \mathcal M(K^{+}, \Xi_c^{(\prime)+}; \rho^{0}) + \mathcal M(K^{+}, \Xi_c^{(\prime)+}; \omega) + \mathcal M(K^{+}, \Xi_c^{(\prime)+}; \phi) + \mathcal M(\pi^{+}, \Lambda_c^{+}/\Sigma_c^{+}; \bar K^{*0}) \\
&\quad + \mathcal M(K^{*+}, \Xi_c^{(\prime)+}; \pi^{0}) + \mathcal M(K^{*+}, \Xi_c^{(\prime)+}; \eta_8) + \mathcal M(K^{*+}, \Xi_c^{\prime+}; \eta_1) + \mathcal M(\rho^{+}, \Lambda_c^{+}/\Sigma_c^{+}; \bar K^{0}) \\
&\quad + \mathcal M(K^{*+}, \Xi_c^{(\prime)+}; \rho^{0}) + \mathcal M(K^{*+}, \Xi_c^{(\prime)+}; \omega) + \mathcal M(K^{*+}, \Xi_c^{(\prime)+}; \phi) + \mathcal M(\rho^{+}, \Lambda_c^{+}/\Sigma_c^{+}; \bar K^{*0}) \\
&\quad + \mathcal M(\Xi_c^{(\prime)+}, K^{+}; \Omega_c^{0}) + \mathcal M(\Lambda_c^{+}/\Sigma_c^{+}, \pi^{+}; \Xi_c^{(\prime)0}) + \mathcal M(\Xi_c^{(\prime)+}, K^{*+}; \Omega_c^{0}) + \mathcal M(\Lambda_c^{+}/\Sigma_c^{+}, \rho^{+}; \Xi_c^{(\prime)0}) \\
&\quad + \mathcal M(\Xi_c^{(\prime)+}, K^{+}; \Omega_c^{*0}) + \mathcal M(\Lambda_c^{+}/\Sigma_c^{+}, \pi^{+}; \Xi_c^{*\prime 0}) + \mathcal M(\Xi_c^{(\prime)+}, K^{*+}; \Omega_c^{*0}) + \mathcal M(\Lambda_c^{+}/\Sigma_c^{+}, \rho^{+}; \Xi_c^{*\prime 0}) .
\end{aligned}
\end{equation}
\begin{equation}
\begin{aligned}
\mathcal A(\Xi_{cc}^{++} \to \Xi_c^{+} K^{*+})
&= T_{SD}(\Xi_{cc}^{++} \to \Xi_c^{+} K^{*+})+C_{SD}(\Xi_{cc}^{++} \to \Xi_c^{+} K^{*+}) \\
&\quad + \mathcal M(K^{+}, \Xi_c^{(\prime)+}; \pi^{0}) + \mathcal M(K^{+}, \Xi_c^{(\prime)+}; \eta_8) + \mathcal M(K^{+}, \Xi_c^{+}; \eta_1) + \mathcal M(\pi^{+}, \Lambda_c^{+}/\Sigma_c^{+}; \bar K^{0}) \\
&\quad + \mathcal M(K^{+}, \Xi_c^{(\prime)+}; \rho^{0}) + \mathcal M(K^{+}, \Xi_c^{(\prime)+}; \omega) + \mathcal M(K^{+}, \Xi_c^{(\prime)+}; \phi) + \mathcal M(\pi^{+}, \Lambda_c^{+}/\Sigma_c^{+}; \bar K^{*0}) \\
&\quad + \mathcal M(K^{*+}, \Xi_c^{(\prime)+}; \pi^{0}) + \mathcal M(K^{*+}, \Xi_c^{(\prime)+}; \eta_8) + \mathcal M(K^{*+}, \Xi_c^{+}; \eta_1) + \mathcal M(\rho^{+}, \Lambda_c^{+}/\Sigma_c^{+}; \bar K^{0}) \\
&\quad + \mathcal M(K^{*+}, \Xi_c^{(\prime)+}; \rho^{0}) + \mathcal M(K^{*+}, \Xi_c^{(\prime)+}; \omega) + \mathcal M(K^{*+}, \Xi_c^{(\prime)+}; \phi) + \mathcal M(\rho^{+}, \Lambda_c^{+}/\Sigma_c^{+}; \bar K^{*0}) \\
&\quad + \mathcal M(\Xi_c^{(\prime)+}, K^{+}; \Omega_c^{0}) + \mathcal M(\Lambda_c^{+}/\Sigma_c^{+}, \pi^{+}; \Xi_c^{(\prime)0}) + \mathcal M(\Xi_c^{(\prime)+}, K^{*+}; \Omega_c^{0}) + \mathcal M(\Lambda_c^{+}/\Sigma_c^{+}, \rho^{+}; \Xi_c^{(\prime)0}) \\
&\quad + \mathcal M(\Xi_c^{(\prime)+}, K^{+}; \Omega_c^{*0}) + \mathcal M(\Lambda_c^{+}/\Sigma_c^{+}, \pi^{+}; \Xi_c^{*\prime 0}) + \mathcal M(\Xi_c^{(\prime)+}, K^{*+}; \Omega_c^{*0}) + \mathcal M(\Lambda_c^{+}/\Sigma_c^{+}, \rho^{+}; \Xi_c^{*\prime 0}) .
\end{aligned}
\end{equation}
\begin{equation}
\begin{aligned}
\mathcal A(\Xi_{cc}^{++} \to \Sigma_c^{+} K^{*+})
&= T_{SD}(\Xi_{cc}^{++} \to \Sigma_c^{+} K^{*+}) +C_{SD}(\Xi_{cc}^{++} \to \Sigma_c^{+} K^{*+})\\
&\quad + \mathcal M(K^{+}, \Lambda_c^{+}; \pi^{0}) + \mathcal M(K^{+}, \Sigma_c^{+}; \eta_8) + \mathcal M(K^{+}, \Sigma_c^{+}; \eta_1) + \mathcal M(K^{+}, \Lambda_c^{+}; \rho^{0}) \\
&\quad + \mathcal M(K^{+}, \Sigma_c^{+}; \omega) + \mathcal M(K^{*+}, \Lambda_c^{+}; \pi^{0}) + \mathcal M(K^{*+}, \Sigma_c^{+}; \eta_8) + \mathcal M(K^{*+}, \Sigma_c^{+}; \eta_1) \\
&\quad + \mathcal M(K^{*+}, \Lambda_c^{+}; \rho^{0}) + \mathcal M(K^{*+}, \Sigma_c^{+}; \omega) + \mathcal M(\Lambda_c^{+}/\Sigma_c^{+}, K^{+}; \Xi_c^{(\prime)0}) + \mathcal M(\Lambda_c^{+}/\Sigma_c^{+}, K^{*+}; \Xi_c^{(\prime)0}) \\
&\quad + \mathcal M(\Lambda_c^{+}/\Sigma_c^{+}, K^{+}; \Xi_c^{*\prime 0}) + \mathcal M(\Lambda_c^{+}/\Sigma_c^{+}, K^{*+}; \Xi_c^{*\prime 0}) .
\end{aligned}
\end{equation}
\begin{equation}
\begin{aligned}
\mathcal A(\Xi_{cc}^{++} \to \Lambda_c^{+} K^{*+})
&= T_{SD}(\Xi_{cc}^{++} \to \Lambda_c^{+} K^{*+}) +C_{SD}(\Xi_{cc}^{++} \to \Lambda_c^{+} K^{*+})\\
&\quad + \mathcal M(K^{+}, \Sigma_c^{+}; \pi^{0}) + \mathcal M(K^{+}, \Lambda_c^{+}; \eta_8) + \mathcal M(K^{+}, \Lambda_c^{+}; \eta_1) + \mathcal M(K^{+}, \Sigma_c^{+}; \rho^{0}) \\
&\quad + \mathcal M(K^{+}, \Lambda_c^{+}; \omega) + \mathcal M(K^{*+}, \Sigma_c^{+}; \pi^{0}) + \mathcal M(K^{*+}, \Lambda_c^{+}; \eta_8) + \mathcal M(K^{*+}, \Lambda_c^{+}; \eta_1) \\
&\quad + \mathcal M(K^{*+}, \Sigma_c^{+}; \rho^{0}) + \mathcal M(K^{*+}, \Lambda_c^{+}; \omega) + \mathcal M(\Lambda_c^{+}/\Sigma_c^{+}, K^{+}; \Xi_c^{(\prime)0}) + \mathcal M(\Lambda_c^{+}/\Sigma_c^{+}, K^{*+}; \Xi_c^{(\prime)0}) \\
&\quad + \mathcal M(\Lambda_c^{+}/\Sigma_c^{+}, K^{+}; \Xi_c^{*\prime 0}) + \mathcal M(\Lambda_c^{+}/\Sigma_c^{+}, K^{*+}; \Xi_c^{*\prime 0}) .
\end{aligned}
\end{equation}
\begin{equation}
\begin{aligned}
\mathcal A(\Xi_{cc}^{+} \to \Xi_c^{0} \rho^{+})
&= T_{SD}(\Xi_{cc}^{+} \to \Xi_c^{0} \rho^{+}) \\
&\quad + \mathcal M(\pi^{+}, \Xi_c^{(\prime)0}; \pi^{0}) 
+ \mathcal M(\pi^{+}, \Xi_c^{(\prime)0}; \eta_8)
 + \mathcal M(\pi^{+}, \Xi_c^{0}; \eta_1) 
  + \mathcal M(\pi^{+}, \Xi_c^{(\prime)0}; \rho^{0})
 \\
&\quad + \mathcal M(\pi^{+}, \Xi_c^{(\prime)0}; \omega) 
 + \mathcal M(\rho^{+}, \Xi_c^{(\prime)0}; \pi^{0}) 
 + \mathcal M(\rho^{+}, \Xi_c^{(\prime)0}; \eta_8) 
 + \mathcal M(\rho^{+}, \Xi_c^{0}; \eta_1) \\
&\quad  + \mathcal M(\rho^{+}, \Xi_c^{(\prime)0}; \rho^{0}) + \mathcal M(\rho^{+}, \Xi_c^{(\prime)0}; \omega)  .
\end{aligned}
\end{equation}
\begin{equation}
\begin{aligned}
\mathcal A(\Xi_{cc}^{+} \to \Xi_c^{\prime 0} \rho^{+})
&= T_{SD}(\Xi_{cc}^{+} \to \Xi_c^{\prime 0} \rho^{+}) \\
&\quad + \mathcal M(\pi^{+}, \Xi_c^{(\prime)0}; \pi^{0}) 
+ \mathcal M(\pi^{+}, \Xi_c^{(\prime)0}; \eta_8)
 + \mathcal M(\pi^{+}, \Xi_c^{\prime 0}; \eta_1) 
  + \mathcal M(\pi^{+}, \Xi_c^{(\prime)0}; \rho^{0})
 \\
&\quad + \mathcal M(\pi^{+}, \Xi_c^{(\prime)0}; \omega) 
 + \mathcal M(\rho^{+}, \Xi_c^{(\prime)0}; \pi^{0}) 
 + \mathcal M(\rho^{+}, \Xi_c^{(\prime)0}; \eta_8) 
 + \mathcal M(\rho^{+}, \Xi_c^{\prime 0}; \eta_1) \\
&\quad  + \mathcal M(\rho^{+}, \Xi_c^{(\prime)0}; \rho^{0}) + \mathcal M(\rho^{+}, \Xi_c^{(\prime)0}; \omega)  .
\end{aligned}
\end{equation}
\begin{equation}
\begin{aligned}
\mathcal A(\Xi_{cc}^{+} \to \Sigma_c^{0} \rho^{+})
&= T_{SD}(\Xi_{cc}^{+} \to \Sigma_c^{0} \rho^{+})  \\
&\quad + \mathcal M(\pi^{+}, \Sigma_c^{0}; \pi^{0}) + \mathcal M(\pi^{+}, \Sigma_c^{0}; \eta_8) + \mathcal M(\pi^{+}, \Sigma_c^{0}; \eta_1) + \mathcal M(K^{+}, \Xi_c^{(\prime)0}; K^{0}) \\
&\quad + \mathcal M(\pi^{+}, \Sigma_c^{0}; \rho^{0}) + \mathcal M(\pi^{+}, \Sigma_c^{0}; \omega) + \mathcal M(K^{+}, \Xi_c^{(\prime)0}; K^{*0}) \\
&\quad  + \mathcal M(\rho^{+}, \Sigma_c^{0}; \pi^{0}) + \mathcal M(\rho^{+}, \Sigma_c^{0}; \eta_8) + \mathcal M(\rho^{+}, \Sigma_c^{0}; \eta_1) + \mathcal M(K^{*+}, \Xi_c^{(\prime)0}; K^{0})\\
&\quad   + \mathcal M(\rho^{+}, \Sigma_c^{0}; \rho^{0}) + \mathcal M(\rho^{+}, \Sigma_c^{0}; \omega)  + \mathcal M(K^{*+}, \Xi_c^{(\prime)0}; K^{*0}).
\end{aligned}
\end{equation}
\begin{equation}
\begin{aligned}
\mathcal A(\Xi_{cc}^{+} \to \Xi_c^{0} K^{*+})
&= T_{SD}(\Xi_{cc}^{+} \to \Xi_c^{0} K^{*+}) \\
&\quad + \mathcal M(\pi^{+}, \Sigma_c^{0}; \bar K^{0}) + \mathcal M(K^{+}, \Xi_c^{(\prime) 0}; \pi^{0}) + \mathcal M(K^{+}, \Xi_c^{(\prime)0}; \eta_8)  + \mathcal M(K^{+}, \Xi_c^{0}; \eta_1)  \\
&\quad + \mathcal M(\pi^{+}, \Sigma_c^{0}; \bar K^{*0})  + \mathcal M(K^{+}, \Xi_c^{(\prime) 0}; \rho^{0}) + \mathcal M(K^{+}, \Xi_c^{(\prime)0}; \omega) 
+ \mathcal M(K^{+}, \Xi_c^{(\prime)0}; \phi)\\
&\quad  + \mathcal M(\rho^{+}, \Sigma_c^{0}; \bar K^{0}) + \mathcal M(K^{*+}, \Xi_c^{(\prime)0}; \pi^{0})  + \mathcal M(K^{*+}, \Xi_c^{(\prime)0}; \eta_8)  + \mathcal M(K^{*+}, \Xi_c^{0}; \eta_1) \\
&\quad + \mathcal M(\rho^{+}, \Sigma_c^{0}; \bar K^{*0}) + \mathcal M(K^{*+}, \Xi_c^{(\prime)0}; \rho^{0})  + \mathcal M(K^{*+}, \Xi_c^{(\prime)0}; \omega)  + \mathcal M(K^{*+}, \Xi_c^{(\prime)0}; \phi).
\end{aligned}
\end{equation}
\begin{equation}
\begin{aligned}
\mathcal A(\Xi_{cc}^{+} \to \Xi_c^{\prime 0} K^{*+})
&= T_{SD}(\Xi_{cc}^{+} \to \Xi_c^{\prime 0} K^{*+}) \\
&\quad + \mathcal M(\pi^{+}, \Sigma_c^{0}; \bar K^{0}) + \mathcal M(K^{+}, \Xi_c^{(\prime) 0}; \pi^{0}) + \mathcal M(K^{+}, \Xi_c^{(\prime)0}; \eta_8)  + \mathcal M(K^{+}, \Xi_c^{\prime 0}; \eta_1)  \\
&\quad + \mathcal M(\pi^{+}, \Sigma_c^{0}; \bar K^{*0})  + \mathcal M(K^{+}, \Xi_c^{(\prime) 0}; \rho^{0}) + \mathcal M(K^{+}, \Xi_c^{(\prime)0}; \omega) 
+ \mathcal M(K^{+}, \Xi_c^{(\prime)0}; \phi)\\
&\quad  + \mathcal M(\rho^{+}, \Sigma_c^{0}; \bar K^{0}) + \mathcal M(K^{*+}, \Xi_c^{(\prime)0}; \pi^{0})  + \mathcal M(K^{*+}, \Xi_c^{(\prime)0}; \eta_8)  + \mathcal M(K^{*+}, \Xi_c^{\prime 0}; \eta_1) \\
&\quad + \mathcal M(\rho^{+}, \Sigma_c^{0}; \bar K^{*0}) + \mathcal M(K^{*+}, \Xi_c^{(\prime)0}; \rho^{0})  + \mathcal M(K^{*+}, \Xi_c^{(\prime)0}; \omega)  + \mathcal M(K^{*+}, \Xi_c^{(\prime)0}; \phi).
\end{aligned}
\end{equation}
\begin{equation}
\begin{aligned}
\mathcal{A}(\Xi_{cc}^{+}\to\Sigma_{c}^{0}K^{*+}) & =\mathcal{T}_{SD}(\Xi_{cc}^{+}\to\Sigma_{c}^{0}K^{*+}).\\
\mathcal{A}(\Omega_{cc}^{+}\to\Omega_{c}^{0}\rho^{+}) & =\mathcal{T}_{SD}(\Omega_{cc}^{+}\to\Omega_{c}^{0}\rho^{+}).\\
\end{aligned}
\end{equation}
\begin{equation}
\begin{aligned}
\mathcal A(\Omega_{cc}^{+} \to \Xi_c^{0} \rho^{+})
&= T_{SD}(\Omega_{cc}^{+} \to \Xi_c^{0} \rho^{+}) \\
&\quad +\mathcal M(\pi^{+}, \Xi_c^{(\prime)0}; \pi^{0}) + \mathcal M(\pi^{+}, \Xi_c^{(\prime) 0}; \eta_8)  + \mathcal M(\pi^{+}, \Xi_c^{0}; \eta_1) + \mathcal M(K^{+}, \Omega_c^{0}; K^{0}) \\
&\quad+ \mathcal M(\pi^{+}, \Xi_c^{(\prime)0}; \rho^{0}) + \mathcal M(\pi^{+}, \Xi_c^{(\prime)0}; \omega)  + \mathcal M(K^{+}, \Omega_c^{0}; K^{*0}) \\
&\quad+ \mathcal M(\rho^{+}, \Xi_c^{(\prime)0}; \pi^{0})  + \mathcal M(\rho^{+}, \Xi_c^{(\prime)0}; \eta_8)  + \mathcal M(\rho^{+}, \Xi_c^{0}; \eta_1)+ \mathcal M(K^{*+}, \Omega_c^{0}; K^{0}) \\
&\quad  + \mathcal M(\rho^{+}, \Xi_c^{(\prime)0}; \rho^{0}) + \mathcal M(\rho^{+}, \Xi_c^{(\prime)0}; \omega) + \mathcal M(K^{*+}, \Omega_c^{0}; K^{*0}) .
\end{aligned}
\end{equation}
\begin{equation}
\begin{aligned}
\mathcal A(\Omega_{cc}^{+} \to \Xi_c^{\prime 0} \rho^{+})
&= T_{SD}(\Omega_{cc}^{+} \to \Xi_c^{\prime 0} \rho^{+}) \\
&\quad +\mathcal M(\pi^{+}, \Xi_c^{(\prime)0}; \pi^{0}) + \mathcal M(\pi^{+}, \Xi_c^{(\prime) 0}; \eta_8)  + \mathcal M(\pi^{+}, \Xi_c^{\prime 0}; \eta_1) + \mathcal M(K^{+}, \Omega_c^{0}; K^{0}) \\
&\quad+ \mathcal M(\pi^{+}, \Xi_c^{(\prime)0}; \rho^{0}) + \mathcal M(\pi^{+}, \Xi_c^{(\prime)0}; \omega)  + \mathcal M(K^{+}, \Omega_c^{0}; K^{*0}) \\
&\quad+ \mathcal M(\rho^{+}, \Xi_c^{(\prime)0}; \pi^{0})  + \mathcal M(\rho^{+}, \Xi_c^{(\prime)0}; \eta_8)  + \mathcal M(\rho^{+}, \Xi_c^{\prime 0}; \eta_1)+ \mathcal M(K^{*+}, \Omega_c^{0}; K^{0}) \\
&\quad  + \mathcal M(\rho^{+}, \Xi_c^{(\prime)0}; \rho^{0}) + \mathcal M(\rho^{+}, \Xi_c^{(\prime)0}; \omega) + \mathcal M(K^{*+}, \Omega_c^{0}; K^{*0}) .
\end{aligned}
\end{equation}
\begin{equation}
\begin{aligned}
\mathcal A(\Omega_{cc}^{+} \to \Omega_c^{0} K^{*+})
&= T_{SD}(\Omega_{cc}^{+} \to \Omega_c^{0} K^{*+}) \\
&\quad  + \mathcal M(\pi^{+}, \Xi_c^{(\prime) 0}; \bar K^{0}) + \mathcal M(K^{+}, \Omega_c^{0}; \eta_8) + \mathcal M(K^{+}, \Omega_c^{0}; \eta_1) + \mathcal M(\pi^{+}, \Xi_c^{(\prime)0}; \bar K^{*0}) \\
&\quad + \mathcal M(K^{+}, \Omega_c^{0}; \phi) + \mathcal M(\rho^{+}, \Xi_c^{(\prime)0}; \bar K^{0})  + \mathcal M(K^{*+}, \Omega_c^{0}; \eta_8) + \mathcal M(K^{*+}, \Omega_c^{0}; \eta_1) \\
&\quad+ \mathcal M(\rho^{+}, \Xi_c^{(\prime)0}; \bar K^{*0})  + \mathcal M(K^{*+}, \Omega_c^{0}; \phi) .
\end{aligned}
\end{equation}
\begin{equation}
\begin{aligned}
\mathcal A(\Omega_{cc}^{+} \to \Xi_c^{0} K^{*+})
&= T_{SD}(\Omega_{cc}^{+} \to \Xi_c^{0} K^{*+}) \\
&\quad + \mathcal M(K^{+}, \Xi_c^{(\prime)0}; \pi^{0}) + \mathcal M(K^{+}, \Xi_c^{(\prime)0}; \eta_8)   + \mathcal M(K^{+}, \Xi_c^{0}; \eta_1) + \mathcal M(K^{+}, \Xi_c^{(\prime)0}; \rho^{0}) \\
&\quad + \mathcal M(K^{+}, \Xi_c^{(\prime)0}; \omega)  + \mathcal M(K^{+}, \Xi_c^{(\prime)0}; \phi) + \mathcal M(K^{*+}, \Xi_c^{(\prime)0}; \pi^{0})  + \mathcal M(K^{*+}, \Xi_c^{(\prime)0}; \eta_8) \\
&\quad  + \mathcal M(K^{*+}, \Xi_c^{0}; \eta_1) + \mathcal M(K^{*+}, \Xi_c^{(\prime)0}; \rho^{0}) + \mathcal M(K^{*+}, \Xi_c^{(\prime)0}; \omega)   + \mathcal M(K^{*+}, \Xi_c^{(\prime)0}; \phi) .
\end{aligned}
\end{equation}
\begin{equation}
\begin{aligned}
\mathcal A(\Omega_{cc}^{+} \to \Xi_c^{\prime 0} K^{*+})
&= T_{SD}(\Omega_{cc}^{+} \to \Xi_c^{\prime 0} K^{*+}) \\
&\quad + \mathcal M(K^{+}, \Xi_c^{(\prime)0}; \pi^{0}) + \mathcal M(K^{+}, \Xi_c^{(\prime)0}; \eta_8)   + \mathcal M(K^{+}, \Xi_c^{\prime 0}; \eta_1) + \mathcal M(K^{+}, \Xi_c^{(\prime)0}; \rho^{0}) \\
&\quad + \mathcal M(K^{+}, \Xi_c^{(\prime)0}; \omega)  + \mathcal M(K^{+}, \Xi_c^{(\prime)0}; \phi) + \mathcal M(K^{*+}, \Xi_c^{(\prime)0}; \pi^{0})  + \mathcal M(K^{*+}, \Xi_c^{(\prime)0}; \eta_8) \\
&\quad  + \mathcal M(K^{*+}, \Xi_c^{\prime 0}; \eta_1) + \mathcal M(K^{*+}, \Xi_c^{(\prime)0}; \rho^{0}) + \mathcal M(K^{*+}, \Xi_c^{(\prime)0}; \omega)   + \mathcal M(K^{*+}, \Xi_c^{(\prime)0}; \phi) .
\end{aligned}
\end{equation}
}}
The amplitudes for long-distance dominated and Cabibbo flavored
modes can be given as follows.
 {\scriptsize{
\begin{equation}
\begin{aligned}
\mathcal A(\Xi_{cc}^{++} \to \Sigma_c^{++} \bar K^{*0})
&= C_{SD}(\Xi_{cc}^{++} \to \Sigma_c^{++} \bar K^{*0}) \\
&\quad + \mathcal M(\pi^{+}, \Xi_c^{(\prime)+}; K^{+}) + \mathcal M(\pi^{+}, \Xi_c^{(\prime)+}; K^{*+}) + \mathcal M(\rho^{+}, \Xi_c^{(\prime)+}; K^{+}) + \mathcal M(\rho^{+}, \Xi_c^{(\prime)+}; K^{*+}) \\
&\quad + \mathcal M(\Xi_c^{(\prime)+}, \pi^{+}; \Lambda_c^{+}/\Sigma_c^{+})  + \mathcal M(\Xi_c^{(\prime)+}, \rho^{+}; \Lambda_c^{+}/\Sigma_c^{+})   + \mathcal M(\Xi_c^{(\prime)+}, \pi^{+}; \Sigma_c^{*+}) + \mathcal M(\Xi_c^{(\prime)+}, \rho^{+}; \Sigma_c^{*+}) .
\end{aligned}
\end{equation}

\begin{equation}
\begin{aligned}
\mathcal A(\Xi_{cc}^{+} \to \Omega_c^{0} K^{*+})
&=  \mathcal M(\pi^{+}, \Xi_c^{(\prime) 0}; \bar K^{0}) + \mathcal M(\pi^{+}, \Xi_c^{(\prime) 0}; \bar K^{*0})  + \mathcal M(\rho^{+}, \Xi_c^{(\prime) 0}; \bar K^{0}) + \mathcal M(\rho^{+}, \Xi_c^{(\prime) 0}; \bar K^{*0}) .
\end{aligned}
\end{equation}

\begin{equation}
\begin{aligned}
\mathcal A(\Xi_{cc}^{+} \to \Sigma_c^{+} \bar K^{*0})
&= C_{SD}(\Xi_{cc}^{+} \to \Sigma_c^{+} \bar K^{*0}) \\
&\quad + \mathcal M(\pi^{+}, \Xi_c^{(\prime) 0}; K^{+}) + \mathcal M(\pi^{+}, \Xi_c^{(\prime) 0}; K^{*+}) + \mathcal M(\rho^{+}, \Xi_c^{(\prime) 0}; K^{+}) + \mathcal M(\rho^{+}, \Xi_c^{(\prime) 0}; K^{*+})  \\
&\quad + \mathcal M(\Xi_c^{(\prime) 0}, \pi^{+}; \Sigma_c^{0})  + \mathcal M(\Xi_c^{(\prime) 0}, \rho^{+}; \Sigma_c^{0})  + \mathcal M(\Xi_c^{(\prime) 0}, \pi^{+}; \Sigma_c^{*0}) + \mathcal M(\Xi_c^{(\prime) 0}, \rho^{+}; \Sigma_c^{*0})  .
\end{aligned}
\end{equation}

\begin{equation}
\begin{aligned}
\mathcal A(\Xi_{cc}^{+} \to \Lambda_c^{+} \bar K^{*0})
&= C_{SD}(\Xi_{cc}^{+} \to \Lambda_c^{+} \bar K^{*0}) \\
&\quad + \mathcal M(\pi^{+}, \Xi_c^{(\prime) 0}; K^{+}) + \mathcal M(\pi^{+}, \Xi_c^{(\prime) 0}; K^{*+}) + \mathcal M(\rho^{+}, \Xi_c^{(\prime) 0}; K^{+}) + \mathcal M(\rho^{+}, \Xi_c^{(\prime) 0}; K^{*+})  \\
&\quad + \mathcal M(\Xi_c^{(\prime) 0}, \pi^{+}; \Sigma_c^{0})  + \mathcal M(\Xi_c^{(\prime) 0}, \rho^{+}; \Sigma_c^{0})  + \mathcal M(\Xi_c^{(\prime) 0}, \pi^{+}; \Sigma_c^{*0}) + \mathcal M(\Xi_c^{(\prime) 0}, \rho^{+}; \Sigma_c^{*0})  .
\end{aligned}
\end{equation}
\begin{equation}
\begin{aligned}
\mathcal A(\Xi_{cc}^{+} \to \Sigma_c^{++} K^{*-})
&= \mathcal M(\Xi_c^{(\prime)0}, \pi^{+}; \Lambda_c^{+}/\Sigma_c^{+}) + \mathcal M(\Xi_c^{(\prime)0}, \rho^{+}; \Lambda_c^{+}/\Sigma_c^{+}) + \mathcal M(\Xi_c^{(\prime)0}, \pi^{+}; \Sigma_c^{*+}) + \mathcal M(\Xi_c^{(\prime)0}, \rho^{+}; \Sigma_c^{*+}) .
\end{aligned}
\end{equation}
\begin{equation}
\begin{aligned}
\mathcal A(\Xi_{cc}^{+} \to \Xi_c^{\prime +} \rho^{0})
&= \mathcal M(\pi^{+}, \Xi_c^{(\prime)0}; \pi^{+})  
+ \mathcal M(\pi^{+}, \Xi_c^{(\prime)0}; \rho^{+})  
+ \mathcal M(\rho^{+}, \Xi_c^{(\prime)0}; \pi^{+})
+ \mathcal M(\rho^{+}, \Xi_c^{(\prime)0}; \rho^{+}) \\
&\quad + \mathcal M(\Xi_c^{(\prime)0}, \pi^{+}; \Xi_c^{(\prime)0}) 
 + \mathcal M(\Xi_c^{(\prime)0}, \rho^{+}; \Xi_c^{(\prime)0})  
  + \mathcal M(\Xi_c^{(\prime)0}, \pi^{+}; \Xi_c^{*\prime 0})
   + \mathcal M(\Xi_c^{(\prime)0}, \rho^{+}; \Xi_c^{*\prime 0}) .
\end{aligned}
\end{equation}
\begin{equation}
\begin{aligned}
\mathcal A(\Xi_{cc}^{+} \to \Xi_c^{\prime +} \phi)
&=  \mathcal M(\Xi_c^{(\prime)0}, \pi^{+}; \Xi_c^{(\prime)0}) 
 + \mathcal M(\Xi_c^{(\prime)0}, \rho^{+}; \Xi_c^{(\prime)0})  
  + \mathcal M(\Xi_c^{(\prime)0}, \pi^{+}; \Xi_c^{*\prime 0})
   + \mathcal M(\Xi_c^{(\prime)0}, \rho^{+}; \Xi_c^{*\prime 0}) .
\end{aligned}
\end{equation}
\begin{equation}
\begin{aligned}
\mathcal A(\Xi_{cc}^{+} \to \Xi_c^{\prime +} \omega)
&= \mathcal M(\pi^{+}, \Xi_c^{(\prime)0}; \pi^{+})  
+ \mathcal M(\pi^{+}, \Xi_c^{(\prime)0}; \rho^{+})  
+ \mathcal M(\rho^{+}, \Xi_c^{(\prime)0}; \pi^{+})
+ \mathcal M(\rho^{+}, \Xi_c^{(\prime)0}; \rho^{+}) \\
&\quad + \mathcal M(\Xi_c^{(\prime)0}, \pi^{+}; \Xi_c^{(\prime)0}) 
 + \mathcal M(\Xi_c^{(\prime)0}, \rho^{+}; \Xi_c^{(\prime)0})  
  + \mathcal M(\Xi_c^{(\prime)0}, \pi^{+}; \Xi_c^{*\prime 0})
   + \mathcal M(\Xi_c^{(\prime)0}, \rho^{+}; \Xi_c^{*\prime 0}) .
\end{aligned}
\end{equation}
\begin{equation}
\begin{aligned}
\mathcal A(\Xi_{cc}^{+} \to \Xi_c^{+} \rho^{0})
&= \mathcal M(\pi^{+}, \Xi_c^{(\prime)0}; \pi^{+})  
+ \mathcal M(\pi^{+}, \Xi_c^{(\prime)0}; \rho^{+})  
+ \mathcal M(\rho^{+}, \Xi_c^{(\prime)0}; \pi^{+})
+ \mathcal M(\rho^{+}, \Xi_c^{(\prime)0}; \rho^{+}) \\
&\quad + \mathcal M(\Xi_c^{(\prime)0}, \pi^{+}; \Xi_c^{(\prime)0}) 
 + \mathcal M(\Xi_c^{(\prime)0}, \rho^{+}; \Xi_c^{(\prime)0})  
  + \mathcal M(\Xi_c^{(\prime)0}, \pi^{+}; \Xi_c^{*\prime 0})
   + \mathcal M(\Xi_c^{(\prime)0}, \rho^{+}; \Xi_c^{*\prime 0}) .
\end{aligned}
\end{equation}
\begin{equation}
\begin{aligned}
\mathcal A(\Xi_{cc}^{+} \to \Xi_c^{+} \phi)
&= \mathcal M(\Xi_c^{(\prime)0}, \pi^{+}; \Xi_c^{(\prime)0}) 
 + \mathcal M(\Xi_c^{(\prime)0}, \rho^{+}; \Xi_c^{(\prime)0})  
  + \mathcal M(\Xi_c^{(\prime)0}, \pi^{+}; \Xi_c^{*\prime 0})
   + \mathcal M(\Xi_c^{(\prime)0}, \rho^{+}; \Xi_c^{*\prime 0}) .
\end{aligned}
\end{equation}
\begin{equation}
\begin{aligned}
\mathcal A(\Xi_{cc}^{+} \to \Xi_c^{+} \omega)
&= \mathcal M(\pi^{+}, \Xi_c^{(\prime)0}; \pi^{+})  
+ \mathcal M(\pi^{+}, \Xi_c^{(\prime)0}; \rho^{+})  
+ \mathcal M(\rho^{+}, \Xi_c^{(\prime)0}; \pi^{+})
+ \mathcal M(\rho^{+}, \Xi_c^{(\prime)0}; \rho^{+}) \\
&\quad + \mathcal M(\Xi_c^{(\prime)0}, \pi^{+}; \Xi_c^{(\prime)0}) 
 + \mathcal M(\Xi_c^{(\prime)0}, \rho^{+}; \Xi_c^{(\prime)0})  
  + \mathcal M(\Xi_c^{(\prime)0}, \pi^{+}; \Xi_c^{*\prime 0})
   + \mathcal M(\Xi_c^{(\prime)0}, \rho^{+}; \Xi_c^{*\prime 0}) .
\end{aligned}
\end{equation}
\begin{equation}
\begin{aligned}
\mathcal A(\Omega_{cc}^{+} \to \Xi_c^{\prime +} \bar K^{*0})
&= C_{SD}(\Omega_{cc}^{+} \to \Xi_c^{\prime +} \bar K^{*0}) \\
&\quad + \mathcal M(\pi^{+}, \Omega_c^{0}; K^{+}) + \mathcal M(\pi^{+}, \Omega_c^{0}; K^{*+}) + \mathcal M(\rho^{+}, \Omega_c^{0}; K^{+}) + \mathcal M(\rho^{+}, \Omega_c^{0}; K^{*+}) \\
&\quad + \mathcal M(\Omega_c^{0}, \pi^{+}; \Xi_c^{(\prime)0})  + \mathcal M(\Omega_c^{0}, \rho^{+}; \Xi_c^{(\prime)0}) + \mathcal M(\Omega_c^{0}, \pi^{+}; \Xi_c^{*\prime 0}) + \mathcal M(\Omega_c^{0}, \rho^{+}; \Xi_c^{*\prime 0}) .
\end{aligned}
\end{equation}
\begin{equation}
\begin{aligned}
\mathcal A(\Omega_{cc}^{+} \to \Xi_c^{+} \bar K^{*0})
&= C_{SD}(\Omega_{cc}^{+} \to \Xi_c^{+} \bar K^{*0}) \\
&\quad + \mathcal M(\pi^{+}, \Omega_c^{0}; K^{+}) + \mathcal M(\pi^{+}, \Omega_c^{0}; K^{*+}) + \mathcal M(\rho^{+}, \Omega_c^{0}; K^{+}) + \mathcal M(\rho^{+}, \Omega_c^{0}; K^{*+}) \\
&\quad + \mathcal M(\Omega_c^{0}, \pi^{+}; \Xi_c^{(\prime)0})  + \mathcal M(\Omega_c^{0}, \rho^{+}; \Xi_c^{(\prime)0}) + \mathcal M(\Omega_c^{0}, \pi^{+}; \Xi_c^{*\prime 0}) + \mathcal M(\Omega_c^{0}, \rho^{+}; \Xi_c^{*\prime 0}) .
\end{aligned}
\end{equation}
}}

The amplitudes for long-distance dominated singly Cabibbo suppressed
modes can be given as follows.
{\scriptsize{
\begin{equation}
\begin{aligned}
\mathcal A(\Xi_{cc}^{++} \to \Sigma_c^{++} \rho^{0})
&= C_{SD}(\Xi_{cc}^{++} \to \Sigma_c^{++} \rho^{0}) \\
&\quad + \mathcal M(\pi^{+}, \Lambda_c^{+}/\Sigma_c^{+}; \pi^{+}) 
 + \mathcal M(K^{+}, \Xi_c^{(\prime)+}; K^{+}) 
 + \mathcal M(\pi^{+}, \Lambda_c^{+}/\Sigma_c^{+}; \rho^{+}) 
 + \mathcal M(K^{+}, \Xi_c^{(\prime)+}; K^{*+})\\
&\quad + \mathcal M(\rho^{+}, \Lambda_c^{+}/\Sigma_c^{+}; \pi^{+}) 
 + \mathcal M(K^{*+}, \Xi_c^{(\prime)+}; K^{+})
  + \mathcal M(\rho^{+}, \Lambda_c^{+}/\Sigma_c^{+}; \rho^{+})  
  + \mathcal M(K^{*+}, \Xi_c^{(\prime)+}; K^{*+}) \\
&\quad + \mathcal M(\Lambda_c^{+}, \pi^{+}; \Sigma_c^{+}) 
+ \mathcal M(\Sigma_c^{+}, \pi^{+}; \Lambda_c^{+}) 
+ \mathcal M(\Xi_c^{(\prime)+}, K^{+}; \Xi_c^{(\prime)+})  
 + \mathcal M(\Lambda_c^{+}, \rho^{+}; \Sigma_c^{+})
 \\
&\quad + \mathcal M(\Sigma_c^{+}, \rho^{+}; \Lambda_c^{+})
 + \mathcal M(\Xi_c^{(\prime)+}, K^{*+}; \Xi_c^{(\prime)+})  
  + \mathcal M(\Lambda_c^{+}, \pi^{+}; \Sigma_c^{*+}) 
  + \mathcal M(\Xi_c^{(\prime)+}, K^{+}; \Xi_c^{*\prime +}) 
  \\
&\quad+ \mathcal M(\Lambda_c^{+}, \rho^{+}; \Sigma_c^{*+}) 
+ \mathcal M(\Xi_c^{(\prime)+}, K^{*+}; \Xi_c^{*\prime +}) .
\end{aligned}
\end{equation}
\begin{equation}
\begin{aligned}
\mathcal A(\Xi_{cc}^{++} \to \Sigma_c^{++} \phi)
&= C_{SD}(\Xi_{cc}^{++} \to \Sigma_c^{++} \phi) \\
&\quad + \mathcal M(K^{+}, \Xi_c^{(\prime)+}; K^{+})  
 + \mathcal M(K^{+}, \Xi_c^{(\prime)+}; K^{*+})  
  + \mathcal M(K^{*+}, \Xi_c^{(\prime)+}; K^{+})
    + \mathcal M(K^{*+}, \Xi_c^{(\prime)+}; K^{*+}) \\
&\quad  + \mathcal M(\Xi_c^{(\prime)+}, K^{+}; \Xi_c^{(\prime)+}) 
  + \mathcal M(\Xi_c^{(\prime)+}, K^{*+}; \Xi_c^{(\prime)+})
    + \mathcal M(\Xi_c^{(\prime)+}, K^{+}; \Xi_c^{*\prime +}) 
    + \mathcal M(\Xi_c^{(\prime)+}, K^{*+}; \Xi_c^{*\prime +}) .
\end{aligned}
\end{equation}
\begin{equation}
\begin{aligned}
\mathcal A(\Xi_{cc}^{++} \to \Sigma_c^{++} \omega)
&= C_{SD}(\Xi_{cc}^{++} \to \Sigma_c^{++} \omega)\\
&\quad + \mathcal M(\pi^{+}, \Lambda_c^{+}/\Sigma_c^{+}; \pi^{+}) 
 + \mathcal M(K^{+}, \Xi_c^{(\prime)+}; K^{+}) 
 + \mathcal M(\pi^{+}, \Lambda_c^{+}/\Sigma_c^{+}; \rho^{+}) 
 + \mathcal M(K^{+}, \Xi_c^{(\prime)+}; K^{*+})\\
&\quad + \mathcal M(\rho^{+}, \Lambda_c^{+}/\Sigma_c^{+}; \pi^{+}) 
 + \mathcal M(K^{*+}, \Xi_c^{(\prime)+}; K^{+})
  + \mathcal M(\rho^{+}, \Lambda_c^{+}/\Sigma_c^{+}; \rho^{+})  
  + \mathcal M(K^{*+}, \Xi_c^{(\prime)+}; K^{*+}) \\
&\quad + \mathcal M(\Lambda_c^{+}, \pi^{+}; \Lambda_c^{+}) + \mathcal M(\Sigma_c^{+}, \pi^{+}; \Sigma_c^{+}) + \mathcal M(\Xi_c^{(\prime)+}, K^{+}; \Xi_c^{(\prime)+})  \\
&\quad + \mathcal M(\Lambda_c^{+}, \rho^{+}; \Lambda_c^{+}) + \mathcal M(\Sigma_c^{+}, \rho^{+}; \Sigma_c^{+}) + \mathcal M(\Xi_c^{(\prime)+}, K^{*+}; \Xi_c^{(\prime)+})  \\
&\quad + \mathcal M(\Sigma_c^{+}, \pi^{+}; \Sigma_c^{*+}) + \mathcal M(\Xi_c^{(\prime)+}, K^{+}; \Xi_c^{*\prime +}) + \mathcal M(\Sigma_c^{+}, \rho^{+}; \Sigma_c^{*+}) + \mathcal M(\Xi_c^{(\prime)+}, K^{*+}; \Xi_c^{*\prime +}) .
\end{aligned}
\end{equation}
\begin{equation}
\begin{aligned}
\mathcal A(\Xi_{cc}^{+} \to \Sigma_c^{+} \rho^{0})
&= C_{SD}(\Xi_{cc}^{+} \to \Sigma_c^{+} \rho^{0}) \\
&\quad + \mathcal M(\pi^{+}, \Sigma_c^{0}; \pi^{+}) 
+ \mathcal M(K^{+}, \Xi_c^{(\prime) 0}; K^{+})
 + \mathcal M(\pi^{+}, \Sigma_c^{0}; \rho^{+}) 
 + \mathcal M(K^{+}, \Xi_c^{(\prime) 0}; K^{*+}) \\
&\quad + \mathcal M(\rho^{+}, \Sigma_c^{0}; \pi^{+}) 
 + \mathcal M(K^{*+}, \Xi_c^{(\prime) 0}; K^{+}) 
+ \mathcal M(\rho^{+}, \Sigma_c^{0}; \rho^{+})
 +  \mathcal M(K^{*+}, \Xi_c^{(\prime) 0}; K^{*+}) \\
&\quad + \mathcal M(\Sigma_c^{0}, \pi^{+}; \Sigma_c^{0}) 
+ \mathcal M(\Xi_c^{(\prime)0}, K^{+}; \Xi_c^{(\prime)0}) 
 + \mathcal M(\Sigma_c^{0}, \rho^{+}; \Sigma_c^{0}) 
+ \mathcal M(\Xi_c^{(\prime)0}, K^{*+}; \Xi_c^{(\prime)0}) \\
&\quad  + \mathcal M(\Sigma_c^{0}, \pi^{+}; \Sigma_c^{*0}) 
+ \mathcal M(\Xi_c^{(\prime)0}, K^{+}; \Xi_c^{*\prime 0}) 
 + \mathcal M(\Sigma_c^{0}, \rho^{+}; \Sigma_c^{*0}) 
 + \mathcal M(\Xi_c^{(\prime)0}, K^{*+}; \Xi_c^{*\prime 0}) .
\end{aligned}
\end{equation}
\begin{equation}
\begin{aligned}
\mathcal A(\Xi_{cc}^{+} \to \Sigma_c^{+} \phi)
&= C_{SD}(\Xi_{cc}^{+} \to \Sigma_c^{+} \phi)\\
&\quad + \mathcal M(K^{+}, \Xi_c^{(\prime)0}; K^{+}) 
 + \mathcal M(K^{+}, \Xi_c^{(\prime)0}; K^{*+})  
 + \mathcal M(K^{*+}, \Xi_c^{(\prime)0}; K^{+}) 
 + \mathcal M(K^{*+}, \Xi_c^{(\prime)0}; K^{*+})  \\
&\quad + \mathcal M(\Xi_c^{(\prime)0}, K^{+}; \Xi_c^{(\prime)0}) 
+ \mathcal M(\Xi_c^{(\prime)0}, K^{*+}; \Xi_c^{(\prime)0}) 
 + \mathcal M(\Xi_c^{(\prime)0}, K^{+}; \Xi_c^{*\prime 0}) 
  + \mathcal M(\Xi_c^{(\prime)0}, K^{*+}; \Xi_c^{*\prime 0}).
\end{aligned}
\end{equation}
\begin{equation}
\begin{aligned}
\mathcal A(\Xi_{cc}^{+} \to \Sigma_c^{+} \omega)
&= C_{SD}(\Xi_{cc}^{+} \to \Sigma_c^{+} \omega) \\
&\quad + \mathcal M(\pi^{+}, \Sigma_c^{0}; \pi^{+}) 
+ \mathcal M(K^{+}, \Xi_c^{(\prime) 0}; K^{+})
 + \mathcal M(\pi^{+}, \Sigma_c^{0}; \rho^{+}) 
 + \mathcal M(K^{+}, \Xi_c^{(\prime) 0}; K^{*+}) \\
&\quad + \mathcal M(\rho^{+}, \Sigma_c^{0}; \pi^{+}) 
 + \mathcal M(K^{*+}, \Xi_c^{(\prime) 0}; K^{+}) 
+ \mathcal M(\rho^{+}, \Sigma_c^{0}; \rho^{+})
 +  \mathcal M(K^{*+}, \Xi_c^{(\prime) 0}; K^{*+}) \\
&\quad + \mathcal M(\Sigma_c^{0}, \pi^{+}; \Sigma_c^{0}) 
+ \mathcal M(\Xi_c^{(\prime)0}, K^{+}; \Xi_c^{(\prime)0}) 
 + \mathcal M(\Sigma_c^{0}, \rho^{+}; \Sigma_c^{0}) 
+ \mathcal M(\Xi_c^{(\prime)0}, K^{*+}; \Xi_c^{(\prime)0}) \\
&\quad  + \mathcal M(\Sigma_c^{0}, \pi^{+}; \Sigma_c^{*0}) 
+ \mathcal M(\Xi_c^{(\prime)0}, K^{+}; \Xi_c^{*\prime 0}) 
 + \mathcal M(\Sigma_c^{0}, \rho^{+}; \Sigma_c^{*0}) 
 + \mathcal M(\Xi_c^{(\prime)0}, K^{*+}; \Xi_c^{*\prime 0}) .
\end{aligned}
\end{equation}
\begin{equation}
\begin{aligned}
\mathcal A(\Xi_{cc}^{+} \to \Lambda_c^{+} \rho^{0})
&= C_{SD}(\Xi_{cc}^{+} \to \Lambda_c^{+} \rho^{0}) \\
&\quad + \mathcal M(\pi^{+}, \Sigma_c^{0}; \pi^{+}) 
+ \mathcal M(K^{+}, \Xi_c^{(\prime) 0}; K^{+})
 + \mathcal M(\pi^{+}, \Sigma_c^{0}; \rho^{+}) 
 + \mathcal M(K^{+}, \Xi_c^{(\prime) 0}; K^{*+}) \\
&\quad + \mathcal M(\rho^{+}, \Sigma_c^{0}; \pi^{+}) 
 + \mathcal M(K^{*+}, \Xi_c^{(\prime) 0}; K^{+}) 
+ \mathcal M(\rho^{+}, \Sigma_c^{0}; \rho^{+})
 +  \mathcal M(K^{*+}, \Xi_c^{(\prime) 0}; K^{*+}) \\
&\quad + \mathcal M(\Sigma_c^{0}, \pi^{+}; \Sigma_c^{0}) 
+ \mathcal M(\Xi_c^{(\prime)0}, K^{+}; \Xi_c^{(\prime)0}) 
 + \mathcal M(\Sigma_c^{0}, \rho^{+}; \Sigma_c^{0}) 
+ \mathcal M(\Xi_c^{(\prime)0}, K^{*+}; \Xi_c^{(\prime)0}) \\
&\quad  + \mathcal M(\Sigma_c^{0}, \pi^{+}; \Sigma_c^{*0}) 
+ \mathcal M(\Xi_c^{(\prime)0}, K^{+}; \Xi_c^{*\prime 0}) 
 + \mathcal M(\Sigma_c^{0}, \rho^{+}; \Sigma_c^{*0}) 
 + \mathcal M(\Xi_c^{(\prime)0}, K^{*+}; \Xi_c^{*\prime 0}) .
\end{aligned}
\end{equation}

\begin{equation}
\begin{aligned}
\mathcal A(\Xi_{cc}^{+} \to \Lambda_c^{+} \phi)
&= C_{SD}(\Xi_{cc}^{+} \to \Lambda_c^{+} \phi) \\
&\quad + \mathcal M(K^{+}, \Xi_c^{(\prime)0}; K^{+}) 
 + \mathcal M(K^{+}, \Xi_c^{(\prime)0}; K^{*+})  
 + \mathcal M(K^{*+}, \Xi_c^{(\prime)0}; K^{+}) 
 + \mathcal M(K^{*+}, \Xi_c^{(\prime)0}; K^{*+})  \\
&\quad + \mathcal M(\Xi_c^{(\prime)0}, K^{+}; \Xi_c^{(\prime)0}) 
+ \mathcal M(\Xi_c^{(\prime)0}, K^{*+}; \Xi_c^{(\prime)0}) 
 + \mathcal M(\Xi_c^{(\prime)0}, K^{+}; \Xi_c^{*\prime 0}) 
  + \mathcal M(\Xi_c^{(\prime)0}, K^{*+}; \Xi_c^{*\prime 0}).
\end{aligned}
\end{equation}
\begin{equation}
\begin{aligned}
\mathcal A(\Xi_{cc}^{+} \to \Lambda_c^{+} \omega)
&= C_{SD}(\Xi_{cc}^{+} \to \Lambda_c^{+} \omega) \\
&\quad + \mathcal M(\pi^{+}, \Sigma_c^{0}; \pi^{+}) 
+ \mathcal M(K^{+}, \Xi_c^{(\prime) 0}; K^{+})
 + \mathcal M(\pi^{+}, \Sigma_c^{0}; \rho^{+}) 
 + \mathcal M(K^{+}, \Xi_c^{(\prime) 0}; K^{*+}) \\
&\quad + \mathcal M(\rho^{+}, \Sigma_c^{0}; \pi^{+}) 
 + \mathcal M(K^{*+}, \Xi_c^{(\prime) 0}; K^{+}) 
+ \mathcal M(\rho^{+}, \Sigma_c^{0}; \rho^{+})
 +  \mathcal M(K^{*+}, \Xi_c^{(\prime) 0}; K^{*+}) \\
&\quad + \mathcal M(\Sigma_c^{0}, \pi^{+}; \Sigma_c^{0}) 
+ \mathcal M(\Xi_c^{(\prime)0}, K^{+}; \Xi_c^{(\prime)0}) 
 + \mathcal M(\Sigma_c^{0}, \rho^{+}; \Sigma_c^{0}) 
+ \mathcal M(\Xi_c^{(\prime)0}, K^{*+}; \Xi_c^{(\prime)0}) \\
&\quad  + \mathcal M(\Sigma_c^{0}, \pi^{+}; \Sigma_c^{*0}) 
+ \mathcal M(\Xi_c^{(\prime)0}, K^{+}; \Xi_c^{*\prime 0}) 
 + \mathcal M(\Sigma_c^{0}, \rho^{+}; \Sigma_c^{*0}) 
 + \mathcal M(\Xi_c^{(\prime)0}, K^{*+}; \Xi_c^{*\prime 0}) .
\end{aligned}
\end{equation}
\begin{equation}
\begin{aligned}
\mathcal A(\Xi_{cc}^{+} \to \Sigma_c^{++} \rho^{-})
&= \mathcal M(\Sigma_c^{0}, \pi^{+}; \Lambda_c^{+}/\Sigma_c^{+}) 
+ \mathcal M(\Xi_c^{(\prime)0}, K^{+}; \Xi_c^{(\prime)+}) 
+ \mathcal M(\Sigma_c^{0}, \rho^{+}; \Lambda_c^{+}/\Sigma_c^{+}) 
 + \mathcal M(\Xi_c^{(\prime)0}, K^{*+}; \Xi_c^{(\prime)+})\\
&\quad + \mathcal M(\Sigma_c^{0}, \pi^{+}; \Sigma_c^{*+}) 
+ \mathcal M(\Xi_c^{(\prime)0}, K^{+}; \Xi_c^{*\prime +}) 
+ \mathcal M(\Sigma_c^{0}, \rho^{+}; \Sigma_c^{*+})
 + \mathcal M(\Xi_c^{(\prime)0}, K^{*+}; \Xi_c^{*\prime +}) .
\end{aligned}
\end{equation}
\begin{equation}
\begin{aligned}
\mathcal A(\Xi_{cc}^{+} \to \Xi_c^{\prime +} K^{*0})
&= \mathcal M(K^{+}, \Xi_c^{(\prime)0}; \pi^{+}) + \mathcal M(K^{+}, \Xi_c^{(\prime)0}; \rho^{+}) 
 + \mathcal M(K^{*+}, \Xi_c^{(\prime)0}; \pi^{+}) + \mathcal M(K^{*+}, \Xi_c^{(\prime)0}; \rho^{+})  \\
&\quad + \mathcal M(\Xi_c^{(\prime)0}, K^{+}; \Omega_c^{0}) + \mathcal M(\Sigma_c^{0}, \pi^{+}; \Xi_c^{(\prime)0}) 
+ \mathcal M(\Xi_c^{(\prime)0}, K^{*+}; \Omega_c^{0}) + \mathcal M(\Sigma_c^{0}, \rho^{+}; \Xi_c^{(\prime)0})  \\
&\quad + \mathcal M(\Xi_c^{(\prime)0}, K^{+}; \Omega_c^{*0}) + \mathcal M(\Sigma_c^{0}, \pi^{+}; \Xi_c^{*\prime 0}) + \mathcal M(\Xi_c^{(\prime)0}, K^{*+}; \Omega_c^{*0}) + \mathcal M(\Sigma_c^{0}, \rho^{+}; \Xi_c^{*\prime 0}) .
\end{aligned}
\end{equation}
\begin{equation}
\begin{aligned}
\mathcal A(\Xi_{cc}^{+} \to \Xi_c^{+} K^{*0})
&= \mathcal M(K^{+}, \Xi_c^{(\prime)0}; \pi^{+}) + \mathcal M(K^{+}, \Xi_c^{(\prime)0}; \rho^{+}) 
 + \mathcal M(K^{*+}, \Xi_c^{(\prime)0}; \pi^{+}) + \mathcal M(K^{*+}, \Xi_c^{(\prime)0}; \rho^{+})  \\
&\quad + \mathcal M(\Xi_c^{(\prime)0}, K^{+}; \Omega_c^{0}) + \mathcal M(\Sigma_c^{0}, \pi^{+}; \Xi_c^{(\prime)0}) 
+ \mathcal M(\Xi_c^{(\prime)0}, K^{*+}; \Omega_c^{0}) + \mathcal M(\Sigma_c^{0}, \rho^{+}; \Xi_c^{(\prime)0})  \\
&\quad + \mathcal M(\Xi_c^{(\prime)0}, K^{+}; \Omega_c^{*0}) + \mathcal M(\Sigma_c^{0}, \pi^{+}; \Xi_c^{*\prime 0}) + \mathcal M(\Xi_c^{(\prime)0}, K^{*+}; \Omega_c^{*0}) + \mathcal M(\Sigma_c^{0}, \rho^{+}; \Xi_c^{*\prime 0}) .
\end{aligned}
\end{equation}
\begin{equation}
\begin{aligned}
\mathcal A(\Omega_{cc}^{+} \to \Sigma_c^{+} \bar K^{*0})
&= \mathcal M(\pi^{+}, \Xi_c^{(\prime)0}; K^{+})  + \mathcal M(\pi^{+}, \Xi_c^{(\prime)0}; K^{*+})  
 + \mathcal M(\rho^{+}, \Xi_c^{(\prime)0}; K^{+}) + \mathcal M(\rho^{+}, \Xi_c^{(\prime)0}; K^{*+}) \\
&\quad + \mathcal M(\Omega_c^{0}, K^{+}; \Xi_c^{(\prime)0}) + \mathcal M(\Xi_c^{(\prime)0}, \rho^{+}; \Sigma_c^{0}) 
 + \mathcal M(\Omega_c^{0}, K^{*+}; \Xi_c^{(\prime)0}) + \mathcal M(\Xi_c^{(\prime)0}, \rho^{+}; \Sigma_c^{0})  \\
&\quad + \mathcal M(\Omega_c^{0}, K^{+}; \Xi_c^{*\prime 0}) + \mathcal M(\Xi_c^{(\prime)0}, \pi^{+}; \Sigma_c^{*0})  + \mathcal M(\Omega_c^{0}, K^{*+}; \Xi_c^{*\prime 0})
 + \mathcal M(\Xi_c^{(\prime)0}, \rho^{+}; \Sigma_c^{*0}) .
\end{aligned}
\end{equation}
\begin{equation}
\begin{aligned}
\mathcal A(\Omega_{cc}^{+} \to \Lambda_c^{+} \bar K^{*0})
&= \mathcal M(\pi^{+}, \Xi_c^{(\prime)0}; K^{+})  + \mathcal M(\pi^{+}, \Xi_c^{(\prime)0}; K^{*+})  
 + \mathcal M(\rho^{+}, \Xi_c^{(\prime)0}; K^{+}) + \mathcal M(\rho^{+}, \Xi_c^{(\prime)0}; K^{*+}) \\
&\quad + \mathcal M(\Omega_c^{0}, K^{+}; \Xi_c^{(\prime)0}) + \mathcal M(\Xi_c^{(\prime)0}, \rho^{+}; \Sigma_c^{0}) 
 + \mathcal M(\Omega_c^{0}, K^{*+}; \Xi_c^{(\prime)0}) + \mathcal M(\Xi_c^{(\prime)0}, \rho^{+}; \Sigma_c^{0})  \\
&\quad + \mathcal M(\Omega_c^{0}, K^{+}; \Xi_c^{*\prime 0}) + \mathcal M(\Xi_c^{(\prime)0}, \pi^{+}; \Sigma_c^{*0})  + \mathcal M(\Omega_c^{0}, K^{*+}; \Xi_c^{*\prime 0})
 + \mathcal M(\Xi_c^{(\prime)0}, \rho^{+}; \Sigma_c^{*0}) .
\end{aligned}
\end{equation}
\begin{equation}
\begin{aligned}
\mathcal A(\Omega_{cc}^{+} \to \Sigma_c^{++} K^{*-})
&= \mathcal M(\Omega_c^{0}, K^{+}; \Xi_c^{(\prime)+})  
+ \mathcal M(\Xi_c^{(\prime)0}, \pi^{+}; \Lambda_c^{+}/\Sigma_c^{+})  
+ \mathcal M(\Omega_c^{0}, K^{*+}; \Xi_c^{\Sigma_c^{+}+}) 
+ \mathcal M(\Xi_c^{\Sigma_c^{+}0}, \rho^{+}; \Lambda_c^{+}/\Sigma_c^{+})  \\
&\quad + \mathcal M(\Omega_c^{0}, K^{+}; \Xi_c^{*\prime +}) 
+ \mathcal M(\Xi_c^{(\prime)0}, \pi^{+}; \Sigma_c^{*+}) 
+ \mathcal M(\Omega_c^{0}, K^{*+}; \Xi_c^{*\prime +}) 
 + \mathcal M(\Xi_c^{(\prime)0}, \rho^{+}; \Sigma_c^{*+}) .
\end{aligned}
\end{equation}
\begin{equation}
\begin{aligned}
\mathcal A(\Omega_{cc}^{+} \to \Xi_c^{\prime +} \rho^{0})
&= \mathcal M(\pi^{+}, \Xi_c^{(\prime)0}; \pi^{+}) 
 + \mathcal M(K^{+}, \Omega_c^{0}; K^{+}) 
 + \mathcal M(\pi^{+}, \Xi_c^{(\prime)0}; \rho^{+})   
 + \mathcal M(K^{+}, \Omega_c^{0}; K^{*+}) \\
&\quad 
 + \mathcal M(\rho^{+}, \Xi_c^{(\prime)0}; \pi^{+}) 
 + \mathcal M(K^{*+}, \Omega_c^{0}; K^{+}) 
 + \mathcal M(\rho^{+}, \Xi_c^{(\prime)0}; \rho^{+}) 
 + \mathcal M(K^{*+}, \Omega_c^{0}; K^{*+}) \\
&\quad + \mathcal M(\Xi_c^{(\prime)0}, \pi^{+}; \Xi_c^{(\prime)0}) 
 + \mathcal M(\Omega_c^{0}, K^{+}; \Omega_c^{0})
  + \mathcal M(\Xi_c^{(\prime)0}, \rho^{+}; \Xi_c^{(\prime)0}) 
   + \mathcal M(\Omega_c^{0}, K^{*+}; \Omega_c^{0}) \\
&\quad+ \mathcal M(\Xi_c^{(\prime)0}, \pi^{+}; \Xi_c^{*\prime 0})  
 + \mathcal M(\Xi_c^{(\prime)0}, \rho^{+}; \Xi_c^{*\prime 0}) .
\end{aligned}
\end{equation}
\begin{equation}
\begin{aligned}
\mathcal A(\Omega_{cc}^{+} \to \Xi_c^{\prime +} \phi)
&= \mathcal M(K^{+}, \Omega_c^{0}; K^{+}) + \mathcal M(K^{+}, \Omega_c^{0}; K^{*+}) + \mathcal M(K^{*+}, \Omega_c^{0}; K^{+}) + \mathcal M(K^{*+}, \Omega_c^{0}; K^{*+}) \\
&\quad + \mathcal M(\Xi_c^{(\prime)0}, \pi^{+}; \Xi_c^{(\prime)0})  
 + \mathcal M(\Omega_c^{0}, K^{+}; \Omega_c^{0}) 
 + \mathcal M(\Xi_c^{(\prime)0}, \rho^{+}; \Xi_c^{(\prime)0}) 
 + \mathcal M(\Omega_c^{0}, K^{*+}; \Omega_c^{0}) \\
&\quad
 + \mathcal M(\Xi_c^{(\prime)0}, \pi^{+}; \Xi_c^{*\prime 0})  
  + \mathcal M(\Omega_c^{0}, K^{+}; \Omega_c^{*0})
   + \mathcal M(\Xi_c^{(\prime)0}, \rho^{+}; \Xi_c^{*\prime 0}) 
   + \mathcal M(\Omega_c^{0}, K^{*+}; \Omega_c^{*0}) .
\end{aligned}
\end{equation}
\begin{equation}
\begin{aligned}
\mathcal A(\Omega_{cc}^{+} \to \Xi_c^{\prime +} \omega)
&= \mathcal M(\pi^{+}, \Xi_c^{0}; \pi^{+}) + \mathcal M(\pi^{+}, \Xi_c^{\prime 0}; \pi^{+}) + \mathcal M(K^{+}, \Omega_c^{0}; K^{+}) + \mathcal M(\pi^{+}, \Xi_c^{0}; \rho^{+}) \\
&\quad + \mathcal M(\pi^{+}, \Xi_c^{\prime 0}; \rho^{+}) + \mathcal M(K^{+}, \Omega_c^{0}; K^{*+}) + \mathcal M(\rho^{+}, \Xi_c^{0}; \pi^{+}) + \mathcal M(\rho^{+}, \Xi_c^{\prime 0}; \pi^{+}) \\
&\quad + \mathcal M(K^{*+}, \Omega_c^{0}; K^{+}) + \mathcal M(\rho^{+}, \Xi_c^{0}; \rho^{+}) + \mathcal M(\rho^{+}, \Xi_c^{\prime 0}; \rho^{+}) + \mathcal M(K^{*+}, \Omega_c^{0}; K^{*+}) \\
&\quad + \mathcal M(\Xi_c^{0}, \pi^{+}; \Xi_c^{0}) + \mathcal M(\Xi_c^{\prime 0}, \pi^{+}; \Xi_c^{0}) + \mathcal M(\Xi_c^{0}, \pi^{+}; \Xi_c^{\prime 0}) + \mathcal M(\Xi_c^{\prime 0}, \pi^{+}; \Xi_c^{\prime 0}) \\
&\quad + \mathcal M(\Xi_c^{0}, \rho^{+}; \Xi_c^{0}) + \mathcal M(\Xi_c^{\prime 0}, \rho^{+}; \Xi_c^{0}) + \mathcal M(\Xi_c^{0}, \rho^{+}; \Xi_c^{\prime 0}) \\
&\quad + \mathcal M(\Xi_c^{\prime 0}, \rho^{+}; \Xi_c^{\prime 0}) + \mathcal M(\Xi_c^{0}, \pi^{+}; \Xi_c^{*\prime 0}) + \mathcal M(\Xi_c^{\prime 0}, \pi^{+}; \Xi_c^{*\prime 0}) \\
&\quad  + \mathcal M(\Xi_c^{0}, \rho^{+}; \Xi_c^{*\prime 0}) + \mathcal M(\Xi_c^{\prime 0}, \rho^{+}; \Xi_c^{*\prime 0}) .
\end{aligned}
\end{equation}
\begin{equation}
\begin{aligned}
\mathcal A(\Omega_{cc}^{+} \to \Xi_c^{+} \rho^{0})
&= \mathcal M(\pi^{+}, \Xi_c^{(\prime)0}; \pi^{+}) 
 + \mathcal M(K^{+}, \Omega_c^{0}; K^{+}) 
 + \mathcal M(\pi^{+}, \Xi_c^{(\prime)0}; \rho^{+})   
 + \mathcal M(K^{+}, \Omega_c^{0}; K^{*+}) \\
&\quad 
 + \mathcal M(\rho^{+}, \Xi_c^{(\prime)0}; \pi^{+}) 
 + \mathcal M(K^{*+}, \Omega_c^{0}; K^{+}) 
 + \mathcal M(\rho^{+}, \Xi_c^{(\prime)0}; \rho^{+}) 
 + \mathcal M(K^{*+}, \Omega_c^{0}; K^{*+}) \\
&\quad + \mathcal M(\Xi_c^{(\prime)0}, \pi^{+}; \Xi_c^{(\prime)0}) 
 + \mathcal M(\Omega_c^{0}, K^{+}; \Omega_c^{0})
  + \mathcal M(\Xi_c^{(\prime)0}, \rho^{+}; \Xi_c^{(\prime)0}) 
   + \mathcal M(\Omega_c^{0}, K^{*+}; \Omega_c^{0}) \\
&\quad+ \mathcal M(\Xi_c^{(\prime)0}, \pi^{+}; \Xi_c^{*\prime 0})  
 + \mathcal M(\Xi_c^{(\prime)0}, \rho^{+}; \Xi_c^{*\prime 0}) .
\end{aligned}
\end{equation}
\begin{equation}
\begin{aligned}
\mathcal A(\Omega_{cc}^{+} \to \Xi_c^{+} \phi)
&= \mathcal M(K^{+}, \Omega_c^{0}; K^{+}) + \mathcal M(K^{+}, \Omega_c^{0}; K^{*+}) + \mathcal M(K^{*+}, \Omega_c^{0}; K^{+}) + \mathcal M(K^{*+}, \Omega_c^{0}; K^{*+}) \\
&\quad + \mathcal M(\Xi_c^{(\prime)0}, \pi^{+}; \Xi_c^{(\prime)0})  
 + \mathcal M(\Omega_c^{0}, K^{+}; \Omega_c^{0}) 
 + \mathcal M(\Xi_c^{(\prime)0}, \rho^{+}; \Xi_c^{(\prime)0}) 
  + \mathcal M(\Omega_c^{0}, K^{*+}; \Omega_c^{0}) \\
&\quad + \mathcal M(\Xi_c^{(\prime)0}, \pi^{+}; \Xi_c^{*\prime 0})  
 + \mathcal M(\Omega_c^{0}, K^{+}; \Omega_c^{*0})
  + \mathcal M(\Xi_c^{(\prime)0}, \rho^{+}; \Xi_c^{*\prime 0}) 
 + \mathcal M(\Omega_c^{0}, K^{*+}; \Omega_c^{*0}) .
\end{aligned}
\end{equation}
\begin{equation}
\begin{aligned}
\mathcal A(\Omega_{cc}^{+} \to \Xi_c^{+} \omega)
&= \mathcal M(\pi^{+}, \Xi_c^{(\prime)0}; \pi^{+}) 
 + \mathcal M(K^{+}, \Omega_c^{0}; K^{+}) 
 + \mathcal M(\pi^{+}, \Xi_c^{(\prime)0}; \rho^{+})   
 + \mathcal M(K^{+}, \Omega_c^{0}; K^{*+}) \\
&\quad 
 + \mathcal M(\rho^{+}, \Xi_c^{(\prime)0}; \pi^{+}) 
 + \mathcal M(K^{*+}, \Omega_c^{0}; K^{+}) 
 + \mathcal M(\rho^{+}, \Xi_c^{(\prime)0}; \rho^{+}) 
 + \mathcal M(K^{*+}, \Omega_c^{0}; K^{*+}) \\
&\quad + \mathcal M(\Xi_c^{(\prime)0}, \pi^{+}; \Xi_c^{(\prime)0}) 
 + \mathcal M(\Omega_c^{0}, K^{+}; \Omega_c^{0})
  + \mathcal M(\Xi_c^{(\prime)0}, \rho^{+}; \Xi_c^{(\prime)0}) 
   + \mathcal M(\Omega_c^{0}, K^{*+}; \Omega_c^{0}) \\
&\quad+ \mathcal M(\Xi_c^{(\prime)0}, \pi^{+}; \Xi_c^{*\prime 0})  
 + \mathcal M(\Xi_c^{(\prime)0}, \rho^{+}; \Xi_c^{*\prime 0}) .
\end{aligned}
\end{equation}
}}

The amplitudes for long-distance dominated doubly Cabibbo suppressed
can be given as follows.
{\scriptsize{
\begin{equation}
\begin{aligned}
\mathcal A(\Xi_{cc}^{++} \to \Sigma_c^{++}  K^{*0})
&= C_{SD}(\Xi_{cc}^{++} \to \Sigma_c^{++}  K^{*0}) \\
&\quad + \mathcal M(K^{+}, \Lambda_c^{+}/\Sigma_c^{+}; \pi^{+})
 + \mathcal M(K^{+}, \Lambda_c^{+}/\Sigma_c^{+}; \rho^{+}) 
 + \mathcal M(K^{*+}, \Lambda_c^{+}/\Sigma_c^{+}; \pi^{+})  
 + \mathcal M(K^{*+}, \Lambda_c^{+}/\Sigma_c^{+}; \rho^{+}) \\
&\quad + \mathcal M(\Lambda_c^{+}/\Sigma_c^{+}, K^{+}; \Xi_c^{(\prime)+}) 
 + \mathcal M(\Lambda_c^{+}/\Sigma_c^{+}, K^{*+}; \Xi_c^{(\prime)+})\\
&\quad
 + \mathcal M(\Lambda_c^{+}/\Sigma_c^{+}, K^{+}; \Xi_c^{*\prime +}) 
  + \mathcal M(\Lambda_c^{+}/\Sigma_c^{+}, K^{*+}; \Xi_c^{*\prime +}) .
\end{aligned}
\end{equation}
\begin{equation}
\begin{aligned}
\mathcal A(\Xi_{cc}^{+} \to \Sigma_c^{+}  K^{*0})
&= C_{SD}(\Xi_{cc}^{+} \to \Sigma_c^{+}  K^{*0}) \\
&\quad + \mathcal M(K^{+}, \Sigma_c^{0}; \pi^{+}) + \mathcal M(K^{+}, \Sigma_c^{0}; \rho^{+}) + \mathcal M(K^{*+}, \Sigma_c^{0}; \pi^{+}) + \mathcal M(K^{*+}, \Sigma_c^{0}; \rho^{+}) \\
&\quad + \mathcal M(\Sigma_c^{0}, K^{+}; \Xi_c^{(\prime)0})
 + \mathcal M(\Sigma_c^{0}, K^{*+}; \Xi_c^{(\prime)0})  
  + \mathcal M(\Sigma_c^{0}, K^{+}; \Xi_c^{*\prime 0})
 + \mathcal M(\Sigma_c^{0}, K^{*+}; \Xi_c^{*\prime 0}) .
\end{aligned}
\end{equation}
\begin{equation}
\begin{aligned}
\mathcal A(\Xi_{cc}^{+} \to \Lambda_c^{+} K^{*0})
&= C_{SD}(\Xi_{cc}^{+} \to \Lambda_c^{+} K^{*0}) \\
&\quad + \mathcal M(K^{+}, \Sigma_c^{0}; \pi^{+}) 
+ \mathcal M(K^{+}, \Sigma_c^{0}; \rho^{+}) 
+ \mathcal M(K^{*+}, \Sigma_c^{0}; \pi^{+}) 
+ \mathcal M(K^{*+}, \Sigma_c^{0}; \rho^{+}) \\
&\quad + \mathcal M(\Sigma_c^{0}, K^{+}; \Xi_c^{(\prime)0})
 + \mathcal M(\Sigma_c^{0}, K^{*+}; \Xi_c^{(\prime)0})  
  + \mathcal M(\Sigma_c^{0}, K^{+}; \Xi_c^{*\prime 0})
 + \mathcal M(\Sigma_c^{0}, K^{*+}; \Xi_c^{*\prime 0}) .
\end{aligned}
\end{equation}
\begin{equation}
\begin{aligned}
\mathcal A(\Omega_{cc}^{+} \to \Sigma_c^{+} \rho^{0})
&= \mathcal M(K^{+}, \Xi_c^{(\prime)0}; K^{+}) 
+ \mathcal M(K^{+}, \Xi_c^{(\prime)0}; K^{*+}) 
 + \mathcal M(K^{*+}, \Xi_c^{(\prime)0}; K^{+}) 
  + \mathcal M(K^{*+}, \Xi_c^{(\prime)0}; K^{*+})  \\
&\quad + \mathcal M(\Xi_c^{(\prime)0}, K^{+}; \Xi_c^{(\prime)0})  
+ \mathcal M(\Xi_c^{(\prime)0}, K^{*+}; \Xi_c^{(\prime)0}) 
 + \mathcal M(\Xi_c^{(\prime)0}, K^{+}; \Xi_c^{*\prime 0}) 
 + \mathcal M(\Xi_c^{(\prime)0}, K^{*+}; \Xi_c^{*\prime 0}).
\end{aligned}
\end{equation}
\begin{equation}
\begin{aligned}
\mathcal A(\Omega_{cc}^{+} \to \Sigma_c^{+} \phi)
&= \mathcal M(K^{+}, \Xi_c^{(\prime)0}; K^{+}) 
+ \mathcal M(K^{+}, \Xi_c^{(\prime)0}; K^{*+}) 
 + \mathcal M(K^{*+}, \Xi_c^{(\prime)0}; K^{+}) 
  + \mathcal M(K^{*+}, \Xi_c^{(\prime)0}; K^{*+})  \\
&\quad + \mathcal M(\Xi_c^{(\prime)0}, K^{+}; \Xi_c^{(\prime)0})  
+ \mathcal M(\Xi_c^{(\prime)0}, K^{*+}; \Xi_c^{(\prime)0}) 
 + \mathcal M(\Xi_c^{(\prime)0}, K^{+}; \Xi_c^{*\prime 0}) 
 + \mathcal M(\Xi_c^{(\prime)0}, K^{*+}; \Xi_c^{*\prime 0}).
\end{aligned}
\end{equation}
\begin{equation}
\begin{aligned}
\mathcal A(\Omega_{cc}^{+} \to \Sigma_c^{+} \omega)
&= \mathcal M(K^{+}, \Xi_c^{(\prime)0}; K^{+}) 
+ \mathcal M(K^{+}, \Xi_c^{(\prime)0}; K^{*+}) 
 + \mathcal M(K^{*+}, \Xi_c^{(\prime)0}; K^{+}) 
  + \mathcal M(K^{*+}, \Xi_c^{(\prime)0}; K^{*+})  \\
&\quad + \mathcal M(\Xi_c^{(\prime)0}, K^{+}; \Xi_c^{(\prime)0})  
+ \mathcal M(\Xi_c^{(\prime)0}, K^{*+}; \Xi_c^{(\prime)0}) 
 + \mathcal M(\Xi_c^{(\prime)0}, K^{+}; \Xi_c^{*\prime 0}) 
 + \mathcal M(\Xi_c^{(\prime)0}, K^{*+}; \Xi_c^{*\prime 0}).
\end{aligned}
\end{equation}

\begin{equation}
\begin{aligned}
\mathcal A(\Omega_{cc}^{+} \to \Lambda_c^{+} \rho^{0})
&= \mathcal M(K^{+}, \Xi_c^{(\prime)0}; K^{+}) 
+ \mathcal M(K^{+}, \Xi_c^{(\prime)0}; K^{*+}) 
 + \mathcal M(K^{*+}, \Xi_c^{(\prime)0}; K^{+}) 
  + \mathcal M(K^{*+}, \Xi_c^{(\prime)0}; K^{*+})  \\
&\quad + \mathcal M(\Xi_c^{(\prime)0}, K^{+}; \Xi_c^{(\prime)0})  
+ \mathcal M(\Xi_c^{(\prime)0}, K^{*+}; \Xi_c^{(\prime)0}) 
 + \mathcal M(\Xi_c^{(\prime)0}, K^{+}; \Xi_c^{*\prime 0}) 
 + \mathcal M(\Xi_c^{(\prime)0}, K^{*+}; \Xi_c^{*\prime 0}).
\end{aligned}
\end{equation}

\begin{equation}
\begin{aligned}
\mathcal A(\Omega_{cc}^{+} \to \Lambda_c^{+} \phi)
&= \mathcal M(K^{+}, \Xi_c^{(\prime)0}; K^{+}) 
+ \mathcal M(K^{+}, \Xi_c^{(\prime)0}; K^{*+}) 
 + \mathcal M(K^{*+}, \Xi_c^{(\prime)0}; K^{+}) 
  + \mathcal M(K^{*+}, \Xi_c^{(\prime)0}; K^{*+})  \\
&\quad + \mathcal M(\Xi_c^{(\prime)0}, K^{+}; \Xi_c^{(\prime)0})  
+ \mathcal M(\Xi_c^{(\prime)0}, K^{*+}; \Xi_c^{(\prime)0}) 
 + \mathcal M(\Xi_c^{(\prime)0}, K^{+}; \Xi_c^{*\prime 0}) 
 + \mathcal M(\Xi_c^{(\prime)0}, K^{*+}; \Xi_c^{*\prime 0}).
\end{aligned}
\end{equation}

\begin{equation}
\begin{aligned}
\mathcal A(\Omega_{cc}^{+} \to \Lambda_c^{+} \omega)
&= \mathcal M(K^{+}, \Xi_c^{(\prime)0}; K^{+}) 
+ \mathcal M(K^{+}, \Xi_c^{(\prime)0}; K^{*+}) 
 + \mathcal M(K^{*+}, \Xi_c^{(\prime)0}; K^{+}) 
  + \mathcal M(K^{*+}, \Xi_c^{(\prime)0}; K^{*+})  \\
&\quad + \mathcal M(\Xi_c^{(\prime)0}, K^{+}; \Xi_c^{(\prime)0})  
+ \mathcal M(\Xi_c^{(\prime)0}, K^{*+}; \Xi_c^{(\prime)0}) 
 + \mathcal M(\Xi_c^{(\prime)0}, K^{+}; \Xi_c^{*\prime 0}) 
 + \mathcal M(\Xi_c^{(\prime)0}, K^{*+}; \Xi_c^{*\prime 0}).
\end{aligned}
\end{equation}
\begin{equation}
\begin{aligned}
\mathcal A(\Omega_{cc}^{+} \to \Sigma_c^{0} \rho^{+})
&= \mathcal M(K^{+}, \Xi_c^{(\prime)0};  K^{0})
 + \mathcal M(K^{+}, \Xi_c^{(\prime)0};  K^{*0}) 
 + \mathcal M(K^{*+}, \Xi_c^{(\prime)0};  K^{0})
  + \mathcal M(K^{*+}, \Xi_c^{(\prime)0};  K^{*0}) .
\end{aligned}
\end{equation}
\begin{equation}
\begin{aligned}
\mathcal A(\Omega_{cc}^{+} \to \Sigma_c^{++} \rho^{-})
&= \mathcal M(\Xi_c^{(\prime)0}, K^{+}; \Xi_c^{(\prime)+}) 
 + \mathcal M(\Xi_c^{(\prime)0}, K^{*+}; \Xi_c^{(\prime)+}) 
 + \mathcal M(\Xi_c^{(\prime)0}, K^{+}; \Xi_c^{*\prime +}) 
 + \mathcal M(\Xi_c^{(\prime)0}, K^{*+}; \Xi_c^{*\prime +}).
\end{aligned}
\end{equation}
\begin{equation}
\begin{aligned}
\mathcal A(\Omega_{cc}^{+} \to \Xi_c^{\prime +} K^{*0})
&= C_{SD}(\Omega_{cc}^{+} \to \Xi_c^{\prime +} K^{*0}) \\
&\quad + \mathcal M(K^{+}, \Xi_c^{(\prime)0}; \pi^{+}) 
 + \mathcal M(K^{+}, \Xi_c^{(\prime)0}; \rho^{+})  
  + \mathcal M(K^{*+}, \Xi_c^{(\prime)0}; \pi^{+}) 
+ \mathcal M(K^{*+}, \Xi_c^{(\prime)0}; \rho^{+})\\
&\quad + \mathcal M(\Xi_c^{(\prime)0}, K^{+}; \Omega_c^{0}) 
+ \mathcal M(\Xi_c^{(\prime)0}, K^{*+}; \Omega_c^{0})  
 + \mathcal M(\Xi_c^{(\prime)0}, K^{+}; \Omega_c^{*0}) 
+ \mathcal M(\Xi_c^{(\prime)0}, K^{*+}; \Omega_c^{*0}).
\end{aligned}
\end{equation}
\begin{equation}
\begin{aligned}
\mathcal A(\Omega_{cc}^{+} \to \Xi_c^{+} K^{*0})
&= C_{SD}(\Omega_{cc}^{+} \to \Xi_c^{+} K^{*0}) \\
&\quad + \mathcal M(K^{+}, \Xi_c^{(\prime)0}; \pi^{+}) 
 + \mathcal M(K^{+}, \Xi_c^{(\prime)0}; \rho^{+})  
  + \mathcal M(K^{*+}, \Xi_c^{(\prime)0}; \pi^{+}) 
+ \mathcal M(K^{*+}, \Xi_c^{(\prime)0}; \rho^{+})\\
&\quad + \mathcal M(\Xi_c^{(\prime)0}, K^{+}; \Omega_c^{0}) 
+ \mathcal M(\Xi_c^{(\prime)0}, K^{*+}; \Omega_c^{0})  
 + \mathcal M(\Xi_c^{(\prime)0}, K^{+}; \Omega_c^{*0}) 
+ \mathcal M(\Xi_c^{(\prime)0}, K^{*+}; \Omega_c^{*0}).
\end{aligned}
\end{equation}
}}
\section{Lagrangian couplings}
\label{app:la}
The hadronic strong interaction Lagrangians~\cite{Aliev:2010yx,Yan:1992gz,Casalbuoni:1996pg,Meissner:1987ge,Li:2012bt,Aliev:2010nh} can be presented as following.

\begin{align}
\mathcal{L}_{VPP} & =ig_{VP_{1}P_{2}}[V^{\mu}P_{1}\partial_{\mu}P_{2}-V^{\mu}P_{2}\partial_{\mu}P_{1}],\label{eq:LVPP}\\
\mathcal{L}_{VVP} & =\frac{g_{V_{1}V_{2}P}}{\sqrt{m_{V_{1}}m_{V_{2}}}}\varepsilon_{\mu\nu\alpha\beta}Tr[\partial^{\mu}V_{1}^{\nu}\partial^{\alpha}PV_{2}^{\beta}],\label{eq:LVVP}\\
\mathcal{L}_{V_{1}V_{2}V_{3}}& = i\frac{g_{V_{1}V_{2}V_{3}}}{\sqrt{2}} \left[ V_{1}^\alpha \left( \partial_\alpha V_{2}^{\beta} V_{3\beta} -V_{2}^{\beta} \partial_\alpha V_{3\beta}\right) \right. \nonumber\\
&+ \left( \partial_\alpha V_{1}^\beta V_{2\beta} - V_{1}^\beta \partial_\alpha V_{2\beta} \right) V_{3}^{\alpha} + \left. V_{2}^{\alpha} \left( V_{1}^\beta \partial_\alpha V_{3\beta} - \partial_\alpha  V_{1}^\beta V_{3\beta} \right) \right],\label{eq:VVV}\\
\mathcal{L}_{P{\mathcal{B}}_{c}{\mathcal{B}}_{c}} & =g_{P{\mathcal{B}}_{c}{\mathcal{B}}_{c}}Tr[\overline{{\mathcal{B}}}_{c}i\gamma_{5}P{\mathcal{B}}_{c}],\label{eq:PBB}\\
\mathcal{L}_{V{\mathcal{B}}_{c}{\mathcal{B}}_{c}} & =f_{1V{\mathcal{B}}_{c}{\mathcal{B}}_{c}}Tr[\overline{\mathcal{B}}_{c}\gamma_{\mu}V^{\mu}{\mathcal{B}}_{c}]+\frac{f_{2V{\mathcal{B}}_{c}{\mathcal{B}}_{c}}}{m_{{\mathcal{B}}_{c}}+m_{{\mathcal{B}}_{c}}}Tr[\overline{\mathcal{B}}_{c}\sigma_{\mu\nu}\partial^{\mu}V^{\nu}{\mathcal{B}}_{c}],\label{eq:VBB}\\
\mathcal{L}_{P{\mathcal{B}}_{c}{\mathcal{B}}_{c}^{*}} & ={g_{P{\mathcal{B}}_{c}{\mathcal{B}}_{c}^{*}}}\epsilon^{ijk}(\overline{\mathcal{B}}_{c})^{j}_{i}({\mathcal{B}}_{c}^{*})^{mkl}_{\mu}\partial^{\mu}P^{i}_{m},\label{eq:PBBS}\\
\mathcal{L}_{V{\mathcal{B}}_{c}{\mathcal{B}}_{c}^{*}} & =-i{g_{V{\mathcal{B}}_{c}\bar{\mathcal{B}}_{c}^{*}}}[\overline{\mathcal{B}}_{c}^{*\mu}\gamma^{5}\gamma^{\nu}{\mathcal{B}}_{c}+\overline{\mathcal{B}}_{c}\gamma^{5}\gamma^{\nu}{\mathcal{B}}_{c}^{*}](\partial_{\mu}V_{\nu}-\partial_{\nu}V_{\mu}).\label{eq:LVBBS}
\end{align}
For estimating the hadronic scattering amplitudes, we need the following Feynman rules derived from the matrix element of the effective hadronic Lagrangian with definite initial and final states,
\begin{align}\label{eq:strong vertex1}
&\langle V(k,\lambda_k)P(p_{2})|i\mathcal{L}_{VPP}|P(p_{1})\rangle=ig_{VVP}\varepsilon^{*\mu}(k,\lambda_k)(p_{1}+p_{2})_{\mu}\,,\\
&\langle\mathcal{B}_{2}(p_{2})P(q)|i\mathcal{L}_{PBB}|\mathcal{B}_{1}(p_{1})\rangle=ig_{BBP}\bar{u}(p_{2})i\gamma_{5}u(p_{1})\,,\\
& \langle\mathcal{B}_{2}(p_{2})V(q,\lambda_q)|i\mathcal{L}_{VBB}|\mathcal{B}_{1}(p_{1})\rangle=i\bar{u}(p_{2})\left[f_{1}\gamma_{\nu}+f_{2}\frac{i}{m_{1}+m_{2}}\sigma_{\mu\nu}q^{\mu}\right]\varepsilon^{*\nu}(q,\lambda_q)u(p_{1})\,,  \\
&\langle V(p_{3},\lambda_{3})V(k,\lambda_{k})|i\mathcal{L}_{VVP}|P(p_{1})\rangle=i\frac{g_{V_{1}V_{2}P}}{\sqrt{m_{V_{1}}m_{V_{2}}}}\epsilon^{\mu\nu\alpha\beta}p_{3\,\mu}\varepsilon^{*}_{\nu}(\lambda_{3},p_{3})p_{1\,\alpha}\varepsilon^{*}_{\beta}(k,\lambda_{k})\,,\\
&\langle B(p_4,\lambda_{4})|i\mathcal{L}_{VBB^{*}}|B^{*}(k,\lambda_{k})V(p_{1},\lambda_{1})\rangle =-i{g_{V{\mathcal{B}}_{c}{\mathcal{B}}_{c}^{*}}}\bar{u}(p_{4},\lambda_{4})\gamma^{5}\gamma^{\nu}u^{\mu}(k,\lambda_{k}) \nonumber\\
&\hspace{6.2cm} \times \left[p_{1\,\mu}\varepsilon_{\nu}(p_1,\lambda_1)-p_{1\,\nu}\varepsilon_{\mu}(p_1,\lambda_1)\right]\,, \\
& \langle B(p_{4,\lambda_{4}})|i\mathcal{L}_{PBB^{*}}|B^{*}(k,\lambda_{k})P(p_{1},\lambda_{1})\rangle
={g_{P{\mathcal{B}}_{c}{\mathcal{B}}_{c}^{*}}} \bar{u}(p_{4},\lambda_{4})p_{1\,\mu}u^{\mu}(k,\lambda_{k})\,,
\end{align}
\begin{align}
\label{eq:strong vertex2}
& \langle V(p_{3},\lambda_{3})V(k,\lambda_{k})|i\mathcal{L}_{VVV}|V(p_{1},\lambda_{1})\rangle
= -\frac{ig_{VVV}}{\sqrt{2}}\varepsilon_{\mu}(p_{1},\lambda_{1})\varepsilon^{\mu*}(p_{3},\lambda_{3})\varepsilon^{*}_{\nu}(k,\lambda_{k})\left(p^{\nu}_{3}+p^{\nu}_{1}\right)\nonumber\\
        & \hspace{6.2cm} -\frac{ig_{VVV}}{\sqrt{2}}\varepsilon^{*}_{\mu}(k,\lambda_{k})\varepsilon^{\mu}(p_{1},\lambda_{1})\varepsilon^{*}_{\nu}(p_{3},\lambda_{3})\left(-p^{\nu}_{1}-p^{\nu}_{k}\right)\nonumber\\
        & \hspace{6.2cm} -\frac{ig_{VVV}}{\sqrt{2}}\varepsilon^{*}_{\mu}(p_{3},\lambda_{3})\varepsilon^{*\mu}(k,\lambda_{k})\varepsilon_{\nu}(p_{1},\lambda_{1})\left(p^{\nu}_{k}-p^{\nu}_{3}\right) .
\end{align}
$\varepsilon^{\alpha\beta\rho\tau}$ is the Levi-Civita tensor.

\section{Strong couplings}
\label{app:strong}
In general, meson-baryon and meson-meson strong couplings can be obtained from the SU(3)-invariant strong
Hamiltonian. In the present work, we use the SU(3) flavor symmetry to analysis the vector-vector-pseudoscalar (VVP), vector-pseudoscalar-pseudoscalar (VPP), baryon-baryon-pseudoscalar ($BBP/BB^{*}P$) and baryon-baryon-vector ($BBV/BB^{*}V$) couplings. 
Under SU(3) flavor symmetry, the pseudoscalar and vector meson can be written as follows,
\begin{align}
P(J^{P}=0^{-})=\left(\begin{array}{ccc}
\frac{\pi^{0}}{\sqrt{2}}+\frac{\eta_{8}}{\sqrt{6}} & \pi^{+} & K^{+}\\
\pi^{-} & -\frac{\pi^{0}}{\sqrt{2}}+\frac{\eta_{8}}{\sqrt{6}} & K^{0}\\
K^{-} & \overline{K}^{0} & -\sqrt{\frac{2}{3}}\eta_{8}
\end{array}\right)+\frac{1}{\sqrt{3}}\left(\begin{array}{ccc}
\eta_{1} & 0 & 0\\
0 & \eta_{1} & 0\\
0 & 0 & \eta_{1}
\end{array}\right),
\end{align}
\begin{align}
V(J^{P}=1^{-})=\left(\begin{array}{ccc}
\frac{\rho^{0}}{\sqrt{2}}+\frac{\omega}{\sqrt{2}} & \rho^{+} & K^{\ast+}\\
\rho^{-} & -\frac{\rho^{0}}{\sqrt{2}}+\frac{\omega}{\sqrt{2}} & K^{\ast0}\\
K^{\ast-} & \overline{K}^{\ast0} & \phi
\end{array}\right).\label{eq:pv}
\end{align}
And the spin-parity $1/2^{+}$ singly charmed baryons include anti-triplet ${\cal B}_{\mathbf{\bar 3}}$ and sextet ${\cal B}_{\mathbf{6}}$, which can given as follows,
\begin{align}
B_{6}(J^{P}=\frac{1}{2}^{+})=\left(\begin{array}{ccc}
\Sigma_{c}^{++} & \frac{\Sigma_{c}^{+}}{\sqrt{2}} & \frac{\Xi_{c}^{\prime+}}{\sqrt{2}}\\
\frac{\Sigma_{c}^{+}}{\sqrt{2}} & \Sigma_{c}^{0} & \frac{\Xi_{c}^{\prime0}}{\sqrt{2}}\\
\frac{\Xi_{c}^{\prime+}}{\sqrt{2}} & \frac{\Xi_{c}^{\prime0}}{\sqrt{2}} & \Omega_{c}^{0}
\end{array}\right),\quad B_{\bar{3}}(J^{P}=\frac{1}{2}^{+})=\left(\begin{array}{ccc}
0 & \Lambda_{c}^{+} & \Xi_{c}^{+}\\
-\Lambda_{c}^{+} & 0 & \Xi_{c}^{0}\\
-\Xi_{c}^{+} & -\Xi_{c}^{0} & 0
\end{array}\right).\label{eq:b36}
\end{align}
In addition, the spin-parity $3/2^{+}$ singly charmed baryons sextet ${\cal B}_{\mathbf{6}}$ can given as follows,
\begin{align}
B_{6}^{*}(J^{P}=\frac{3}{2}^{+})=\left(\begin{array}{ccc}
\Sigma_{c}^{*++} & \frac{\Sigma_{c}^{*+}}{\sqrt{2}} & \frac{\Xi_{c}^{*\prime+}}{\sqrt{2}}\\
\frac{\Sigma_{c}^{*+}}{\sqrt{2}} & \Sigma_{c}^{*0} & \frac{\Xi_{c}^{*\prime0}}{\sqrt{2}}\\
\frac{\Xi_{c}^{*\prime+}}{\sqrt{2}} & \frac{\Xi_{c}^{*\prime0}}{\sqrt{2}} & \Omega_{c}^{*0}
\end{array}\right).\label{eq:bs6}
\end{align}
Then with the help of Eqs.~(\ref{eq:pv})-(\ref{eq:bs6}), we can obtain the relationships of these strong couplings, as shown in Tabs.~\ref{tab:VPP}-\ref{tab:VBB}.

\begin{table}
  \caption{The strong coupling constants between the vector mesons and two pseudoscalar mesons $g_{VPP}$ can be obtained by using the coupling $g_{\rho\pi\pi}=6.05=-\frac{g_{VPP}}{\sqrt{2}}$~\cite{Cheng:2004ru} and SU(3) relations.}
  \label{tab:VPP}
  \begin{tabular}{cc|cc|cc|cc}
    \hline\hline
  Vertex&g&Vertex&g&Vertex&g&Vertex&g\\\hline
  $\rho^0    \pi^+   \pi^-  $ & $ -\frac{g_{VPP}}{\sqrt{2}}$&
  $\rho^0    \pi^0   \eta_8  $ & $ \frac{g_{VPP}}{\sqrt{6}}$&
  $\rho^0    \pi^0   \eta_{1}  $ & $ \frac{g_{VPP}}{\sqrt{3}}$&
  $\rho^0    \pi^-   \pi^+  $ & $ \frac{g_{VPP}}{\sqrt{2}}$\\\hline
  $\rho^0    \overline K^0   K^0  $ & $ -\frac{g_{VPP}}{\sqrt{2}}$&
  $\rho^0    K^-   K^+  $ & $ \frac{g_{VPP}}{\sqrt{2}}$&
  $\rho^0    \eta_8   \pi^0  $ & $ \frac{g_{VPP}}{\sqrt{6}}$&
  $\rho^0    \eta_{1}   \pi^0  $ & $ \frac{g_{VPP}}{\sqrt{3}}$\\\hline
  $\omega    \pi^+   \pi^-  $ & $ \frac{g_{VPP}}{\sqrt{2}}$&
  $\omega    \pi^0   \pi^0  $ & $ \frac{g_{VPP}}{\sqrt{2}}$&
  $\omega    \pi^-   \pi^+  $ & $ \frac{g_{VPP}}{\sqrt{2}}$&
  $\omega    \overline K^0   K^0  $ & $ \frac{g_{VPP}}{\sqrt{2}}$\\\hline
  $\omega    K^-   K^+  $ & $ \frac{g_{VPP}}{\sqrt{2}}$&
  $\omega    \eta_8   \eta_8  $ & $ \frac{g_{VPP}}{3 \sqrt{2}}$&
  $\omega    \eta_8   \eta_{1}  $ & $ \frac{g_{VPP}}{3}$&
  $\omega    \eta_{1}   \eta_8  $ & $ \frac{g_{VPP}}{3}$\\\hline
  $\omega    \eta_{1}   \eta_{1}  $ & $ \frac{\sqrt{2} g_{VPP}}{3}$&
  $\rho^+    \pi^+   \pi^0  $ & $ \frac{g_{VPP}}{\sqrt{2}}$&
  $\rho^+    \pi^+   \eta_8  $ & $ \frac{g_{VPP}}{\sqrt{6}}$&
  $\rho^+    \pi^+   \eta_{1}  $ & $ \frac{g_{VPP}}{\sqrt{3}}$\\\hline
  $\rho^+    \pi^0   \pi^+  $ & $ -\frac{g_{VPP}}{\sqrt{2}}$&
  $\rho^+    \overline K^0   K^+  $ & $ g_{VPP}$&
  $\rho^+    \eta_8   \pi^+  $ & $ \frac{g_{VPP}}{\sqrt{6}}$&
  $\rho^+    \eta_{1}   \pi^+  $ & $ \frac{g_{VPP}}{\sqrt{3}}$\\\hline
  $K^{*+}    K^+   \pi^0  $ & $ \frac{g_{VPP}}{\sqrt{2}}$&
  $K^{*+}    K^+   \eta_8  $ & $ \frac{g_{VPP}}{\sqrt{6}}$&
  $K^{*+}    K^+   \eta_{1}  $ & $ \frac{g_{VPP}}{\sqrt{3}}$&
  $K^{*+}    K^0   \pi^+  $ & $ g_{VPP}$\\\hline
  $K^{*+}    \eta_8   K^+  $ & $ -\sqrt{\frac{2}{3}} g_{VPP}$&
  $K^{*+}    \eta_{1}   K^+  $ & $ \frac{g_{VPP}}{\sqrt{3}}$&
  $\rho^-    \pi^0   \pi^-  $ & $ \frac{g_{VPP}}{\sqrt{2}}$&
  $\rho^-    \pi^-   \pi^0  $ & $ -\frac{g_{VPP}}{\sqrt{2}}$\\\hline
  $\rho^-    \pi^-   \eta_8  $ & $ \frac{g_{VPP}}{\sqrt{6}}$&
  $\rho^-    \pi^-   \eta_{1}  $ & $ \frac{g_{VPP}}{\sqrt{3}}$&
  $\rho^-    K^-   K^0  $ & $ g_{VPP}$&
  $\rho^-    \eta_8   \pi^-  $ & $ \frac{g_{VPP}}{\sqrt{6}}$\\\hline
  $\rho^-    \eta_{1}   \pi^-  $ & $ \frac{g_{VPP}}{\sqrt{3}}$&
  $K^{*0}    K^+   \pi^-  $ & $ g_{VPP}$&
  $K^{*0}    K^0   \pi^0  $ & $ -\frac{g_{VPP}}{\sqrt{2}}$&
  $K^{*0}    K^0   \eta_8  $ & $ \frac{g_{VPP}}{\sqrt{6}}$\\\hline
  $K^{*0}    K^0   \eta_{1}  $ & $ \frac{g_{VPP}}{\sqrt{3}}$&
  $K^{*0}    \eta_8   K^0  $ & $ -\sqrt{\frac{2}{3}} g_{VPP}$&
  $K^{*0}    \eta_{1}   K^0  $ & $ \frac{g_{VPP}}{\sqrt{3}}$&
  $K^{*-}    \pi^0   K^-  $ & $ \frac{g_{VPP}}{\sqrt{2}}$\\\hline
  $K^{*-}    \pi^-   \overline K^0  $ & $ g_{VPP}$&
  $K^{*-}    K^-   \eta_8  $ & $ -\sqrt{\frac{2}{3}} g_{VPP}$&
  $K^{*-}    K^-   \eta_{1}  $ & $ \frac{g_{VPP}}{\sqrt{3}}$&
  $K^{*-}    \eta_8   K^-  $ & $ \frac{g_{VPP}}{\sqrt{6}}$\\\hline
  $K^{*-}    \eta_{1}   K^-  $ & $ \frac{g_{VPP}}{\sqrt{3}}$&
  $\overline{K}^{*0}   \pi^+   K^-  $ & $ g_{VPP}$&
  $\overline{K}^{*0}   \pi^0   \overline K^0  $ & $ -\frac{g_{VPP}}{\sqrt{2}}$&
  $\overline{K}^{*0}   \overline K^0   \eta_8  $ & $ -\sqrt{\frac{2}{3}} g_{VPP}$\\\hline
  $\overline{K}^{*0}   \overline K^0   \eta_{1}  $ & $ \frac{g_{VPP}}{\sqrt{3}}$&
  $\overline{K}^{*0}   \eta_8   \overline K^0  $ & $ \frac{g_{VPP}}{\sqrt{6}}$&
  $\overline{K}^{*0}   \eta_{1}   \overline K^0  $ & $ \frac{g_{VPP}}{\sqrt{3}}$&
  $\phi   K^+   K^-  $ & $ g_{VPP}$\\\hline
  $\phi   K^0   \overline K^0  $ & $ g_{VPP}$&
  $\phi   \eta_8   \eta_8  $ & $ \frac{2 g_{VPP}}{3}$&
  $\phi   \eta_8   \eta_{1}  $ & $ -\frac{1}{3} \sqrt{2} g_{VPP}$&
  $\phi   \eta_{1}   \eta_8  $ & $ -\frac{1}{3} \sqrt{2} g_{VPP}$\\\hline
  $\phi   \eta_{1}   \eta_{1}  $ & $ \frac{g_{VPP}}{3}$\\
  \hline\hline
  \end{tabular}
\end{table}
\begin{table}
  \caption{The strong couplings between the pseudoscalar meson (P) and two vector mesons (V) can be obtained by using the coupling constant $g_{\omega \rho \pi}=g_{VVP}/\sqrt{2}=-10$~\cite{Nakayama:2006ps} and SU(3) relationships.}
  \label{tab:VVP}
 \begin{tabular}{cc|cc|cc|cc}
\hline\hline
Vertex & g & Vertex & g & Vertex & g & Vertex & g \\
\hline
$\rho^0 \rho^0 \eta_8$ & $\frac{g_{VVP}}{\sqrt{6}}$ & $\rho^0 \rho^0 \eta_1$ & $\frac{g_{VVP}}{\sqrt{3}}$ & $\rho^0 \omega \pi^0$ & $\frac{g_{VVP}}{\sqrt{2}}$ & $\rho^0 \rho^+ \pi^-$ & $-\frac{g_{VVP}}{\sqrt{2}}$ \\\hline
$\rho^0 \rho^- \pi^+$ & $\frac{g_{VVP}}{\sqrt{2}}$ & $\rho^0 K^{*-} K^+$ & $\frac{g_{VVP}}{\sqrt{2}}$ & $\rho^0 \overline{K}^{*0} K^0$ & $-\frac{g_{VVP}}{\sqrt{2}}$ & $\omega \rho^0 \pi^0$ & $\frac{g_{VVP}}{\sqrt{2}}$ \\\hline
$\omega \omega \eta_8$ & $\frac{g_{VVP}}{\sqrt{6}}$ & $\omega \omega \eta_1$ & $\frac{g_{VVP}}{\sqrt{3}}$ & $\omega \rho^+ \pi^-$ & $\frac{g_{VVP}}{\sqrt{2}}$ & $\omega \rho^- \pi^+$ & $\frac{g_{VVP}}{\sqrt{2}}$ \\\hline
$\omega K^{*-} K^+$ & $\frac{g_{VVP}}{\sqrt{2}}$ & $\omega \overline{K}^{*0} K^0$ & $\frac{g_{VVP}}{\sqrt{2}}$ & $\rho^+ \rho^0 \pi^+$ & $-\frac{g_{VVP}}{\sqrt{2}}$ & $\rho^+ \omega \pi^+$ & $\frac{g_{VVP}}{\sqrt{2}}$ \\\hline
$\rho^+ \rho^+ \pi^0$ & $\frac{g_{VVP}}{\sqrt{2}}$ & $\rho^+ \rho^+ \eta_8$ & $\frac{g_{VVP}}{\sqrt{6}}$ & $\rho^+ \rho^+ \eta_1$ & $\frac{g_{VVP}}{\sqrt{3}}$ & $\rho^+ \overline{K}^{*0} K^+$ & ${g_{VVP}}$ \\
$K^{*+} K^{*+} \pi^0$ & $\frac{g_{VVP}}{\sqrt{2}}$ & $K^{*+} K^{*+} \eta_8$ & $\frac{g_{VVP}}{\sqrt{6}}$ & $K^{*+} K^{*+} \eta_1$ & $\frac{g_{VVP}}{\sqrt{3}}$ & $K^{*+} K^{*0} \pi^+$ & ${g_{VVP}}$ \\\hline
$K^{*+} \phi K^+$ & ${g_{VVP}}$ & $\rho^- \rho^0 \pi^-$ & $\frac{g_{VVP}}{\sqrt{2}}$ & $\rho^- \omega \pi^-$ & $\frac{g_{VVP}}{\sqrt{2}}$ & $\rho^- \rho^- \pi^0$ & $-\frac{g_{VVP}}{\sqrt{2}}$ \\\hline
$\rho^- \rho^- \eta_8$ & $\frac{g_{VVP}}{\sqrt{6}}$ & $\rho^- \rho^- \eta_1$ & $\frac{g_{VVP}}{\sqrt{3}}$ & $\rho^- K^{*-} K^0$ & ${g_{VVP}}$ & $K^{*0} K^{*+} \pi^-$ & ${g_{VVP}}$ \\\hline
$K^{*0} K^{*0} \pi^0$ & $-\frac{g_{VVP}}{\sqrt{2}}$ & $K^{*0} K^{*0} \eta_8$ & $\frac{g_{VVP}}{\sqrt{6}}$ & $K^{*0} K^{*0} \eta_1$ & $\frac{g_{VVP}}{\sqrt{3}}$ & $K^{*0} \phi K^0$ & ${g_{VVP}}$ \\\hline
$K^{*-} \rho^0 K^-$ & $\frac{g_{VVP}}{\sqrt{2}}$ & $K^{*-} \omega K^-$ & $\frac{g_{VVP}}{\sqrt{2}}$ & $K^{*-} \rho^- \overline K^0$ & ${g_{VVP}}$ & $K^{*-} K^{*-} \eta_8$ & $-\sqrt{\frac{2}{3}} {g_{VVP}}$ \\\hline
$K^{*-} K^{*-} \eta_1$ & $\frac{g_{VVP}}{\sqrt{3}}$ & $\overline{K}^{*0} \rho^0 \overline K^0$ & $-\frac{g_{VVP}}{\sqrt{2}}$ & $\overline{K}^{*0} \omega \overline K^0$ & $\frac{g_{VVP}}{\sqrt{2}}$ & $\overline{K}^{*0} \rho^+ K^-$ & ${g_{VVP}}$ \\\hline
$\overline{K}^{*0} \overline{K}^{*0} \eta_8$ & $-\sqrt{\frac{2}{3}} {g_{VVP}}$ & $\overline{K}^{*0} \overline{K}^{*0} \eta_1$ & $\frac{g_{VVP}}{\sqrt{3}}$ & $\phi K^{*+} K^-$ & ${g_{VVP}}$ & $\phi K^{*0} \overline K^0$ & $g_{VVP}$ \\\hline
$\phi \phi \eta_8$ & $-\sqrt{\frac{2}{3}} {g_{VVP}}$ & $\phi \phi \eta_1$ & $\frac{g_{VVP}}{\sqrt{3}}$  \\\hline
\hline
\end{tabular}
\end{table}
\begin{table}
\caption{The strong coupling constants between three vector mesons $g_{VVV}$ can be obtained by using the coupling $g_{\rho\rho\rho}=4.17=-\frac{g_{VVV}}{\sqrt{2}}$~\cite{Oset:2010tof} and SU(3) relations.}\label{tab:1}
\begin{tabular}{cc|cc|cc|cc}
\hline\hline
Vertex&g&Vertex&g&Vertex&g&Vertex&g\\\hline
$\rho^0 \rho^0 \omega$ & $\frac{g_{VVV}}{\sqrt{2}}$ & $\rho^0 \omega \rho^0$ & $\frac{g_{VVV}}{\sqrt{2}}$ & $\rho^0 \rho^+ \rho^-$ & $-\frac{g_{VVV}}{\sqrt{2}}$ & $\rho^0 \rho^- \rho^+$ & $\frac{g_{VVV}}{\sqrt{2}}$ \\
\hline
$\rho^0 K^{*-} K^{*+}$ & $\frac{g_{VVV}}{\sqrt{2}}$ & $\rho^0 \overline{K}^{*0} K^{*0}$ & $-\frac{g_{VVV}}{\sqrt{2}}$ & $\omega \rho^0 \rho^0$ & $\frac{g_{VVV}}{\sqrt{2}}$ & $\omega \omega \omega$ & $\frac{g_{VVV}}{\sqrt{2}}$ \\
\hline
$\omega \rho^+ \rho^-$ & $\frac{g_{VVV}}{\sqrt{2}}$ & $\omega \rho^- \rho^+$ & $\frac{g_{VVV}}{\sqrt{2}}$ & $\omega K^{*-} K^{*+}$ & $\frac{g_{VVV}}{\sqrt{2}}$ & $\omega \overline{K}^{*0} K^{*0}$ & $\frac{g_{VVV}}{\sqrt{2}}$ \\
\hline
$\rho^+ \rho^0 \rho^+$ & $-\frac{g_{VVV}}{\sqrt{2}}$ & $\rho^+ \omega \rho^+$ & $\frac{g_{VVV}}{\sqrt{2}}$ & $\rho^+ \rho^+ \rho^0$ & $\frac{g_{VVV}}{\sqrt{2}}$ & $\rho^+ \rho^+ \omega$ & $\frac{g_{VVV}}{\sqrt{2}}$ \\
\hline
$\rho^+ \overline{K}^{*0} K^{*+}$ & ${g_{VVV}}$ & $K^{*+} K^{*+} \rho^0$ & $\frac{g_{VVV}}{\sqrt{2}}$ & $K^{*+} K^{*+} \omega$ & $\frac{g_{VVV}}{\sqrt{2}}$ & $K^{*+} K^{*0} \rho^+$ & ${g_{VVV}}$ \\
\hline
$K^{*+} \phi K^{*+}$ & ${g_{VVV}}$ & $\rho^- \rho^0 \rho^-$ & $\frac{g_{VVV}}{\sqrt{2}}$ & $\rho^- \omega \rho^-$ & $\frac{g_{VVV}}{\sqrt{2}}$ & $\rho^- \rho^- \rho^0$ & $-\frac{g_{VVV}}{\sqrt{2}}$ \\
\hline
$\rho^- \rho^- \omega$ & $\frac{g_{VVV}}{\sqrt{2}}$ & $\rho^- K^{*-} K^{*0}$ & ${g_{VVV}}$ & $K^{*0} K^{*+} \rho^-$ & ${g_{VVV}}$ & $K^{*0} K^{*0} \rho^0$ & $-\frac{g_{VVV}}{\sqrt{2}}$ \\
\hline
$K^{*0} K^{*0} \omega$ & $\frac{g_{VVV}}{\sqrt{2}}$ & $K^{*0} \phi K^{*0}$ & ${g_{VVV}}$ & $K^{*-} \rho^0 K^{*-}$ & $\frac{g_{VVV}}{\sqrt{2}}$ & $K^{*-} \omega K^{*-}$ & $\frac{g_{VVV}}{\sqrt{2}}$ \\
\hline
$K^{*-} \rho^- \overline{K}^{*0}$ & ${g_{VVV}}$ & $K^{*-} K^{*-} \phi$ & ${g_{VVV}}$ & $\overline{K}^{*0} \rho^0 \overline{K}^{*0}$ & $-\frac{g_{VVV}}{\sqrt{2}}$ & $\overline{K}^{*0} \omega \overline{K}^{*0}$ & $\frac{g_{VVV}}{\sqrt{2}}$ \\
\hline
$\overline{K}^{*0} \rho^+ K^{*-}$ & ${g_{VVV}}$ & $\overline{K}^{*0} \overline{K}^{*0} \phi$ & ${g_{VVV}}$ & $\phi K^{*+} K^{*-}$ & ${g_{VVV}}$ & $\phi K^{*0} \overline{K}^{*0}$ & ${g_{VVV}}$ \\
\hline
$\phi \phi \phi$ & ${g_{VVV}}$  \\
\hline
\end{tabular}
\end{table}
\begin{table}
    \caption{The spin $1/2$ singly charmed baryons ground states include anti-triplet ${\cal B}_{\mathbf{\bar 3}}$ and sextet ${\cal B}_{\mathbf{6}}$. The strong couplings of ${\cal B}_{\mathbf{\bar 3}}{\cal B}_{\mathbf{\bar 3}}P$, ${\cal B}_{\mathbf{6}}{\cal B}_{\mathbf{\bar 3}}P$ and ${\cal B}_{\mathbf{6}}{\cal B}_{\mathbf{6}}P$ can be obtained by using the coupling constant $g_{\Xi_{c}^{+}\Xi_{c}^{+}\pi^{0}} = 0.7=\frac{g_{PBB1}}{\sqrt{2}}$, $g_{\Xi_{c}^{\prime+}\Xi_{c}^{+}\pi^{0}} = 3.1=\frac{g_{PBB2}}{2}$ and ${\Sigma_{c}^{+}\Sigma_{c}^{0}\pi^{+}} = 8.0=\frac{g_{PBB3}}{\sqrt{2}}$~\cite{Aliev:2010yx}, respectively. Since the matrix forms of the spin-3/2 singly charmed baryon ground state ${\cal B}_{\mathbf{6}}^{*}$ and the 1/2 sextet ${\cal B}_{\mathbf{6}}$ under symmetry are similar, the derived relations are identical. Then the couplings ${\cal B}_{\mathbf{6}}^{*}{\cal B}_{\mathbf{\bar 3}}P$ and ${\cal B}_{\mathbf{6}}^{*}{\cal B}_{\mathbf{6}}P$ can be get from $g_{\Sigma_{c}^{*0}\Lambda_{c}^{+}\pi^{-}} = 3.9={g_{PB^{*}B2}}$ and $g_{\Sigma_{c}^{*+}\Sigma_{c}^{0}\pi^{+}} = 4.3=\frac{g_{PB^{*}B3}}{\sqrt{2}}$~\cite{Aliev:2010ev}, respectively.}
    \label{tab:PBB}
    \begin{tabular}{cc|cc|cc|cc}\hline\hline
    Vertex&g&Vertex&g&Vertex&g&Vertex&g\\\hline
    $\Lambda_c^+   \Lambda_c^+  \eta_8  $ & $ \sqrt{\frac{2}{3}} g_{PBB1}$&
    $\Lambda_c^+   \Lambda_c^+  \eta_{1}  $ & $ \frac{2 g_{PBB1}}{\sqrt{3}}$&
    $\Lambda_c^+   \Xi_c^+  K^0  $ & $ g_{PBB1}$&
    $\Lambda_c^+   \Xi_c^0  K^+  $ & $ -g_{PBB1}$\\\hline
    $\Xi_c^+   \Lambda_c^+  \overline K^0  $ & $ g_{PBB1}$&
    $\Xi_c^+   \Xi_c^+  \pi^0  $ & $ \frac{g_{PBB1}}{\sqrt{2}}$&
    $\Xi_c^+   \Xi_c^+  \eta_8  $ & $ -\frac{g_{PBB1}}{\sqrt{6}}$&
    $\Xi_c^+   \Xi_c^+  \eta_{1}  $ & $ \frac{2 g_{PBB1}}{\sqrt{3}}$\\\hline
    $\Xi_c^+   \Xi_c^0  \pi^+  $ & $ g_{PBB1}$&
    $\Xi_c^0   \Lambda_c^+  K^-  $ & $ -g_{PBB1}$&
    $\Xi_c^0   \Xi_c^+  \pi^-  $ & $ g_{PBB1}$&
    $\Xi_c^0   \Xi_c^0  \pi^0  $ & $ -\frac{g_{PBB1}}{\sqrt{2}}$\\\hline
    $\Xi_c^0   \Xi_c^0  \eta_8  $ & $ -\frac{g_{PBB1}}{\sqrt{6}}$&
    $\Xi_c^0   \Xi_c^0  \eta_{1}  $ & $ \frac{2 g_{PBB1}}{\sqrt{3}}$&\\
    \hline
    $\Sigma_{c}^{++}   \Lambda_c^+  \pi^+  $ & $ -g_{PBB2}$&
    $\Sigma_{c}^{++}   \Xi_c^+  K^+  $ & $ -g_{PBB2}$&
    $\Sigma_{c}^{+}   \Lambda_c^+  \pi^0  $ & $ g_{PBB2}$&
    $\Sigma_{c}^{+}   \Xi_c^+  K^0  $ & $ -\frac{g_{PBB2}}{\sqrt{2}}$\\\hline
    $\Sigma_{c}^{+}   \Xi_c^0  K^+  $ & $ -\frac{g_{PBB2}}{\sqrt{2}}$&
    $\Sigma_{c}^{0}   \Lambda_c^+  \pi^-  $ & $ g_{PBB2}$&
    $\Sigma_{c}^{0}   \Xi_c^0  K^0  $ & $ -g_{PBB2}$&
    $\Xi_{c}^{\prime+}   \Lambda_c^+  \overline K^0  $ & $ -\frac{g_{PBB2}}{\sqrt{2}}$\\\hline
    $\Xi_{c}^{\prime+}   \Xi_c^+  \pi^0  $ & $ \frac{g_{PBB2}}{2}$&
    $\Xi_{c}^{\prime+}   \Xi_c^+  \eta_8  $ & $ \frac{\sqrt{3} g_{PBB2}}{2}$&
    $\Xi_{c}^{\prime+}   \Xi_c^0  \pi^+  $ & $ \frac{g_{PBB2}}{\sqrt{2}}$&
    $\Xi_{c}^{\prime0}   \Lambda_c^+  K^-  $ & $ \frac{g_{PBB2}}{\sqrt{2}}$\\\hline
    $\Xi_{c}^{\prime0}   \Xi_c^+  \pi^-  $ & $ \frac{g_{PBB2}}{\sqrt{2}}$&
    $\Xi_{c}^{\prime0}   \Xi_c^0  \pi^0  $ & $ -\frac{g_{PBB2}}{2}$&
    $\Xi_{c}^{\prime0}   \Xi_c^0  \eta_8  $ & $ \frac{\sqrt{3} g_{PBB2}}{2}$&
    $\Omega_{c}^{0}   \Xi_c^+  K^-  $ & $ g_{PBB2}$\\\hline
    $\Omega_{c}^{0}   \Xi_c^0  \overline K^0  $ & $ g_{PBB2}$\\
    \hline
    $\Sigma_{c}^{++}   \Sigma_{c}^{++}  \pi^0  $ & $ \frac{g_{PBB3}}{\sqrt{2}}$&
    $\Sigma_{c}^{++}   \Sigma_{c}^{++}  \eta_8  $ & $ \frac{g_{PBB3}}{\sqrt{6}}$&
    $\Sigma_{c}^{++}   \Sigma_{c}^{++}  \eta_{1}  $ & $ \frac{g_{PBB3}}{\sqrt{3}}$&
    $\Sigma_{c}^{++}   \Sigma_{c}^{+}  \pi^+  $ & $ \frac{g_{PBB3}}{\sqrt{2}}$\\\hline
    $\Sigma_{c}^{++}   \Xi_{c}^{\prime+}  K^+  $ & $ \frac{g_{PBB3}}{\sqrt{2}}$&
    $\Sigma_{c}^{+}   \Sigma_{c}^{++}  \pi^-  $ & $ \frac{g_{PBB3}}{\sqrt{2}}$&
    $\Sigma_{c}^{+}   \Sigma_{c}^{+}  \eta_8  $ & $ \frac{g_{PBB3}}{\sqrt{6}}$&
    $\Sigma_{c}^{+}   \Sigma_{c}^{+}  \eta_{1}  $ & $ \frac{g_{PBB3}}{\sqrt{3}}$\\\hline
    $\Sigma_{c}^{+}   \Sigma_{c}^{0}  \pi^+  $ & $ \frac{g_{PBB3}}{\sqrt{2}}$&
    $\Sigma_{c}^{+}   \Xi_{c}^{\prime+}  K^0  $ & $ \frac{g_{PBB3}}{2}$&
    $\Sigma_{c}^{+}   \Xi_{c}^{\prime0}  K^+  $ & $ \frac{g_{PBB3}}{2}$&
    $\Sigma_{c}^{0}   \Sigma_{c}^{+}  \pi^-  $ & $ \frac{g_{PBB3}}{\sqrt{2}}$\\\hline
    $\Sigma_{c}^{0}   \Sigma_{c}^{0}  \pi^0  $ & $ -\frac{g_{PBB3}}{\sqrt{2}}$&
    $\Sigma_{c}^{0}   \Sigma_{c}^{0}  \eta_8  $ & $ \frac{g_{PBB3}}{\sqrt{6}}$&
    $\Sigma_{c}^{0}   \Sigma_{c}^{0}  \eta_{1}  $ & $ \frac{g_{PBB3}}{\sqrt{3}}$&
    $\Sigma_{c}^{0}   \Xi_{c}^{\prime0}  K^0  $ & $ \frac{g_{PBB3}}{\sqrt{2}}$\\\hline
    $\Xi_{c}^{\prime+}   \Sigma_{c}^{++}  K^-  $ & $ \frac{g_{PBB3}}{\sqrt{2}}$&
    $\Xi_{c}^{\prime+}   \Sigma_{c}^{+}  \overline K^0  $ & $ \frac{g_{PBB3}}{2}$&
    $\Xi_{c}^{\prime+}   \Xi_{c}^{\prime+}  \pi^0  $ & $ \frac{g_{PBB3}}{2 \sqrt{2}}$&
    $\Xi_{c}^{\prime+}   \Xi_{c}^{\prime+}  \eta_8  $ & $ -\frac{g_{PBB3}}{2 \sqrt{6}}$\\\hline
    $\Xi_{c}^{\prime+}   \Xi_{c}^{\prime+}  \eta_{1}  $ & $ \frac{g_{PBB3}}{\sqrt{3}}$&
    $\Xi_{c}^{\prime+}   \Xi_{c}^{\prime0}  \pi^+  $ & $ \frac{g_{PBB3}}{2}$&
    $\Xi_{c}^{\prime+}   \Omega_{c}^{0}  K^+  $ & $ \frac{g_{PBB3}}{\sqrt{2}}$&
    $\Xi_{c}^{\prime0}   \Sigma_{c}^{+}  K^-  $ & $ \frac{g_{PBB3}}{2}$\\\hline
    $\Xi_{c}^{\prime0}   \Sigma_{c}^{0}  \overline K^0  $ & $ \frac{g_{PBB3}}{\sqrt{2}}$&
    $\Xi_{c}^{\prime0}   \Xi_{c}^{\prime+}  \pi^-  $ & $ \frac{g_{PBB3}}{2}$&
    $\Xi_{c}^{\prime0}   \Xi_{c}^{\prime0}  \pi^0  $ & $ -\frac{g_{PBB3}}{2 \sqrt{2}}$&
    $\Xi_{c}^{\prime0}   \Xi_{c}^{\prime0}  \eta_8  $ & $ -\frac{g_{PBB3}}{2 \sqrt{6}}$\\\hline
    $\Xi_{c}^{\prime0}   \Xi_{c}^{\prime0}  \eta_{1}  $ & $ \frac{g_{PBB3}}{\sqrt{3}}$&
    $\Xi_{c}^{\prime0}   \Omega_{c}^{0}  K^0  $ & $ \frac{g_{PBB3}}{\sqrt{2}}$&
    $\Omega_{c}^{0}   \Xi_{c}^{\prime+}  K^-  $ & $ \frac{g_{PBB3}}{\sqrt{2}}$&
    $\Omega_{c}^{0}   \Xi_{c}^{\prime0}  \overline K^0  $ & $ \frac{g_{PBB3}}{\sqrt{2}}$\\\hline
    $\Omega_{c}^{0}   \Omega_{c}^{0}  \eta_8  $ & $ -\sqrt{\frac{2}{3}} g_{PBB3}$&
    $\Omega_{c}^{0}   \Omega_{c}^{0}  \eta_{1}  $ & $ \frac{g_{PBB3}}{\sqrt{3}}$\\\hline
    \hline
    \end{tabular}
\end{table}
\begin{table}
\caption{similar with Tab.~\ref{tab:PBB}, the strong couplings of ${\cal B}_{\mathbf{\bar 3}}{\cal B}_{\mathbf{\bar 3}}V$, ${\cal B}_{\mathbf{6}}{\cal B}_{\mathbf{\bar 3}}V$ and ${\cal B}_{\mathbf{6}}{\cal B}_{\mathbf{6}}V$ can be obtained by using the coupling constant $g_{\Xi_{c}^{0}\Lambda_{c}^{+}K^{*-}} = \{4.6,6\}= g_{VBB1}$, $g_{\Sigma_{c}^{0}\Lambda_{c}^{+}\rho^{-}} = \{2.6,16\}= g_{VBB2}$ and $g_{\Sigma_{c}^{+}\Sigma_{c}^{0}\rho^{+}} = \{4.0,27\}=\frac{g_{VBB3}}{\sqrt{2}}$~\cite{Aliev:2010nh}, respectively. Then the couplings ${\cal B}_{\mathbf{6}}^{*}{\cal B}_{\mathbf{\bar 3}}V$ and ${\cal B}_{\mathbf{6}}^{*}{\cal B}_{\mathbf{6}}V$ can be get from $g_{\Sigma_{c}^{*+}\Lambda_{c}^{+}\rho^{0}} = 10=g_{VB^{*}B2}$ and $g_{\Sigma_{c}^{*0}\Sigma_{c}^{+}\rho^{-}} = 5.77=\frac{g_{VB^{*}B3}}{\sqrt{2}}$~\cite{Lin:2017mtz}, respectively.}
\label{tab:VBB}
\begin{tabular}{cc|cc|cc|cc}
\hline\hline
Vertex&g&Vertex&g&Vertex&g&Vertex&g\\\hline
    $\Lambda_c^+   \Lambda_c^+  \omega  $ & $ -\sqrt{2} g_{VBB1}$&
    $\Lambda_c^+   \Xi_c^+  K^{*0}  $ & $ -g_{VBB1}$&
    $\Lambda_c^+   \Xi_c^0  K^{*+}  $ & $ g_{VBB1}$&
    $\Xi_c^+   \Lambda_c^+  \overline{K}^{*0} $ & $ -g_{VBB1}$\\\hline
    $\Xi_c^+   \Xi_c^+  \rho^0  $ & $ -\frac{g_{VBB1}}{\sqrt{2}}$&
    $\Xi_c^+   \Xi_c^+  \omega  $ & $ -\frac{g_{VBB1}}{\sqrt{2}}$&
    $\Xi_c^+   \Xi_c^+  \phi $ & $ -g_{VBB1}$&
    $\Xi_c^+   \Xi_c^0  \rho^+  $ & $ -g_{VBB1}$\\\hline
    $\Xi_c^0   \Lambda_c^+  K^{*-}  $ & $ g_{VBB1}$&
    $\Xi_c^0   \Xi_c^+  \rho^-  $ & $ -g_{VBB1}$&
    $\Xi_c^0   \Xi_c^0  \rho^0  $ & $ \frac{g_{VBB1}}{\sqrt{2}}$&
    $\Xi_c^0   \Xi_c^0  \omega  $ & $ -\frac{g_{VBB1}}{\sqrt{2}}$\\\hline
    $\Xi_c^0   \Xi_c^0  \phi $ & $ -g_{VBB1}$
    \\\hline
    $\Sigma_{c}^{++}   \Lambda_c^+  \rho^+  $ & $ -g_{VBB2}$&
    $\Sigma_{c}^{++}   \Xi_c^+  K^{*+}  $ & $ -g_{VBB2}$&
    $\Sigma_{c}^{+}   \Lambda_c^+  \rho^0  $ & $ g_{VBB2}$&
    $\Sigma_{c}^{+}   \Xi_c^+  K^{*0}  $ & $ -\frac{g_{VBB2}}{\sqrt{2}}$\\\hline
    $\Sigma_{c}^{+}   \Xi_c^0  K^{*+}  $ & $ -\frac{g_{VBB2}}{\sqrt{2}}$&
    $\Sigma_{c}^{0}   \Lambda_c^+  \rho^-  $ & $ g_{VBB2}$&
    $\Sigma_{c}^{0}   \Xi_c^0  K^{*0}  $ & $ -g_{VBB2}$&
    $\Xi_{c}^{\prime+}   \Lambda_c^+  \overline{K}^{*0} $ & $ -\frac{g_{VBB2}}{\sqrt{2}}$\\\hline
    $\Xi_{c}^{\prime+}   \Xi_c^+  \rho^0  $ & $ \frac{g_{VBB2}}{2}$&
    $\Xi_{c}^{\prime+}   \Xi_c^+  \omega  $ & $ \frac{g_{VBB2}}{2}$&
    $\Xi_{c}^{\prime+}   \Xi_c^+  \phi $ & $ -\frac{g_{VBB2}}{\sqrt{2}}$&
    $\Xi_{c}^{\prime+}   \Xi_c^0  \rho^+  $ & $ \frac{g_{VBB2}}{\sqrt{2}}$\\\hline
    $\Xi_{c}^{\prime0}   \Lambda_c^+  K^{*-}  $ & $ \frac{g_{VBB2}}{\sqrt{2}}$&
    $\Xi_{c}^{\prime0}   \Xi_c^+  \rho^-  $ & $ \frac{g_{VBB2}}{\sqrt{2}}$&
    $\Xi_{c}^{\prime0}   \Xi_c^0  \rho^0  $ & $ -\frac{g_{VBB2}}{2}$&
    $\Xi_{c}^{\prime0}   \Xi_c^0  \omega  $ & $ \frac{g_{VBB2}}{2}$\\\hline
    $\Xi_{c}^{\prime0}   \Xi_c^0  \phi $ & $ -\frac{g_{VBB2}}{\sqrt{2}}$&
    $\Omega_{c}^{0}   \Xi_c^+  K^{*-}  $ & $ g_{VBB2}$&
    $\Omega_{c}^{0}   \Xi_c^0  \overline{K}^{*0} $ & $ g_{VBB2}$
    \\\hline
    $\Sigma_{c}^{++}   \Sigma_{c}^{++}  \rho^0  $ & $ \frac{g_{VBB3}}{\sqrt{2}}$&
    $\Sigma_{c}^{++}   \Sigma_{c}^{++}  \omega  $ & $ \frac{g_{VBB3}}{\sqrt{2}}$&
    $\Sigma_{c}^{++}   \Sigma_{c}^{+}  \rho^+  $ & $ \frac{g_{VBB3}}{\sqrt{2}}$&
    $\Sigma_{c}^{++}   \Xi_{c}^{\prime+}  K^{*+}  $ & $ \frac{g_{VBB3}}{\sqrt{2}}$\\\hline
    $\Sigma_{c}^{+}   \Sigma_{c}^{++}  \rho^-  $ & $ \frac{g_{VBB3}}{\sqrt{2}}$&
    $\Sigma_{c}^{+}   \Sigma_{c}^{+}  \omega  $ & $ \frac{g_{VBB3}}{\sqrt{2}}$&
    $\Sigma_{c}^{+}   \Sigma_{c}^{0}  \rho^+  $ & $ \frac{g_{VBB3}}{\sqrt{2}}$&
    $\Sigma_{c}^{+}   \Xi_{c}^{\prime+}  K^{*0}  $ & $ \frac{g_{VBB3}}{2}$\\\hline
    $\Sigma_{c}^{+}   \Xi_{c}^{\prime0}  K^{*+}  $ & $ \frac{g_{VBB3}}{2}$&
    $\Sigma_{c}^{0}   \Sigma_{c}^{+}  \rho^-  $ & $ \frac{g_{VBB3}}{\sqrt{2}}$&
    $\Sigma_{c}^{0}   \Sigma_{c}^{0}  \rho^0  $ & $ -\frac{g_{VBB3}}{\sqrt{2}}$&
    $\Sigma_{c}^{0}   \Sigma_{c}^{0}  \omega  $ & $ \frac{g_{VBB3}}{\sqrt{2}}$\\\hline
    $\Sigma_{c}^{0}   \Xi_{c}^{\prime0}  K^{*0}  $ & $ \frac{g_{VBB3}}{\sqrt{2}}$&
    $\Xi_{c}^{\prime+}   \Sigma_{c}^{++}  K^{*-}  $ & $ \frac{g_{VBB3}}{\sqrt{2}}$&
    $\Xi_{c}^{\prime+}   \Sigma_{c}^{+}  \overline{K}^{*0} $ & $ \frac{g_{VBB3}}{2}$&
    $\Xi_{c}^{\prime+}   \Xi_{c}^{\prime+}  \rho^0  $ & $ \frac{g_{VBB3}}{2 \sqrt{2}}$\\\hline
    $\Xi_{c}^{\prime+}   \Xi_{c}^{\prime+}  \omega  $ & $ \frac{g_{VBB3}}{2 \sqrt{2}}$&
    $\Xi_{c}^{\prime+}   \Xi_{c}^{\prime+}  \phi $ & $ \frac{g_{VBB3}}{2}$&
    $\Xi_{c}^{\prime+}   \Xi_{c}^{\prime0}  \rho^+  $ & $ \frac{g_{VBB3}}{2}$&
    $\Xi_{c}^{\prime+}   \Omega_{c}^{0}  K^{*+}  $ & $ \frac{g_{VBB3}}{\sqrt{2}}$\\\hline
    $\Xi_{c}^{\prime0}   \Sigma_{c}^{+}  K^{*-}  $ & $ \frac{g_{VBB3}}{2}$&
    $\Xi_{c}^{\prime0}   \Sigma_{c}^{0}  \overline{K}^{*0} $ & $ \frac{g_{VBB3}}{\sqrt{2}}$&
    $\Xi_{c}^{\prime0}   \Xi_{c}^{\prime+}  \rho^-  $ & $ \frac{g_{VBB3}}{2}$&
    $\Xi_{c}^{\prime0}   \Xi_{c}^{\prime0}  \rho^0  $ & $ -\frac{g_{VBB3}}{2 \sqrt{2}}$\\\hline
    $\Xi_{c}^{\prime0}   \Xi_{c}^{\prime0}  \omega  $ & $ \frac{g_{VBB3}}{2 \sqrt{2}}$&
    $\Xi_{c}^{\prime0}   \Xi_{c}^{\prime0}  \phi $ & $ \frac{g_{VBB3}}{2}$&
    $\Xi_{c}^{\prime0}   \Omega_{c}^{0}  K^{*0}  $ & $ \frac{g_{VBB3}}{\sqrt{2}}$&
    $\Omega_{c}^{0}   \Xi_{c}^{\prime+}  K^{*-}  $ & $ \frac{g_{VBB3}}{\sqrt{2}}$\\\hline
    $\Omega_{c}^{0}   \Xi_{c}^{\prime0}  \overline{K}^{*0} $ & $ \frac{g_{VBB3}}{\sqrt{2}}$&
    $\Omega_{c}^{0}   \Omega_{c}^{0}  \phi $ & $ g_{VBB3}$\\\hline
    \hline
    \end{tabular}
\end{table}


\begin{thebibliography}{200}
\bibitem{Mattson:2002vu}
M.~Mattson \textit{et al.} [SELEX],
Phys. Rev. Lett. \textbf{89}, 112001 (2002)
doi:10.1103/PhysRevLett.89.112001
[arXiv:hep-ex/0208014 [hep-ex]].

\bibitem{SELEX:2004lln}
A.~Ocherashvili \textit{et al.} [SELEX],
Phys. Lett. B \textbf{628}, 18-24 (2005)
doi:10.1016/j.physletb.2005.09.043
[arXiv:hep-ex/0406033 [hep-ex]].

\bibitem{Ratti:2003ez}
S.~P.~Ratti,
Nucl. Phys. B Proc. Suppl. \textbf{115}, 33-36 (2003)
doi:10.1016/S0920-5632(02)01948-5

\bibitem{Aubert:2006qw}
B.~Aubert \textit{et al.} [BaBar],
Phys. Rev. D \textbf{74}, 011103 (2006)
doi:10.1103/PhysRevD.74.011103
[arXiv:hep-ex/0605075 [hep-ex]].

\bibitem{Chistov:2006zj}
R.~Chistov \textit{et al.} [Belle],
Phys. Rev. Lett. \textbf{97}, 162001 (2006)
doi:10.1103/PhysRevLett.97.162001
[arXiv:hep-ex/0606051 [hep-ex]].

\bibitem{Aaij:2013voa}
R.~Aaij \textit{et al.} [LHCb],
JHEP \textbf{12}, 090 (2013)
doi:10.1007/JHEP12(2013)090
[arXiv:1310.2538 [hep-ex]].

\bibitem{Aaij:2017ueg}
R.~Aaij \textit{et al.} [LHCb],
Phys. Rev. Lett. \textbf{119}, no.11, 112001 (2017)
doi:10.1103/PhysRevLett.119.112001
[arXiv:1707.01621 [hep-ex]].

\bibitem{Aaij:2018gfl}
R.~Aaij \textit{et al.} [LHCb],
Phys. Rev. Lett. \textbf{121}, no.16, 162002 (2018)
doi:10.1103/PhysRevLett.121.162002
[arXiv:1807.01919 [hep-ex]].

\bibitem{LHCb:2018zpl}
R.~Aaij \textit{et al.} [LHCb],
Phys. Rev. Lett. \textbf{121}, no.5, 052002 (2018)
doi:10.1103/PhysRevLett.121.052002
[arXiv:1806.02744 [hep-ex]].

\bibitem{LHCb:2019qed}
R.~Aaij \textit{et al.} [LHCb],
Chin. Phys. C \textbf{44}, no.2, 022001 (2020)
doi:10.1088/1674-1137/44/2/022001
[arXiv:1910.11316 [hep-ex]].

\bibitem{LHCb:2022rpd}
R.~Aaij \textit{et al.} [LHCb],
JHEP \textbf{05}, 038 (2022)
doi:10.1007/JHEP05(2022)038
[arXiv:2202.05648 [hep-ex]].

\bibitem{He:2024ehg}
J.~He and H.~Yin,
Chin. Sci. Bull. \textbf{69}, no.31, 4550-4557 (2024)
doi:10.1360/TB-2024-0123

\bibitem{LHCb:2021eaf}
R.~Aaij \textit{et al.} [LHCb],
JHEP \textbf{12}, 107 (2021)
doi:10.1007/JHEP12(2021)107
[arXiv:2109.07292 [hep-ex]].

\bibitem{LHCb:2019gqy}
R.~Aaij \textit{et al.} [LHCb],
Sci. China Phys. Mech. Astron. \textbf{63}, no.2, 221062 (2020)
doi:10.1007/s11433-019-1471-8
[arXiv:1909.12273 [hep-ex]].

\bibitem{LHCb:2026pxn}
R.~Aaij \textit{et al.} [LHCb],
[arXiv:2603.28456 [hep-ex]].

\bibitem{LHCb:2021rkb}
R.~Aaij \textit{et al.} [LHCb],
Sci. China Phys. Mech. Astron. \textbf{64}, no.10, 101062 (2021)
doi:10.1007/s11433-021-1742-7
[arXiv:2105.06841 [hep-ex]].

\bibitem{LHCb:2022fbu}
R.~Aaij \textit{et al.} [LHCb],
Chin. Phys. C \textbf{47}, no.9, 093001 (2023)
doi:10.1088/1674-1137/ace9c8
[arXiv:2204.09541 [hep-ex]].

\bibitem{LHCb:2020iko}
R.~Aaij \textit{et al.} [LHCb],
JHEP \textbf{11}, 095 (2020)
doi:10.1007/JHEP11(2020)095
[arXiv:2009.02481 [hep-ex]].

\bibitem{LHCb:2021xba}
R.~Aaij \textit{et al.} [LHCb],
Chin. Phys. C \textbf{45}, no.9, 093002 (2021)
doi:10.1088/1674-1137/ac0c70
[arXiv:2104.04759 [hep-ex]].

\bibitem{LHCb:2021vvq}
R.~Aaij \textit{et al.} [LHCb],
Nature Phys. \textbf{18}, no.7, 751-754 (2022)
doi:10.1038/s41567-022-01614-y
[arXiv:2109.01038 [hep-ex]].

\bibitem{LHCb:2021auc}
R.~Aaij \textit{et al.} [LHCb],
Nature Commun. \textbf{13}, no.1, 3351 (2022)
doi:10.1038/s41467-022-30206-w
[arXiv:2109.01056 [hep-ex]].

\bibitem{Yu:2017zst}
F.~S.~Yu, H.~Y.~Jiang, R.~H.~Li, C.~D.~L{\"u}, W.~Wang and Z.~X.~Zhao,
Chin. Phys. C \textbf{42}, no.5, 051001 (2018)
doi:10.1088/1674-1137/42/5/051001
[arXiv:1703.09086 [hep-ph]].

\bibitem{Kiselev:2001fw}
V.~V.~Kiselev and A.~K.~Likhoded,
Phys. Usp. \textbf{45}, 455-506 (2002)
doi:10.1070/PU2002v045n05ABEH000958
[arXiv:hep-ph/0103169 [hep-ph]].

\bibitem{Ebert:1996ec}
D.~Ebert, R.~N.~Faustov, V.~O.~Galkin, A.~P.~Martynenko and V.~A.~Saleev,
Z. Phys. C \textbf{76}, 111-115 (1997)
doi:10.1007/s002880050534
[arXiv:hep-ph/9607314 [hep-ph]].

\bibitem{Tong:1999qs}
S.~P.~Tong, Y.~B.~Ding, X.~H.~Guo, H.~Y.~Jin, X.~Q.~Li, P.~N.~Shen and R.~Zhang,
Phys. Rev. D \textbf{62}, 054024 (2000)
doi:10.1103/PhysRevD.62.054024
[arXiv:hep-ph/9910259 [hep-ph]].

\bibitem{Ebert:2002ig}
D.~Ebert, R.~N.~Faustov, V.~O.~Galkin and A.~P.~Martynenko,
Phys. Rev. D \textbf{66}, 014008 (2002)
doi:10.1103/PhysRevD.66.014008
[arXiv:hep-ph/0201217 [hep-ph]].

\bibitem{Gershtein:2000nx}
S.~S.~Gershtein, V.~V.~Kiselev, A.~K.~Likhoded and A.~I.~Onishchenko,
Phys. Rev. D \textbf{62}, 054021 (2000)
doi:10.1103/PhysRevD.62.054021

\bibitem{Roberts:2007ni}
W.~Roberts and M.~Pervin,
Int. J. Mod. Phys. A \textbf{23}, 2817-2860 (2008)
doi:10.1142/S0217751X08041219
[arXiv:0711.2492 [nucl-th]].

\bibitem{Valcarce:2008dr}
A.~Valcarce, H.~Garcilazo and J.~Vijande,
Eur. Phys. J. A \textbf{37}, 217-225 (2008)
doi:10.1140/epja/i2008-10616-4
[arXiv:0807.2973 [hep-ph]].

\bibitem{Lu:2017meb}
Q.~F.~L{\"u}, K.~L.~Wang, L.~Y.~Xiao and X.~H.~Zhong,
Phys. Rev. D \textbf{96}, no.11, 114006 (2017)
doi:10.1103/PhysRevD.96.114006
[arXiv:1708.04468 [hep-ph]].

\bibitem{Yu:2022lel}
G.~L.~Yu, Z.~Y.~Li, Z.~G.~Wang, J.~Lu and M.~Yan,
Eur. Phys. J. A \textbf{59}, no.6, 126 (2023)
doi:10.1140/epja/s10050-023-01044-1
[arXiv:2211.00510 [hep-ph]].

\bibitem{Li:2017ndo}
R.~H.~Li, C.~D.~L{\"u}, W.~Wang, F.~S.~Yu and Z.~T.~Zou,
Phys. Lett. B \textbf{767}, 232-235 (2017)
doi:10.1016/j.physletb.2017.02.003
[arXiv:1701.03284 [hep-ph]].

\bibitem{Ebert:2004ck}
D.~Ebert, R.~N.~Faustov, V.~O.~Galkin and A.~P.~Martynenko,
Phys. Rev. D \textbf{70}, 014018 (2004)
[erratum: Phys. Rev. D \textbf{77}, 079903 (2008)]
doi:10.1103/PhysRevD.70.014018
[arXiv:hep-ph/0404280 [hep-ph]].

\bibitem{Roberts:2008wq}
W.~Roberts and M.~Pervin,
Int. J. Mod. Phys. A \textbf{24}, 2401-2413 (2009)
doi:10.1142/S0217751X09043407
[arXiv:0803.3350 [nucl-th]].

\bibitem{Branz:2010pq}
T.~Branz, A.~Faessler, T.~Gutsche, M.~A.~Ivanov, J.~G.~Korner, V.~E.~Lyubovitskij and B.~Oexl,
Phys. Rev. D \textbf{81}, 114036 (2010)
doi:10.1103/PhysRevD.81.114036
[arXiv:1005.1850 [hep-ph]].

\bibitem{Albertus:2010hi}
C.~Albertus, E.~Hernandez and J.~Nieves,
Phys. Lett. B \textbf{690}, 265-271 (2010)
doi:10.1016/j.physletb.2010.05.042
[arXiv:1004.3154 [hep-ph]].

\bibitem{Qin:2021zqx}
Q.~Qin, Y.~J.~Shi, W.~Wang, G.~H.~Yang, F.~S.~Yu and R.~Zhu,
Phys. Rev. D \textbf{105}, no.3, L031902 (2022)
doi:10.1103/PhysRevD.105.L031902
[arXiv:2108.06716 [hep-ph]].

\bibitem{Bahtiyar:2018vub}
H.~Bahtiyar, K.~U.~Can, G.~Erkol, M.~Oka and T.~T.~Takahashi,
Phys. Rev. D \textbf{98}, no.11, 114505 (2018)
doi:10.1103/PhysRevD.98.114505
[arXiv:1807.06795 [hep-lat]].

\bibitem{Chen:2016spr}
H.~X.~Chen, W.~Chen, X.~Liu, Y.~R.~Liu and S.~L.~Zhu,
Rept. Prog. Phys. \textbf{80}, no.7, 076201 (2017)
doi:10.1088/1361-6633/aa6420
[arXiv:1609.08928 [hep-ph]].

\bibitem{Cheng:2021qpd}
H.~Y.~Cheng,
Chin. J. Phys. \textbf{78}, 324-362 (2022)
doi:10.1016/j.cjph.2022.06.021
[arXiv:2109.01216 [hep-ph]].

\bibitem{Chen:2020aos}
H.~X.~Chen, W.~Chen, R.~R.~Dong and N.~Su,
Chin. Phys. Lett. \textbf{37}, no.10, 101201 (2020)
doi:10.1088/0256-307X/37/10/101201
[arXiv:2008.07516 [hep-ph]].

\bibitem{Shi:2019fph}
Y.~J.~Shi, Y.~Xing and Z.~X.~Zhao,
Eur. Phys. J. C \textbf{79}, no.6, 501 (2019)
doi:10.1140/epjc/s10052-019-7014-y
[arXiv:1903.03921 [hep-ph]].

\bibitem{Hu:2019bqj}
X.~H.~Hu and Y.~J.~Shi,
Eur. Phys. J. C \textbf{80}, no.1, 56 (2020)
doi:10.1140/epjc/s10052-020-7635-1
[arXiv:1910.07909 [hep-ph]].

\bibitem{Hu:2022xzu}
X.~H.~Hu and Y.~J.~Shi,
Phys. Rev. D \textbf{107}, no.3, 036007 (2023)
doi:10.1103/PhysRevD.107.036007
[arXiv:2202.07540 [hep-ph]].

\bibitem{Aliev:2022tvs}
T.~M.~Aliev, S.~Bilmis and M.~Savci,
Eur. Phys. J. C \textbf{82}, no.10, 862 (2022)
doi:10.1140/epjc/s10052-022-10845-5
[arXiv:2206.08253 [hep-ph]].

\bibitem{Aliev:2022maw}
T.~M.~Aliev, M.~Savci and S.~Bilmis,
Phys. Rev. D \textbf{106}, no.3, 034017 (2022)
doi:10.1103/PhysRevD.106.034017
[arXiv:2205.14012 [hep-ph]].

\bibitem{Jiang:2018oak}
L.~J.~Jiang, B.~He and R.~H.~Li,
Eur. Phys. J. C \textbf{78}, no.11, 961 (2018)
doi:10.1140/epjc/s10052-018-6445-1
[arXiv:1810.00541 [hep-ph]].

\bibitem{Lu:2009cm}
C.~D.~Lu, Y.~M.~Wang, H.~Zou, A.~Ali and G.~Kramer,
Phys. Rev. D \textbf{80}, 034011 (2009)
doi:10.1103/PhysRevD.80.034011
[arXiv:0906.1479 [hep-ph]].

\bibitem{Li:2012cfa}
H.~n.~Li, C.~D.~Lu and F.~S.~Yu,
Phys. Rev. D \textbf{86}, 036012 (2012)
doi:10.1103/PhysRevD.86.036012
[arXiv:1203.3120 [hep-ph]].

\bibitem{Li:2002pj}
J.~W.~Li, M.~Z.~Yang and D.~S.~Du,
HEPNP \textbf{27}, 665-672 (2003)
[arXiv:hep-ph/0206154 [hep-ph]].

\bibitem{Ablikim:2002ep}
M.~Ablikim, D.~S.~Du and M.~Z.~Yang,
Phys. Lett. B \textbf{536}, 34-42 (2002)
doi:10.1016/S0370-2693(02)01812-9
[arXiv:hep-ph/0201168 [hep-ph]].

\bibitem{Li:1996cj}
X.~Q.~Li and B.~S.~Zou,
Phys. Lett. B \textbf{399}, 297-302 (1997)
doi:10.1016/S0370-2693(97)00308-0
[arXiv:hep-ph/9611223 [hep-ph]].

\bibitem{Dai:1999cs}
Y.~S.~Dai, D.~S.~Du, X.~Q.~Li, Z.~T.~Wei and B.~S.~Zou,
Phys. Rev. D \textbf{60}, 014014 (1999)
doi:10.1103/PhysRevD.60.014014
[arXiv:hep-ph/9903204 [hep-ph]].

\bibitem{Locher:1993cc}
M.~P.~Locher, Y.~Lu and B.~S.~Zou,
Z. Phys. A \textbf{347}, 281-284 (1994)
doi:10.1007/BF01289796
[arXiv:nucl-th/9311021 [nucl-th]].

\bibitem{Cheng:2004ru}
H.~Y.~Cheng, C.~K.~Chua and A.~Soni,
Phys. Rev. D \textbf{71}, 014030 (2005)
doi:10.1103/PhysRevD.71.014030
[arXiv:hep-ph/0409317 [hep-ph]].

\bibitem{Lu:2005mx}
C.~D.~Lu, Y.~L.~Shen and W.~Wang,
Phys. Rev. D \textbf{73}, 034005 (2006)
doi:10.1103/PhysRevD.73.034005
[arXiv:hep-ph/0511255 [hep-ph]].

\bibitem{Han:2021azw}
J.~J.~Han, H.~Y.~Jiang, W.~Liu, Z.~J.~Xiao and F.~S.~Yu,
Chin. Phys. C \textbf{45}, no.5, 053105 (2021)
doi:10.1088/1674-1137/abec68
[arXiv:2101.12019 [hep-ph]].

\bibitem{Aliev:2010yx}
T.~M.~Aliev, K.~Azizi and M.~Savci,
Phys. Lett. B \textbf{696}, 220-226 (2011)
doi:10.1016/j.physletb.2010.12.027
[arXiv:1009.3658 [hep-ph]].

\bibitem{Yan:1992gz}
T.~M.~Yan, H.~Y.~Cheng, C.~Y.~Cheung, G.~L.~Lin, Y.~C.~Lin and H.~L.~Yu,
Phys. Rev. D \textbf{46}, 1148-1164 (1992)
[erratum: Phys. Rev. D \textbf{55}, 5851 (1997)]
doi:10.1103/PhysRevD.46.1148

\bibitem{Casalbuoni:1996pg}
R.~Casalbuoni, A.~Deandrea, N.~Di Bartolomeo, R.~Gatto, F.~Feruglio and G.~Nardulli,
Phys. Rept. \textbf{281}, 145-238 (1997)
doi:10.1016/S0370-1573(96)00027-0
[arXiv:hep-ph/9605342 [hep-ph]].

\bibitem{Meissner:1987ge}
U.~G.~Meissner,
Phys. Rept. \textbf{161}, 213 (1988)
doi:10.1016/0370-1573(88)90090-7

\bibitem{Li:2012bt}
N.~Li and S.~L.~Zhu,
Phys. Rev. D \textbf{86}, 014020 (2012)
doi:10.1103/PhysRevD.86.014020
[arXiv:1204.3364 [hep-ph]].

\bibitem{Aliev:2010nh}
T.~M.~Aliev, K.~Azizi and M.~Savci,
Nucl. Phys. A \textbf{852}, 141-154 (2011)
doi:10.1016/j.nuclphysa.2011.01.011
[arXiv:1011.0086 [hep-ph]].

\bibitem{Hu:2024uia}
X.~H.~Hu, C.~P.~Jia, Y.~Xing and F.~S.~Yu,
Phys. Rev. D \textbf{111}, no.7, 076002 (2025)
doi:10.1103/PhysRevD.111.076002
[arXiv:2403.09511 [hep-ph]].

\bibitem{LHCb:2013hzx}
R.~Aaij \textit{et al.} [LHCb],
Phys. Lett. B \textbf{724}, 27-35 (2013)
doi:10.1016/j.physletb.2013.05.041
[arXiv:1302.5578 [hep-ex]].

\bibitem{Gutsche:2018utw}
T.~Gutsche, M.~A.~Ivanov, J.~G.~K{\"o}rner and V.~E.~Lyubovitskij,
Phys. Rev. D \textbf{98}, no.7, 074011 (2018)
doi:10.1103/PhysRevD.98.074011
[arXiv:1806.11549 [hep-ph]].

\bibitem{Yu:2018com}
Q.~X.~Yu and X.~H.~Guo,
Nucl. Phys. B \textbf{947}, 114727 (2019)
doi:10.1016/j.nuclphysb.2019.114727
[arXiv:1810.00437 [hep-ph]].

\bibitem{Berezhnoy:2018bde}
A.~V.~Berezhnoy, A.~K.~Likhoded and A.~V.~Luchinsky,
Phys. Rev. D \textbf{98}, no.11, 113004 (2018)
doi:10.1103/PhysRevD.98.113004
[arXiv:1809.10058 [hep-ph]].

\bibitem{Cheng:2026mlv}
H.~Y.~Cheng and C.~W.~Liu,
[arXiv:2604.10939 [hep-ph]].

\bibitem{ParticleDataGroup:2024cfk}
S.~Navas \textit{et al.} [Particle Data Group],
Phys. Rev. D \textbf{110}, no.3, 030001 (2024)
doi:10.1103/PhysRevD.110.030001

\bibitem{Choi:2015ywa}
H.~M.~Choi, C.~R.~Ji, Z.~Li and H.~Y.~Ryu,
Phys. Rev. C \textbf{92}, no.5, 055203 (2015)
doi:10.1103/PhysRevC.92.055203
[arXiv:1502.03078 [hep-ph]].

\bibitem{Feldmann:1998vh}
T.~Feldmann, P.~Kroll and B.~Stech,
Phys. Rev. D \textbf{58}, 114006 (1998)
doi:10.1103/PhysRevD.58.114006
[arXiv:hep-ph/9802409 [hep-ph]].

\bibitem{Wang:2017mqp}
W.~Wang, F.~S.~Yu and Z.~X.~Zhao,
Eur. Phys. J. C \textbf{77}, no.11, 781 (2017)
doi:10.1140/epjc/s10052-017-5360-1
[arXiv:1707.02834 [hep-ph]].

\bibitem{Nakayama:2006ps}
K.~Nakayama, Y.~Oh, J.~Haidenbauer and T.~S.~H.~Lee,
Phys. Lett. B \textbf{648}, 351-356 (2007)
doi:10.1016/j.physletb.2007.03.034
[arXiv:nucl-th/0611101 [nucl-th]].

\bibitem{Aliev:2010ev}
T.~M.~Aliev, K.~Azizi and M.~Savci,
Eur. Phys. J. C \textbf{71}, 1675 (2011)
doi:10.1140/epjc/s10052-011-1675-5
[arXiv:1012.5935 [hep-ph]].

\bibitem{Lin:2017mtz}
Y.~H.~Lin, C.~W.~Shen, F.~K.~Guo and B.~S.~Zou,
Phys. Rev. D \textbf{95}, no.11, 114017 (2017)
doi:10.1103/PhysRevD.95.114017
[arXiv:1703.01045 [hep-ph]].

\bibitem{Oset:2010tof}
E.~Oset and A.~Ramos,
Eur. Phys. J. A \textbf{44}, 445-454 (2010)
doi:10.1140/epja/i2010-10957-3
[arXiv:0905.0973 [hep-ph]].

\bibitem{Berezhnoy:1998aa}
A.~V.~Berezhnoy, V.~V.~Kiselev, A.~K.~Likhoded and A.~I.~Onishchenko,
Phys. Rev. D \textbf{57}, 4385-4392 (1998)
doi:10.1103/PhysRevD.57.4385
[arXiv:hep-ph/9710339 [hep-ph]].

\bibitem{Ma:2003zk}
J.~P.~Ma and Z.~G.~Si,
Phys. Lett. B \textbf{568}, 135-145 (2003)
doi:10.1016/j.physletb.2003.06.064
[arXiv:hep-ph/0305079 [hep-ph]].

\bibitem{Chang:2006xp}
C.~H.~Chang, J.~P.~Ma, C.~F.~Qiao and X.~G.~Wu,
J. Phys. G \textbf{34}, 845 (2007)
doi:10.1088/0954-3899/34/5/006
[arXiv:hep-ph/0610205 [hep-ph]].

\bibitem{Chang:2006eu}
C.~H.~Chang, C.~F.~Qiao, J.~X.~Wang and X.~G.~Wu,
Phys. Rev. D \textbf{73}, 094022 (2006)
doi:10.1103/PhysRevD.73.094022
[arXiv:hep-ph/0601032 [hep-ph]].

\bibitem{Zhang:2011hi}
J.~W.~Zhang, X.~G.~Wu, T.~Zhong, Y.~Yu and Z.~Y.~Fang,
Phys. Rev. D \textbf{83}, 034026 (2011)
doi:10.1103/PhysRevD.83.034026
[arXiv:1101.1130 [hep-ph]].

\bibitem{Chang:2005bf}
C.~H.~Chang, C.~F.~Qiao, J.~X.~Wang and X.~G.~Wu,
Phys. Rev. D \textbf{71}, 074012 (2005)
doi:10.1103/PhysRevD.71.074012
[arXiv:hep-ph/0502155 [hep-ph]].

\bibitem{Belle:2019bgi}
Y.~B.~Li \textit{et al.} [Belle],
Phys. Rev. D \textbf{100}, no.3, 031101 (2019)
doi:10.1103/PhysRevD.100.031101
[arXiv:1904.12093 [hep-ex]].

\bibitem{LHCb:2020gge}
R.~Aaij \textit{et al.} [LHCb],
Phys. Rev. D \textbf{102}, no.7, 071101 (2020)
doi:10.1103/PhysRevD.102.071101
[arXiv:2007.12096 [hep-ex]].

\bibitem{Zeng:2024yiv}
S.~Zeng, F.~Xu and Y.~Gu,
Eur. Phys. J. C \textbf{85}, no.2, 184 (2025)
doi:10.1140/epjc/s10052-025-13886-8
[arXiv:2406.02097 [hep-ph]].

\bibitem{Hsiao:2020gtc}
Y.~K.~Hsiao, L.~Yang, C.~C.~Lih and S.~Y.~Tsai,
Eur. Phys. J. C \textbf{80}, no.11, 1066 (2020)
doi:10.1140/epjc/s10052-020-08619-y
[arXiv:2009.12752 [hep-ph]].

\bibitem{LHCbRICHGroup:2012mgd}
M.~Adinolfi \textit{et al.} [LHCb RICH Group],
Eur. Phys. J. C \textbf{73}, 2431 (2013)
doi:10.1140/epjc/s10052-013-2431-9
[arXiv:1211.6759 [physics.ins-det]].

\bibitem{LHCb:2014nio}
R.~Aaij \textit{et al.} [LHCb],
JINST \textbf{10}, no.02, P02007 (2015)
doi:10.1088/1748-0221/10/02/P02007
[arXiv:1408.1251 [hep-ex]].

\bibitem{Belyaev:2015nlt}
I.~M.~Belyaev, E.~M.~Govorkova, V.~Y.~Egorychev and D.~V.~Savrina,
Moscow Univ. Phys. Bull. \textbf{70}, no.6, 497-503 (2015)
doi:10.3103/S0027134915060077

\bibitem{Govorkova:2015vqa}
E.~Govorkova,
Phys. Atom. Nucl. \textbf{79}, no.11-12, 1474-1476 (2016)
doi:10.1134/S1063778816100070
[arXiv:1505.02960 [hep-ex]].
\end{thebibliography}
\end{document}